\documentclass[a4paper,11pt]{JHEP3}

\usepackage{amsmath}
\usepackage{amssymb}
\usepackage{citesort}
\usepackage{graphics,graphicx}
\usepackage{epsf}

\def\bbra#1{\left\langle\!\left\langle#1\right|\right. }
\def\kket#1{\left. \left|#1\right\rangle\!\right\rangle}
\def\pprod#1#2{\langle\hspace*{-2pt}\langle #1| #2\rangle\hspace*{-2pt}\rangle}

\author{
Damian Chor\c{a}\.zkiewicz\footnote{e-mail: damian.ch@ift.uni.wroc.pl}\\
    Institute of Theoretical Physics,
    University of Wroc{\l}aw,
    pl. M. Borna 1, \\ 95-204~Wroc{\l}aw, Poland
    }
\author{
Leszek Hadasz\footnote{e-mail: hadasz@th.if.uj.edu.pl;\ corresponding author} \\
    M. Smoluchowski Institute of Physics,
    Jagiellonian University,
    W.~Reymonta 4,
    30-059~Krak\'ow, Poland
    \\
    \\}

\abstract{We construct, generalizing appropriately the method applied by J.~Teschner in the case of
the Virasoro conformal blocks,
the braiding and fusion matrices of the Neveu-Schwarz super-conformal blocks. Their properties
allow for an explicit verification of the bootstrap equation in the NS sector of the
$N=1$ supersymmetric Liouville field theory.
\vspace*{1cm}}

\title{Braiding and fusion properties of the Neveu-Schwarz super-conformal blocks}

\preprint{}

\begin{document}
\section{Introduction}
In the BPZ approach to the two dimensional Conformal Field Theory \cite{Belavin:1984vu} the r\^ole of a dynamical
principle is played by the associativity of the operator algebra involved or, equivalently,
by the bootstrap (crossing symmetry) equation imposed on the correlation functions. In every specific
CFT with known spectrum and three point coupling constants,
validity of the bootstrap equation can be regarded as the basic consistency condition.

Analysis of the bootstrap equation becomes particulary interesting (even if difficult) for the CFT-s with
continuous spectrum. One of a few interacting and solvable models  of this kind
is the Liouville Field Theory.
Its three point coupling constants have been found independently
by Dorn and Otto \cite{Dorn:1994xn} and by A.\ Zamolodchikov and Al.\ Zamolodchikov
\cite{Zamolodchikov:1995aa}. The authors of \cite{Zamolodchikov:1995aa}
also performed some numerical checks of the bootstrap equation in the LFT
using a recursive representation of conformal blocks developed in a series of papers by Al.\ Zamolodchikov
\cite{Zamolodchikov:1,Zamolodchikov:2,Zamolodchikov:3}.

An analytic proof of this equation which combined  a Moore-Seiberg formalism of CFT
\cite{Moore:1988qv} with a representation theory of quantum groups
has been presented in \cite{Ponsot:1999uf,Ponsot:2000mt}.
Using the results on fusion of degenerate representation of the Virasoro algebra
the authors of \cite{Ponsot:1999uf,Ponsot:2000mt}
derived from the consistency conditions of the Moore-Seiberg type a set of
functional equations for the fusion matrix of the conformal blocks. These equations were then shown
to be satisfied by the Racah-Wiegner coefficients
for an appropriate continuous series of representations of U${}_q({\rm sl}(2,{\mathbb R})).$
Another proof of the validity of bootstrap equation for the Liouville field theory,
which relies on an explicit calculation of the fusion matrix for the conformal blocks
appearing in the LFT by relating it to the braiding matrix of Virasoro chiral vertex operators,
was presented in \cite{Teschner:2001rv,Teschner:2003en}.

Conformal field theory  with $N=1$ supersymmetry
\cite{Friedan:1984rv,Bershadsky:1985dq,Zamolodchikov:1988nm}
is in some sense the simplest generalization of the ``ordinary'' CFT.
Also here the bulk three point coupling constants of the basic interacting model ---
supersymmetric extension of the Liouville theory --- are known
\cite{Poghosian:1996dw} and some numerical checks of the
bootstrap equation in the Neveu-Schwarz sector of the theory (which employed a recursive representation
of $N=1$ NS blocks developed in \cite{Hadasz:2006qb,Belavin:2006zr})
have been performed \cite{Belavin:2007gz,Belavin:2007eq}.  However, an analytical
proof of the consistency of the $N=1$ supersymmetric Liouville theory
is still missing.

A step towards such a proof was taken in \cite{Hadasz:2007wi}, where the form of
the fusion matrix for the Neveu-Schwarz superconformal blocks was postulated.
If correct, it implies  the bootstrap equation for four super-primary NS Liouville fields.
The basic goal of the preset work is to justify the results of \cite{Hadasz:2007wi} by
calculating explicitly (in the spirit of \cite{Teschner:2001rv,Teschner:2003en}) the
braiding matrices of the NS chiral operators and relating them to the fusion matrices of
NS superconformal blocks.

The paper is organized as follows. In the Section \ref{Bootstrap} we briefly discuss the Neveu-Schwarz algebra
and recall the basic facts on the $N=1$ supersymmetric Liouville field theory. This section is in a sense
introductory and  mainly meant to establish a convenient notation. In the Section \ref{Chiral:ops}
we discuss the structure of the highest weight NS super-moduli, define the NS chiral vertex operators
and postulate for them the existence of a braiding relation.  Section \ref{section:chiral:scalar}, which
constitutes the main part of the present work, is devoted to an explicit construction of the
braiding matrix of the NS chiral vertex operators by relating it to an exchange matrix of the screened, normal
ordered exponential build up from the modes of the chiral superscalar field. In  the Section \ref{sect:braid:fus}
we introduce the NS blocks and discuss their braiding and fusion properties.
Coincidence of the result  with the fusion
matrix ``guessed'' in \cite{Hadasz:2007wi} may be viewed as a proof of the validity of the bootstrap equation in
the NS sector of the supersymmetric Liouville field theory.
Finally, Appendix \ref{Appendix:Barnes}
contains some relevant properties of the Barnes double gamma function and Appendix \ref{Appendix:Weyl}
is devoted to a derivation of a Weyl-type representation of the screening charge operators.

\section{Bootstrap in the $N=1$ supersymmetric Liouville field theory}
\label{Bootstrap}
The $N=1$ supersymmetric Liouville field theory
(see \cite{Nakayama:2004vk} for an exhaustive review)
may be defined by the action
\begin{equation*}
\label{L:SLFT}
{\cal S}_{\rm\scriptscriptstyle SLFT}=
\int\!d^2z
\left(\frac{1}{2\pi}\left|\partial\phi\right|^2
+
\frac{1}{2\pi}\left(\psi\bar\partial\psi + \bar\psi\partial\bar\psi\right)
+
2i\mu b^2\bar\psi\psi {\rm e}^{b\phi}
+
2\pi b^2\mu^2{\rm e}^{2b\phi}\right),
\end{equation*}
where $\phi$ is a bosonic and $\psi$ a fermionic field,
$\mu$ denotes a two-dimensional cosmological constant and $b$ is a Liouville coupling constant.

The superconformal symmetry of the SLFT (as well an of any other $N=1$ superconformal filed theory)
is generated by a pair of holomorphic currents
$T(z),\, S(z)$ and their anti-holomorphic counterparts ${\overline T}(\bar z),\,{\overline S}(\bar z),$
where $T$ and ${\overline T}$ are components of the energy-momentum tensor while $S$  and ${\overline S}$
have dimensions $(3/2,0)$ and $(0,3/2),$ respectively.
The algebra of the modes of $T(z)$ and $S(z)$ is
determined by the operator product expansion (OPE):
\begin{eqnarray}
\label{OPE:TS}
\nonumber
T(z)T(0) & = & \frac{c}{2z^4} + \frac{2}{z^2}T(z) + \frac{1}{z}\partial T(0) + \ldots,
\\
T(z)S(0) & = & \frac{3}{2z^2}S(0) + \frac{1}{z}\partial S(0)+ \ldots,
\\
\nonumber
S(z)S(0) & = & \frac{2c}{3z^3} + \frac{2}{z}T(0) + \ldots\,.
\end{eqnarray}
The fields local with respect to $S(z),$  i.e.\ with the OPE
\[
S(z)\phi_{\rm NS}(0,0) \; = \; \sum\limits_{k\in {\mathbb Z} + \frac12} z^{k-\frac32}S_{-k}\phi_{\rm NS}(0,0)
\]
form the Neveu-Schwarz (or NS for brevity) subspace in a space of fields of a given SCFT on which we shall focus in the present paper.
Together with the usual Virasoro generators $L_n$ defined by the OPE
\[
T(z)\phi_{\rm NS}(0,0) \; = \; \sum\limits_{n\in {\mathbb Z}} z^{n-2}L_{-n}\phi_{\rm NS}(0,0),
\]
$S_k$ form the Neveu-Schwarz algebra determined by (\ref{OPE:TS}),
\begin{eqnarray}
\label{NS}
\nonumber
\left[L_m,L_n\right] & = & (m-n)L_{m+n} +\frac{c}{12}m\left(m^2-1\right)\delta_{m+n},
\\
\left[L_m,S_k\right] & = &\frac{m-2k}{2}S_{m+k},
\\
\nonumber
\left\{S_k,S_l\right\} & = & 2L_{k+l} + \frac{c}{3}\left(k^2 -\frac14\right)\delta_{k+l}.
\end{eqnarray}
It is convenient to parameterize the central charge $c$ of the NS algebra as
\begin{equation*}
\label{c:through:Q}
c = \frac32 + 3Q^2,
\end{equation*}
where in the case of SLFT the ``background charge'' $Q$ my be expressed through the Liouville coupling constant as
\[
Q = b + b^{-1}.
\]

In the space of NS fields  there exist ``super-primary'' fields,
realized in
the supersymmetric Liouville theory as an appropriately normal ordered exponents
$V_{a}(z,\bar z) =  {\rm e}^{a\varphi(z,\bar z)}.$
By definition they satisfy:
\begin{eqnarray*}
\label{primary}
\nonumber
\left[L_n,V_{a}(0,0)\right]
& = &
\left[S_k,V_a(0,0)\right]
\; = \; 0,
\hskip 1cm
n,k > 0,
\\[6pt]
\left[L_0,V_{a}(0,0)\right]
& = &
\Delta_{a} V_{a}(0,0),
\hskip 5mm
\Delta_a = \frac12a(Q-a).
\end{eqnarray*}
Each super-primary
field is the ``lowest'' component of the superfield
\begin{equation}
\label{NS:superfield}
\Phi_{a}(z,\theta;\bar z,\bar\theta)
\; = \;
V_{a}(z,\bar z)
+
\theta\,\Lambda_{a}(z,\bar z)
+
\bar\theta\,\bar\Lambda_{a}(z,\bar z)
-
\theta\bar\theta\,\widetilde V_{a}(z,\bar z),
\end{equation}
where
\[
\Lambda_{a}
\; = \;
\left[S_{-1/2},V_{a}\right],
\hskip 5mm
\bar\Lambda_{a}
\; = \;
\left[\bar S_{-1/2},V_{a}\right],
\hskip 5mm
\widetilde V_{a}
\; = \;
\left\{S_{-1/2},\left[\bar S_{-1/2},V_{a}\right]\right\},
\]
and $\theta,\bar\theta$ are Grassman numbers. Global superconformal
transformations
(generated by $L_0,S_{\pm\frac12},L_{\pm 1}$ and their right counterparts)
 allow to express three-point function of primary superfields in the form:
\begin{eqnarray*}
\nonumber
&&
\hspace*{-1.5cm}
\Big\langle
\Phi_{a_3}(z_3,\theta_3;\bar z_3,\bar\theta_3)
\Phi_{a_2}(z_2,\theta_2;\bar z_2,\bar\theta_2)
\Phi_{a_1}(z_1,\theta_1;\bar z_1,\bar\theta_1)
\Big\rangle
\\
&& = \;
Z_{32}^{\gamma_1}\bar Z_{32}^{\bar \gamma_1}\,
Z_{31}^{\gamma_2}\bar Z_{31}^{\bar \gamma_2}\,
Z_{21}^{\gamma_3}\bar Z_{21}^{\bar \gamma_3}\
\Big\langle
\Phi_{a_3}(\infty,0;\infty,0)
\Phi_{a_2}(1,\Theta;1,\bar\Theta)
\Phi_{a_1}(0,0;0,0)
\Big\rangle,
\end{eqnarray*}
where $\gamma_i = 2\Delta_i-(\Delta_1 +\Delta_2 +\Delta_3),\ $  $Z_{ij} = z_i-z_j - \theta_i\theta_j \equiv z_{ij} - \theta_i\theta_j,$
\[
\Theta \; = \; \frac{1}{\sqrt{z_{12}z_{13}z_{23}}}
\left(\theta_1 z_{23} +\theta_2 z_{31} + \theta_3 z_{12} - \frac12\theta_1\theta_2\theta_3\right),
\]
is an odd invariant of the global superconformal group and
\[
\Phi_{a_3}(\infty,0;\infty,0) \equiv \;  \lim_{R\to\infty} R^{2\Delta_3+2\bar\Delta_3}\Phi_{a_3}(R,0;R,0).
\]
The three point function is thus determined
by the superconformal symmetry up to
two independent constants,
\begin{eqnarray*}
\label{constant1}
C(a_3,a_2,a_1) & = & \big\langle V_{a_3}(\infty,\infty) V_{a_2}(1,1) V_{a_1}(0,0)\big\rangle,
\\[6pt]
\label{constant2}
\widetilde C(a_3,a_2,a_1) & = & \big\langle V_{a_3}(\infty,\infty) \widetilde V_{a_2}(1,1) V_{a_1}(0,0)\big\rangle.
\end{eqnarray*}
Their form in the  $N=1$ supersymmetric
Liouville field theory,
\begin{eqnarray*}
C(a_3,a_2,a_1)
& = &
C_0(a)\,
\frac{\Upsilon_{\rm NS}(2a_3)\Upsilon_{\rm NS}(2a_2)\Upsilon_{\rm NS}(2a_1)}
{\Upsilon_{\rm NS}(a-Q)\Upsilon_{\rm NS}(a_{1+2-3})\Upsilon_{\rm NS}(a_{2+3-1})\Upsilon_{\rm NS}(a_{3+1-2})},
\\[10pt]
\widetilde C(a_3,a_2,a_1)
& = &
2i\,C_0(a)\,
\frac{\Upsilon_{\rm NS}(2a_3)\Upsilon_{\rm NS}(2a_2)\Upsilon_{\rm NS}(2a_1)}
{\Upsilon_{\rm R}(a-Q)\Upsilon_{\rm R}(a_{1+2-3})\Upsilon_{\rm R}(a_{2+3-1})\Upsilon_{\rm R}(a_{3+1-2})},
\end{eqnarray*}
with
\[
C_0(a) =
\left(\pi\mu\gamma\left(\frac{bQ}{2}\right)b^{1-b^2}\right)^{\frac{Q-a}{b}}\Upsilon'_{\rm NS}(0),
\]
was first derived in \cite{Poghosian:1996dw}. Here
$a \equiv a_1 + a_2+a_3,\ a_{1+2-3} \equiv a_1 +a_2 -a_3,$ etc.\
and the special functions involved ($\Upsilon_{\rm NS,R}(x), G_{\rm NS,R}(x)$ below etc.)
are defined in Appendix \ref{Appendix:Barnes}.

The super-projective transformations also
allow to express a generic function of
four superfields (\ref{NS:superfield})
through the four-point functions of the form
\begin{equation}
\label{four:point:superfields}
\Big\langle
\Phi_{a_4}(\infty,0;\infty,0)
\Phi_{a_3}(1,\theta_3;1,\bar\theta_3)
\Phi_{a_2}(z,\theta_2;\bar z,\bar\theta_2)
\Phi_{a_1}(0,0;0,0)
\Big\rangle,
\end{equation}
If we denote by
${\cal F}_{\!a_s}^{\rm e}\!\left[^{a_3\: a_2}_{a_4\: a_1}\right]\!(z)$
and
${\cal F}_{\!a_s}^{\rm o}\!\left[^{a_3\: a_2}_{a_4\: a_1}\right]\!(z)$
a pair (out of four) of an even and an odd $N=1$ Neveu-Schwarz blocks\footnote{
See \cite{Belavin:2007gz,Hadasz:2006qb} or Section \ref{sect:braid:fus} for the definitions.}
then a special case of (\ref{four:point:superfields}), the four point function of super-primary fields
\[
G_4(z,\bar z)
=
\big\langle
V_{a_4}(\infty,\infty)
V_{a_3}(1,1)
V_{a_2}(z,\bar z)
V_{a_1}(0,0)
\big\rangle,
\]
can be presented
either in the ``$s-$channel'':
\begin{eqnarray}
\label{s:channel:decomposition}
\nonumber
G_4(z,\bar z)
& = &
\int\limits_{\mathbb S} \!\frac{da_s}{i}
\left[
C(a_4,a_3,a_s)C(\bar a_s,a_2,a_1)
\left|{\cal F}_{\!a_s}^{\rm e}\!\left[^{a_3\: a_2}_{a_4\: a_1}\right]\!(z)\right|^2
\right.
\\[-12pt]
\\[-6pt]
\nonumber
&&
\hskip 1cm
\left.
-\
\widetilde C(a_4,a_3,a_s)\widetilde C(\bar a_s,a_2,a_1)
\left|{\cal F}_{\!a_s}^{\rm o}\!\left[^{a_3\: a_2}_{a_4\: a_1}\right]\!(z)\right|^2
\right]
\end{eqnarray}
with ${\mathbb S} = \frac{Q}{2} + i{\mathbb R}_+,$ or in the ``$t-$channel'' decomposition:
\begin{eqnarray}
\label{t:channel:decomposition}
\nonumber
G_4(z,\bar z)
& = &
\int\limits_{\mathbb S} \!\frac{da_t}{i}
\left[
C(a_4,a_t,a_1)C(\bar a_t,a_2,a_3)
\left|{\cal F}_{\!a_t}^{\rm e}\!\left[^{a_1\: a_2}_{a_4\: a_3}\right]\!(1-z)\right|^2
\right.
\\[-12pt]
\\[-6pt]
\nonumber
&&
\hskip 1cm
\left.
-\
\widetilde C(a_4,a_t,a_1)\widetilde C(\bar a_t,a_2,a_3)
\left|{\cal F}_{\!a_t}^{\rm o}\!\left[^{a_1\: a_2}_{a_4\: a_3}\right]\!(1-z)\right|^2
\right].
\end{eqnarray}
Here and in what follows we use a convenient notation $\bar a = Q-a$ (notice that for $a \in \frac{Q}{2} + i{\mathbb R}$
it is indeed the complex conjugate of $a$). If we now define a fusion matrix $\sf F$ by assuming the existence of a relation
between the blocks appearing in the decompositions above,
\[
{\cal F}^{\eta}_{a_s}\!\left[^{a_3\:a_2}_{a_4\:a_1}\right]\!(z)
=
\int\limits_{\mathbb S}\!\frac{da_t}{2i}\!
\sum\limits_{\rho = {\rm e,o}}
{\sf F}_{a_sa_t}\!\left[^{a_3\:a_2}_{a_4\:a_1}\right]^{\eta}_{\hskip 5pt\rho}
{\cal F}^{\rho}_{a_t}\!\left[^{a_1\:a_2}_{a_4\:a_3}\right]\!(1-z),
\hskip 5mm
\eta = {\rm e,o},
\]
then the coincidence of the decompositions (\ref{s:channel:decomposition}) and (\ref{t:channel:decomposition})
may be recast in the form
\begin{eqnarray}
\label{orthogonality:F}
\nonumber
&&
\hskip -2cm
\int\limits_{{\mathbb S}}\!
\frac{da_s}{i}\
\left({\sf F}_{a_sa_t}\!\left[^{a_3\: a_2}_{a_4\: a_1}\right]\right)^\dagger
\cdot
{\sf C}(a_4,a_3,a_s) \cdot \tau_3 \cdot {\sf C}(\bar a_s,a_2,a_1)
\cdot
{\sf F}_{a_sa'_t}\!\left[^{a_3\: a_2}_{a_4\: a_1}\right]
\\[-6pt]
&&
\hskip 1.1cm = \;\
{\sf C}(a_4,a_t,a_1) \cdot \tau_3 \cdot {\sf C}(\bar a_t,a_3,a_2)\
i\delta(a_t-a'_t).
\end{eqnarray}
where we denoted
\[
{\sf C}(a_3,a_2,a_1)
\;\ = \;\
\left(
\begin{array}{cc}
C(a_3,a_2,a_1) & 0
\\[4pt]
0 & \widetilde C(a_s,a_2,a_1)
\end{array}
\right),
\hskip 1cm
\tau_3 =
\left(
\begin{array}{rr}
1 & 0 \\ 0 & -1
\end{array}
\right).
\]
Coincidence of the $s-$ and $t-$channel representations of the four-point
correlation function, or (equivalently)  Eq.\ (\ref{orthogonality:F}), constitutes the bootstrap equation
for the super-primary fields. Analogous procedure
can be applied also to the other four-point correlation functions
appearing in (\ref{four:point:superfields}), what results in the remaining bootstrap equations
for NS sector of the supersymmetric Liouville field theory.

\section{Chiral vertex operators}
\label{Chiral:ops}

\subsection{Neveu-Schwarz super-module}
Let $\varphi_{a}(0,0)$ denotes a NS superprimary field.
States
\begin{equation}
\label{HWS:def}
\nu_a = \varphi_{a}(0,0)|0\rangle,
\end{equation}
obtained through its action on the invariant vacuum $|0\rangle,$ with
\[
L_n|0\rangle = S_k|0\rangle = 0,
\hskip 1cm
n \geqslant -1,
\hskip 3mm
k \geqslant -{\textstyle \frac12},
\]
are of the highest weight with respect to the NS
algebra (\ref{NS}):
\begin{equation}
\label{highest}
L_0\nu_{a}=\Delta_a\nu_{a}\,,
\hskip 5mm
L_n\nu_{a}= S_k\nu_{a} =0,
\hskip 5mm
n, k  > 0.
\end{equation}
We shall frequently write $\nu_0$ instead of $|0\rangle;$ this is consistent with (\ref{HWS:def}) since
the unique super-primary field with the conformal weight $0$ is the identity operator, $\varphi_0(0,0) = {\mathbf 1}.$

Denote by ${\cal V}^{f}_a$ the free vector space  generated by all vectors
of the form
\begin{equation}
\label{basis}
\nu_{a,NK}
\; = \;
L_{-N}S_{-K} \nu_{a}
\; \equiv \;
L_{-n_j}\ldots L_{-n_1}S_{-k_i}\ldots S_{-k_1}\nu_{a}\,,
\end{equation}
where
 $K = \{k_1,k_2,\ldots,k_i\}$ and
 $N = \{n_1,n_2,\ldots,n_j\}$ are
arbitrary ordered sets of  indices
\[
k_i > \ldots > k_2 > k_1 ,
\hskip 1cm
n_j \geqslant \ldots \geqslant n_2 \geqslant n_1,
\]
such that
$
|K|+|N|
\equiv
k_1+\dots+k_i+n_1+\dots+n_j = f.
$

The $\frac12\mathbb{Z}$-graded representation of the NS  algebra,
determined on the space
$$
{\cal V}_a\
=
\bigoplus\limits_{{f}\in \frac12\mathbb{N}}
{\cal V}^{f}_a\,,
\hskip 5mm
{\cal V}^0_a=\mathbb{C}\, \nu_a\,,
$$
by the relations
(\ref{NS}) and (\ref{highest}), is called the NS supermodule
of the highest weight $\Delta_a$ and the central charge $c$
(to avoid making the notation overloaded we omit the subscript $c$ at $\cal V$).
Each ${\cal V}^{f}_a$ is an eigenspace of $L_0$ with the eigenvalue
$\Delta_a +{f}$. The space ${\cal V}_a$ has also a natural $\mathbb{Z}_2$-grading:
$$
{\cal V}_a
\; = \;
{\cal V}^+_a \oplus
{\cal V}^-_a\,,
\hskip 5mm
{\cal V}^+_a
\; =
\bigoplus\limits_{n\in \mathbb{N}}
{\cal V}^n_a\,,
\hskip 5mm
{\cal V}^-_a
\; = \hskip -5pt
\bigoplus\limits_{k\in  \mathbb{N}+\frac12} \hskip -5pt
{\cal V}^k_a\,,
$$
where ${\cal V}^\pm_a$ are eigenspaces of the
parity operator $(-1)^{\sf F}= (-1)^{2(L_0-\Delta_a)}$.
Finally, there exists on ${\cal V}_a$ a natural, symmetric, bilinear form
\[
\langle\,\cdot\,|\,\cdot\,\rangle_{a}
\hskip 5pt: \hskip 5pt
{\cal V}_a \times {\cal V}_a \ \to \ {\mathbb C},
\]
uniquely determined  by the algebra (\ref{NS}), normalization
$\langle\nu_{a}|\nu_{a}\rangle_{a} =1$ and the relations
$(L_{n})^{\dagger}=L_{-n},\ (S_{k})^{\dag}=S_{-k}.$ In what follows we
will usually suppress the index $a$ at $\langle\,\cdot\,|\,\cdot\,\rangle_{a}.$

\subsection{The vertex}
\label{the:vertex}

The NS chiral vertex operator ${}^{\mathbf N\!}V_{\mathbb A}(z),$ where ${\mathbb A} = \left(^{\;\: a_2}_{a_3\:a_1}\!\right)$
and $z\in \mathbb C,$
is a  linear map from ${\cal V}_{a_1} \equiv {\cal V}_1$ to ${\cal V}_3$ which may be defined by the following conditions\footnote{
The basic facts on the vertex operators can be found in \cite{Moore:1988qv};
\cite{Frenkel} and \cite{Kac2} can be consulted for a clear and extensive
introduction to the subject. The presented formulation parallels the one used in \cite{Teschner:2001rv}.}:
\begin{enumerate}
\item
${}^{\mathbf N\!}V_{\mathbb A}(z)$ is a sum of an even (i.e.\ parity preserving) and an odd (i.e.\ parity reversing) operators,
\[
{}^{\mathbf N\!}V_{\mathbb A}(z) = {}^{\mathbf N\!}V^{\rm e}_{\mathbb A}(z)+{}^{\mathbf N\!}V^{\rm o}_{\mathbb A}(z),
\]
where
\(
{}^{\mathbf N\!}V^{\rm e}_{\mathbb A}(z)\;:\;{\cal V}_1^{\pm} \; \to \; {\cal V}_3^{\pm}
\)
and
\(
{}^{\mathbf N\!}V^{\rm o}_{\mathbb A}(z)\;:\;{\cal V}_1^{\pm} \; \to \; {\cal V}_3^{\mp}.
\)
\item
Let $\theta$ be an anticommuting variable,
\[
\{\theta,\theta\} = \{\theta,S_k\} = \{\theta,{}^{\mathbf N\!}V^{\rm o}_{\mathbb A}(z)\} =
[\theta,L_n] = \left[\theta,{}^{\mathbf N\!}V^{\rm e}_{\mathbb A}(z)\right] = 0.
\]
Then
\begin{eqnarray}
\label{vertex:def}
\nonumber
[L_n,{}^{\mathbf N\!}V_{\mathbb A}(z)]
& = &
z^n\left(z\partial_z + (n+1)\Delta_2\right)\,{}^{\mathbf N\!}V_{\mathbb A}(z),
\\[-6pt]
\\[-6pt]
\nonumber
\left[\theta S_k,{}^{\mathbf N\!}V_{\mathbb A}(z)\right]
& = &
z^{k+\frac12}\left[\theta S_{-1/2},{}^{\mathbf N\!}V_{\mathbb A}(z)\right].
\end{eqnarray}
\item
The commutation relations (\ref{vertex:def}) determine ${}^{\mathbf N\!}V_{\mathbb A}(z)$ up to two
arbitrary functions of the parameters $a_1,a_2,a_3.$ For $z \to 0:$
\begin{equation}
\label{chiral:norm:1}
{}^{\mathbf N\!}V_{\mathbb A}(z)\nu_1
\; = \;
z^{\Delta_3-\Delta_2-\Delta_1}
\left(
N^{\rm e}(a_3,a_2,a_1)\nu_3 + z^{\frac12}\frac{N^{\rm o}(a_3,a_2,a_1)}{2\Delta_3}S_{-\frac12}\nu_3  + {\cal O}(z)
\right),
\end{equation}
so that
\begin{equation*}
\label{chiral:norm:2}
N^{\rm e}(a_3,a_2,a_1) \; = \;  \left\langle\nu_3\big|{}^{\mathbf N\!}V_{\mathbb A}(z)\nu_1\right\rangle,
\hskip 3mm
N^{\rm o}(a_3,a_2,a_1) \; = \;  \big\langle S_{-\frac12}\nu_3\big|{}^{\mathbf N\!}V_{\mathbb A}(z)\nu_1\big\rangle.
\end{equation*}
\end{enumerate}
We shall define a {\em normalized} NS chiral vertex operator $V_{\mathbb A}(z)$ to be the NS chiral vertex operator with
$N^{\rm e}(a_3,a_2,a_1) =N^{\rm o}(a_3,a_2,a_1) = 1.$ Equivalently, for any ${}^{\mathbf N\!}V_{\mathbb A}(z):$
\begin{eqnarray}
\label{normalization:simple}
{}^{\mathbf N\!}V^{\rho}_{\mathbb A}(z)
& = &
N^{\rho}(a_3,a_2,a_1)\,V^{\rho}_{\mathbb A}(z),
\hskip 1cm
\rho = {\rm e,\,o}.
\end{eqnarray}

The operator ${}^{\mathbf N\!}V_{\mathbb A}(z)$ is naturally associated with the ground state
$\nu_2\in {\cal V}_{2}.$ It is useful to define a family of (generalized) NS chiral vertex operators
enumerated by  (and linear in) vectors $\xi \in {\cal V}_2.$ If we denote ${}^{\mathbf N\!}V_{a_3a_1}\!(\nu_2|z) =  {}^{\mathbf N\!}V_{\mathbb A}(z)$
then, first of all,
\begin{eqnarray}
\label{decendans:def:1}
\nonumber
\theta\, {}^{\mathbf N\!}V_{a_3a_1}\!(S_{-1/2}\nu_2|z) & = & \left[\theta S_{-1/2},{}^{\mathbf N\!}V_{a_3a_1}\!(\nu_2|z)\right],
\\[-6pt]
\\[-6pt]
\nonumber
{}^{\mathbf N\!}V_{a_3a_1}\!(L_{-1}\xi|z)
& = & \frac{\partial}{\partial z}\, {}^{\mathbf N\!}V_{a_3a_1}\!(\xi|z), \hskip 5mm \xi \in {\cal V}_2.
\end{eqnarray}
Moreover,
for $k \in {\mathbb Z}+\frac12,\ k \geqslant 3/2:$
\begin{eqnarray*}
\theta\, {}^{\mathbf N\!}V_{a_3a_1}\!(S_{-k}\xi|z)
& = &
\frac{1}{\left(k-\frac32\right)!}
\left(\frac{\partial}{\partial w}\right)^{k-\frac32}
:\!\theta S(w)\,{}^{\mathbf N\!}V_{a_3a_1}(\xi|z)\!:\Big|_{w=z}
\end{eqnarray*}
where
\begin{eqnarray*}
:\!\theta S(w)\,{}^{\mathbf N\!}V_{a_3a_1}\!(\xi|z)\!:
& = &
\Big(\sum\limits_{l\leqslant -\frac32} \theta S_l\, w^{-l-\frac32}\Big){}^{\mathbf N\!}V_{a_3a_1}\!(\xi|z)
+
{}^{\mathbf N\!}V_{a_3a_1}\!(\xi|z)\Big(\sum\limits_{l\geqslant -\frac12}\theta S_l\, w^{-l-\frac32}\Big),
\end{eqnarray*}
with $l \in {\mathbb Z}+\frac12$ and, for $m \geqslant 2:$
\begin{eqnarray*}
{}^{\mathbf N\!}V_{a_3a_1}\!(L_{-m}\xi|z)
& = &
\frac{1}{(m-2)!}
\left(\frac{\partial}{\partial w}\right)^{m-2}
:\!
T(w)\, {}^{\mathbf N\!}V_{a_3a_1}\!(\xi|z)
\!:
\Big|_{w=z}
\end{eqnarray*}
with
\[
:\!
T(w)\, {}^{\mathbf N\!}V_{a_3a_1}\!(\xi|z)
\!:
\; = \;
\Big(\sum\limits_{n\leqslant -2} L_n\, w^{-n-2}\Big) {}^{\mathbf N\!}V_{a_3a_1}\!(\xi|z)
+
{}^{\mathbf N\!}V_{a_3a_1}\!(\xi|z)\Big(\sum\limits_{n\geqslant -1}L_n\, w^{-n-2}\Big).
\]
The state-operator correspondence above is build in a way which ensures that also the operator-state
correspondence,
\[
\forall\, \xi\in {\cal V}_a: \;\;
\lim_{z\to 0}V_{a0}(\xi|z)\nu_0 = \xi,
\]
holds.

Since we shall frequently use the generalized chiral vertex operator
associated with the vector $S_{-1/2}\nu_2$ it is convenient to reserve for it a special notation and write
\begin{eqnarray*}
\label{star:definition}
{}^{\mathbf N\!}V_{a_3a_1}\!(S_{-1/2}\nu_2|z)
& \equiv &
{}^{\mathbf N\!}V_{a_3a_1}\!(*\nu_2|z).
\end{eqnarray*}
It then follows from (\ref{vertex:def}) that
\begin{eqnarray}
\label{comm:with:star}
\nonumber
  \left[ \mathrm{L}_m, {}^{\mathbf N\!}V_{a_3a_1}(\nu_2 | z)  \right]
  &=& z^m \left( z \partial_z + (m+1) \Delta_2 \right){}^{\mathbf N\!}V_{a_3a_1}(\nu_2 | z),
\\[4pt]
\nonumber
 \left[ \mathrm{L}_m, {}^{\mathbf N\!}V_{a_3a_1}(*\nu_2 | z)  \right]
    &=& z^m \left( z \partial_z + (m+1) \left(\Delta_2 + \textstyle \frac{1}{2}\right)  \right)
    {}^{\mathbf N\!}V_{a_3a_1}(*\nu_2 | z),
\\[10pt]
 \nonumber
  \left[ \mathrm{S}_k, {}^{\mathbf N\!}V^{\rm  e}_{a_3a_1}(\nu_2 | z)  \right]
  &=& z^{k+ \frac{1}{2}}\ {}^{\mathbf N\!}V^{\rm o}_{a_3a_1}(*\nu_2 | z),
\\[4pt]
\left\{ \mathrm{S}_k, {}^{\mathbf N\!}V^{\rm o}_{a_3a_1}(\nu_2 | z)  \right\}
  &=& z^{k+ \frac{1}{2}}\ {}^{\mathbf N\!}V^{\rm e}_{a_3a_1}(*\nu_2 | z),
\\[10pt]
\nonumber
 \left[ \mathrm{S}_k, {}^{\mathbf N\!}V^{\rm e}_{a_3a_1}(*\nu_2 | z)  \right]
    &=& z^{k - \frac{1}{2}} \left( z \partial_z + \Delta_2 (2k+1) \right)
   \ {}^{\mathbf N\!}V^{\rm o}_{a_3a_1}(\nu_2 | z),
\\[4pt]
\nonumber
 \left\{ \mathrm{S}_k, {}^{\mathbf N\!}V^{\rm o}_{a_3a_1}(*\nu_2 | z)  \right\}
    &=& z^{k - \frac{1}{2}} \left( z \partial_z + \Delta_2 (2k+1) \right)
   \ {}^{\mathbf N\!}V^{\rm e}_{a_3a_1}(\nu_2 | z).
\end{eqnarray}
Consequently
\begin{eqnarray*}
&&
\hskip -.7cm
{}^{\mathbf N\!}V_{a_3a_1}(*\nu_2 | z)\nu_1\; =
\\[6pt]
\nonumber
&&
z^{\Delta_3-\Delta_2-\Delta_1-\frac12}
\left(
N^{\rm o}(a_3,a_2,a_1)\nu_3
+
z^{\frac12}(\Delta_3+\Delta_2-\Delta_1)N^{\rm e}(a_3,a_2,a_1)S_{-\frac12}\nu_3
+
{\cal O}(z)
\right),
\end{eqnarray*}
so that
\begin{eqnarray}
\label{normalization:star}
{}^{\mathbf N\!}V^{\rho}_{a_3a_1}(*\nu_2 | z)
& = &
N^{\bar\rho}(a_3,a_2,a_1)\,V^{\rho}_{a_3a_1}(*\nu_2 | z),
\end{eqnarray}
where
\(
\bar{\rm e} = {\rm o},\
\bar{\rm o} = {\rm e}.
\)

Let us define a braiding matrix to be an integral kernel which appears in
the operator identity\footnote{
For the rational CFT, where the integral in (\ref{braiding:definition}) is replaced by a finite sum,
the existence of the braiding matrix can be  proven, \cite{Moore:1988qv,Felder:1989wr}.
For the $N=1$ SLFT the assumption that a braiding matrix satisfying (\ref{braiding:definition}) does exists will be
justified by it explicit construction --- the primary goal of this work.}
\begin{equation}
\label{braiding:definition}
V^{\rho}_{a_4a_s}(\xi_3|z_3)V^{\eta}_{a_sa_1}(\xi_2|z_2)
=
\int\limits_{\mathbb S}\!\frac{da_u}{2i}\hskip -4pt
\sum\limits_{\lambda,\delta =\, {\rm e,o}}\hskip -4pt
{\sf B}^{\epsilon}_{a_sa_u}\!\!\left[^{\xi_3\:\xi_2}_{a_4\:a_1}\right]^{\rho\eta}_{\hskip 8pt \lambda\delta}
V^{\lambda}_{a_4a_u}(\xi_2|z_2)V^{\delta}_{a_ua_1}(\xi_3|z_3),
\end{equation}
where $\epsilon = {\rm sign(Arg}\,z_{32})$ with ${\rm sign}(x) = +1$ for $x > 0$ and $-1$ for $x<0,$ and
in the case relevant for the supersymmetric Liouville field theory ${\mathbb S} = \frac{Q}{2} + i{\mathbb R}$.
It follows from (\ref{comm:with:star}) and the definition of $V^\rho_{a_3a_1}(\xi|z)$ that every matrix element of
\(
V^{\rho}_{a_4a_k}(\xi_i|z_i)V^{\eta}_{a_ka_1}(\xi_j|z_j)
\)
can be expressed as a linear differential operator (in $z_i$ and $z_j$) acting on a
matrix element of the product
\(
V^{\rho}_{a_4a_k}(\underline{\hspace*{6pt}}\nu_i|z_i)V^{\eta}_{a_ka_1}(\underline{\hspace*{6pt}}\nu_j|z_j),
\)
where $\underline{\hspace*{6pt}}\nu_l$ denotes
$\nu_l$ or $*\nu_l \equiv S_{-1/2}\nu_l.$ Because the braiding matrix in (\ref{braiding:definition})
does not depend on $z_2$ and $z_3,$ it is then equal to one of the matrices appearing in the
braiding relations for the operators $V^{\rho}_{a_4a_k}(\underline{\hspace*{6pt}}\nu_i|z_i)$ and
$V^{\eta}_{a_ka_1}(\underline{\hspace*{6pt}}\nu_j|z_j):$
\begin{equation}
\label{braiding:definition:2}
V^{\rho}_{a_4a_s}(\underline{\hspace*{6pt}}\nu_3|z_3)V^{\eta}_{a_sa_1}(\underline{\hspace*{6pt}}\nu_2|z_2)
=
\int\limits_{\mathbb S}\!\frac{da_u}{2i}\hskip -4pt
\sum\limits_{\lambda,\delta =\, {\rm e,o}}\hskip -4pt
{\sf B}^{\epsilon}_{a_sa_u}\!\!
\left[^{\underline{\hspace*{5pt}}a_3\:\underline{\hspace*{5pt}}a_2}_{\hspace*{5pt}a_4\:\hspace*{5pt}a_1}
\right]^{\rho\eta}_{\hskip 8pt \lambda\delta}
V^{\lambda}_{a_4a_u}(\underline{\hspace*{6pt}}\nu_2|z_2)V^{\delta}_{a_ua_1}(\underline{\hspace*{6pt}}\nu_3|z_3),
\end{equation}
where, in order to make the notation uniform, we have chosen as the arguments of the braiding
matrix $\underline{\hspace*{6pt}}a_i$ rather then $\underline{\hspace*{6pt}}\nu_i.$

Graphically, we shall denote the braiding ``move'' as

\vskip 5mm

\noindent
\centerline{
\includegraphics*[width=.5\textwidth]{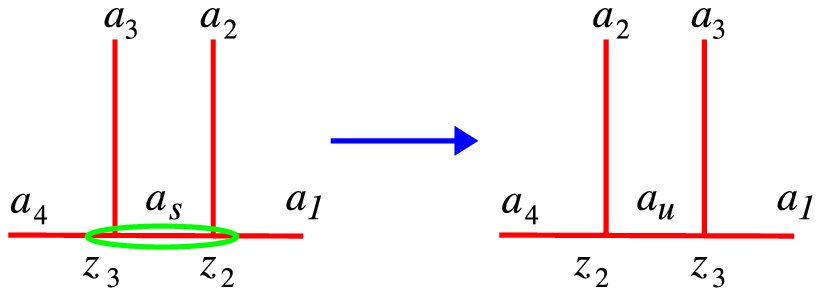}
}
\vskip 2mm

\centerline{\small{\bf Fig.~1}~ Graphical notation for the braiding transformation}

\section{Chiral superscalar}
\label{section:chiral:scalar}

It turns out to be  difficult to derive an explicit form or the braiding matrix just from
algebraic properties of the NS chiral vertex operators.
We shall therefore construct in this section (appropriately modifying the
Teschner's procedure for the Virasoro case) a rather explicit realization of the NS
chiral  vertex on the tensor product of a free chiral scalar
and a free chiral fermion Hilbert spaces. With the properties of the braiding matrix
derived we will then proceed in
Section \ref{sect:braid:fus} to a discussion of the braiding and fusion
properties of the Neveu-Schwarz conformal blocks.

\subsection{Chiral fields}
Let us define for $\sigma\in {\mathbb R}$ the chiral scalar field
\begin{eqnarray*}
\label{chiral:scalar}
\varphi(\sigma)
& = &
{\sf q}
+
 \sigma\hskip .5pt{\sf p}
+
\varphi_<(\sigma)
+
\varphi_>(\sigma)
\end{eqnarray*}
where
\begin{eqnarray*}
\varphi_<(\sigma)
& = &
i \sum\limits_{n = -\infty}^{-1} \frac{{\sf a}_n}{n}\, {\rm e}^{-in\sigma},
\hskip 1cm
\varphi_>(\sigma)
\; = \;
i\sum\limits_{n = 1}^{\infty} \frac{{\sf a}_n}{n}\, {\rm e}^{-in\sigma}.
\end{eqnarray*}
The modes enjoy the usual hermitian conjugation
\begin{equation}
\label{modes:conjugation:bosonic}
{\sf p}^{\dagger} = {\sf p},
\hskip 5mm
{\sf q}^{\dagger} = {\sf q},
\hskip 5mm
{\sf a}^{\dagger}_{n} = {\sf a}_{-n},
\end{equation}
as well as commutation
\[
[{\sf q}, {\sf p}] = i,
\hskip 1cm
[{\sf a}_m,{\sf a}_n] = m\delta_{m+n},
\]
properties, realized on the Hilbert space ${\cal H}_{\rm B} = L^2({\mathbb R})\otimes {\cal F}_{\rm B},$
where ${\cal F}_{\rm B}$ is the Fock space generated by the action of the creation operators ${\sf a}_{-n},\, n > 0,$
on the ground state $\Omega_{\rm B}$ annihilated by ${\sf a}_n,\, n > 0.$

The second ingredient we shall need is the (antiperiodic for $\sigma\to \sigma + 2\pi$) chiral fermion field
\begin{eqnarray}
\label{chiral:fermion}
\psi(\sigma)
& = &
\sum\limits_{k \in {\mathbb Z}+\frac12} \psi_k\, {\rm e}^{-ik\sigma},
\end{eqnarray}
with the conjugation
\begin{equation}
\label{modes:conjugation:fermionic}
\psi_{k}^{\dagger} \; = \; \psi_{-k}
\end{equation}
and anticommutation relation
\(
\{\psi_k,\psi_l\} \; = \; \delta_{k+l}
\)
realized on the Fock space ${\cal F}_{\rm F}$ generated by the action of $\psi_{-k}$ on the fermionic
vacuum $\Omega_{\rm F},\ \psi_k\Omega_{\rm F} = 0,\, k > 0.$

Let ${\cal H}_p$ be the vector space generated by the action of the
creation operators $a_{-n}$ and $\psi_{-k}$ on the ``ground state'' $\tilde\nu_p = |p\rangle\otimes\Omega_{\rm B}\otimes{\Omega}_{\rm F},$
where ${\sf p}|p\rangle = p |p\rangle,\ p \in {\mathbb R}.$ One can define on ${\cal H}_p$ a
standard free field representation
of the Neveu-Schwarz algebra,
\begin{eqnarray*}
\label{NS:alebra:rep}
\nonumber
L_0 & = & \frac18Q^2 + \frac12 {\sf p}^2 +
\sum\limits_{m \geqslant 1}{\sf a}_{-m}{\sf a}_m +
\sum\limits_{k\geqslant \frac12}k\,\psi_{-k}\psi_{k},
\\
L_n & = & \left({\sf p}+\frac{inQ}{2}\right){\sf a}_n +
\frac12\sum\limits_{m\neq 0,n}\!{\sf a}_{n-m}{\sf a}_m +
\frac12\sum\limits_{k\in {\mathbb Z} +\frac12}\hskip -4pt k\,\psi_{n-k}\psi_{k},
\hskip 5mm  n \neq 0,
\\
\nonumber
S_k & = & ({\sf p}+ iQk)\psi_k +\sum\limits_{m\neq 0}{\sf a}_{m}\psi_{k-m},
\end{eqnarray*}
with the central charge $c = \frac32 + 3Q^2$ and the highest weight vector $\tilde\nu_p $
having the conformal weight $\frac18Q^2 + \frac12p^2.$
If we further define on ${\cal H}_p$ a bilinear form $\langle\!\langle \cdot\,|\,\cdot \rangle\!\rangle_p$ such that
a normalization condition
\(
\pprod{\tilde\nu_p}{\tilde\nu_p} =1
\)
and the conjugation properties (\ref{modes:conjugation:bosonic}),  (\ref{modes:conjugation:fermionic}) hold, then
$({\cal H}_p,\langle\!\langle \cdot\,|\,\cdot \rangle\!\rangle_p)$
becomes isomorphic to the NS moduli
$({\cal V}_a, \langle \cdot\,|\,\cdot \rangle_a)$ with $a= \frac{Q}{2} + ip.$

Define now:
\begin{itemize}
\item
a normal ordered exponentials:
\begin{equation*}
\label{exponent} {\sf E}^\alpha(\sigma)
\; = \;
{\rm e}^{\frac12\alpha{\sf q}}\,
{\rm e}^{\alpha\varphi_{<}(\sigma)}\,
{\rm e}^{\alpha\sigma{\sf p}}\,
{\rm e}^{\alpha\varphi_{>}(\sigma)}\,
{\rm e}^{\frac12\alpha{\sf q}},
\end{equation*}
built from the chiral scalar in a way which assures their hermiticity for real $\alpha,$ and
\item
the screening charge:
\begin{eqnarray*}
\label{screening:charge}
{\sf Q}(\sigma)
& = &
\int\limits_\sigma^{\sigma+2\pi}\!\!\!dx\ \psi(x){\sf E}^b(x).
\end{eqnarray*}
\end{itemize}
In the present paper we are interested in algebraic properties of normal ordered exponentials and screening charges. Therefore, we shall not
discuss some of the properties -- such as their domain or self-adjointnes issues -- which would allow to establish them as
true operators  (see however a discussion on the parallel issues for the non-supersymmetric case in
\cite{Teschner:2003en}).

Explicit calculations give:
\begin{eqnarray}
\nonumber
\left[L_n,{\sf E}^\alpha(\sigma)\right]
& = &
{\rm e}^{in\sigma}\left(
-i\frac{d}{d\sigma}
+
n\Delta_\alpha
\right){\sf E}^\alpha(\sigma),
\hskip 1cm
\Delta_\alpha=\frac12\alpha(Q-\alpha),
\\[-6pt]
\label{comm:Ln:Epsi}
\\[-6pt]
\nonumber
\left[L_n,\psi(\sigma)\right]
& = &
{\rm e}^{in\sigma}\left(-i\frac{d}{d\sigma}  + n\frac12\right)\psi(\sigma),
\end{eqnarray}
and
\begin{equation}
\label{acom:Sk:E}
\left[S_k,{\sf E}^\alpha(\sigma)\right]
\; = \;
-i\alpha\, {\rm e}^{ik\sigma}\,\psi(\sigma)\,{\sf E}^\alpha(\sigma).
\end{equation}
Moreover, if we take
\begin{equation}
\label{Q:b}
Q = b +  b^{-1},
\end{equation}
then
\begin{equation}
\label{com:Ln:Q}
\left[L_n,\psi(\sigma){\sf E}^b(\sigma)\right]
=
-i\frac{d}{d\sigma}
\left(
{\rm e}^{in\sigma}
\psi(\sigma){\sf E}^b(\sigma)
\right)
\end{equation}
and
\begin{equation}
\label{acom:Ln:Q}
\left\{
S_k,\psi(\sigma){\sf E}^b(\sigma)
\right\}
\; = \;
\frac{1}{b}\frac{d}{d\sigma}\left({\rm e}^{ik\sigma}\,{\sf E}^b(\sigma)\right),
\end{equation}
what shows that ${\sf Q}(\sigma)$ is a scalar under the transformations generated by $L_n$ and $S_k.$

For real $b$ the screening charge ${\sf Q}(\sigma)$ is  hermitian, its square is therefore positive and
$\left[{\sf Q}(\sigma)^2\right]^t$ may be uniquely defined for complex $t.$ This motivates
the following definition of an ``even'' and an ``odd'' complex powers of the screening charge:
for $s \in {\mathbb C}\setminus{\mathbb Z}:$
\begin{eqnarray}
\label{complex:powers}
\left({\sf Q}(\sigma)\right)^{s}_{\rm e}
& = &
\left({\sf Q}^2(\sigma)\right)^{\frac{s}{2}},
\hskip 1cm
\left({\sf Q}(\sigma)\right)^{s}_{\rm o}
\; = \;
{\sf Q}(\sigma)\left({\sf Q}^2(\sigma)\right)^{\frac{s-1}{2}}
\end{eqnarray}
and
\begin{equation*}
\label{complex:powers:integers}
\left({\sf Q}(\sigma)\right)^{s}_{\rm e}
=
\left({\sf Q}(\sigma)\right)^{s}_{\rm o}
=
\left({\sf Q}(\sigma)\right)^{s}
\hskip .5cm
{\rm for}\; s \in {\mathbb N}.
\end{equation*}

Relations (\ref{comm:Ln:Epsi}) -- (\ref{acom:Ln:Q}) then imply that chiral fields
\begin{eqnarray*}
\label{chiral:fields}
{\sf g}_s^{\alpha,\,\rho}(\sigma)
& = &
{\sf E}^\alpha(\sigma)\left({\sf Q}(\sigma)\right)^{s}_{\rho},
\hskip 1cm \rho= {\rm e,\,o},
\end{eqnarray*}
transform covariantly under superconformal transformations.

As in the non-supersymmetric case
we can define the Euclidean fields by analytic con\-ti\-nu\-a\-tion to imaginary time
\begin{equation}
\label{Euclidean:fields}
{\sf g}_s^{\alpha,\,\rho}(w)
=
{\rm e}^{\tau L_0}\,{\sf g}_s^{\alpha,\,\rho}(\sigma)\,{\rm e}^{-\tau L_0},
\hskip 1cm
w = \tau+i\sigma.
\end{equation}
The fields (\ref{Euclidean:fields}) on a complex cylinder parameterized by the coordinate $w$
are related to the fields ${\sf g}_s^{\alpha,\,\rho}(z)$ on a complex plane $z={\rm e}^w$ via
\begin{equation*}
\label{complex:plane:fields}
{\sf g}_s^{\alpha,\,\rho}(w)
=
{z}^{\Delta_\alpha}\,{\sf g}_s^{\alpha,\,\rho}(z),
\hskip 1cm
\Delta_\alpha = \frac12 \alpha(Q-\alpha).
\end{equation*}
Thanks to a simple dependence on $\sf q,$ the fields ${\sf g}_s^{\alpha,\,\rho}(w)$ have simple
commutation properties with functions of $\sf p,$
\begin{equation*}
\label{commutation:with:p}
{\sf g}_s^{\alpha,\,\rho}(w)f({\sf p})
=
f\big({\sf p}-i(\alpha+bs)\big){\sf g}_s^{\alpha,\,\rho}(w).
\end{equation*}
This relation, the isomorphism
\(
({\cal H}_p,\langle\!\langle \cdot\,|\,\cdot \rangle\!\rangle_p)
\simeq
({\cal V}_a, \langle \cdot\,|\,\cdot \rangle_{\frac{Q}{2} + ip})
\)
and equations (\ref{comm:Ln:Epsi}) -- (\ref{acom:Ln:Q})  show that a restriction
of ${\sf g}_s^{\alpha,\,\rho}(w)$ to ${\cal H}_{p}$ provides a realization of a (unnormalized)
superconformal vertex operator
\(
{}^{\mathbf g}V^{\rho}_{a_3a_1}(\nu_\alpha|w)
\)
with
$a_1 = \frac{Q}{2} +ip$ and
$a_3 = a_1 +\alpha + bs.$

\subsection{Matrix elements}
For the chiral field on the complex $z$ plane let us denote:
\begin{eqnarray}
\nonumber
{\cal M}(a_2,s|a_1) & \equiv & \pprod{\tilde\nu_q}{{\sf g}^{a_2, {\rm e}}_{s}(1)\tilde\nu_p}_q
=
\langle \nu_3|\,{}^{\mathbf g}V^{\rm e}_{a_3a_1}\!(\nu_2|1)\nu_1\rangle_{a_3},
\\[-6pt]
\label{matrix:elements}
\\[-6pt]
\nonumber
{\cal M}^{*}(a_2,s|a_1)& \equiv &\pprod{\tilde\nu_q}{*\!{\sf g}^{a_2,{\rm e}}_{s}(1)\tilde\nu_p}_q
=
\langle \nu_3|\,{}^{\mathbf g}V^{\rm e}_{a_3a_1}\!(*\nu_2|1)\nu_1\rangle_{a_3},
\end{eqnarray}
where
\[
*{\sf g}^{a,{\rm e}}_{s}(1)=\left\{S_{-1/2},{\sf g}^{a,{\rm o}}_{s}(1) \right\},
\hskip 5mm
p = \frac{iQ}{2} - ia_1,
\hskip 5mm
q = \frac{iQ}{2} - i(a_1+a_2+bs) \equiv  \frac{iQ}{2} - ia_3.
\]
To compute these matrix elements we shall use a strategy similar to
the one applied by J.~Teschner in calculating the matrix element
of the chiral primary field in the non-supersymmetric CFT
\cite{Teschner:2003en}.

Introduce two auxiliary fields:
\begin{eqnarray*}
\nonumber
 {\sf g}_{0}(z)&\equiv& {\sf g}_{0}^{-b}(z)= {\sf E}^{-b}(z),
\\[-10pt]
\\[-10pt]
\nonumber
{\sf g}_{1}(z)&\equiv& {\sf g}_{1}^{-b}(z)= {\sf E}^{-b}(z){\sf Q}(z),
\end{eqnarray*}
together with their descendants $*{\sf g}_{r}(z),\ r=0,1.$
We shall need the following simple matrix elements\footnote{
As before, we shall suppress the subscript $q$ in the form $\pprod{\cdot\,}{\,\cdot}_{q};$
since in the adapted notation it is the same as an argument of the ``bra'' it shouldn't result in any confusion.}:
\begin{eqnarray}
\nonumber
\pprod{\tilde\nu_{p+ib}}{{\sf g}_{0}(1)\tilde\nu_{p}}_{p+ib}
& \equiv &
\pprod{p+ib}{{\sf g}_{0}(1)|p}
= 1,
\hskip 2cm
\pprod{p+ib}{{\sf *g}_{0}(1)|\!*\!p}
\; = \;\
b\beta,
\end{eqnarray}
and
\begin{eqnarray}
\label{simple:matrix:elem}
\pprod{p}{{\sf g}_{1}(1)|\!*\!p}
& = &
\pprod{p}{{\sf *g}_{1}(1)|p}
\; = \;
\;\frac{2\pi i\beta\,{\rm e}^{-i\pi b \beta}\; \Gamma(1+b^2)}{\Gamma(1+b\beta)\Gamma(1+b^2-b\beta)}
\; \equiv \;
{\sf M}_{1}(\beta),
\end{eqnarray}
with $\beta=\frac{Q}{2}+ip$ and where, in order to derive (\ref{simple:matrix:elem}), we  used
(after a suitable deformation of the integration contour) the integral representation of the Euler beta function.

Consider now conformal blocks:
\begin{eqnarray}
\nonumber
\Psi_{r}(z)&\equiv&\bbra{p_{4}^{r}}{\sf g}_{r}(z){\sf g}^{a_2}_{s}(1)\kket{p_{1}}\!,
\\[-8pt]
\label{blocks}
\\[-8pt]
\nonumber
\Psi_{r}^{*}(z)&\equiv&\bbra{p_{4}^{r}}{\sf *g}_{r}(z){\sf g}^{a_2}_{s}(1)\kket{p_{1}}\!,
\end{eqnarray}
where $p_{4}^{r}=p_{1}-i(a_{2}+bs+b\;(r-1))$ and where we have suppressed  the ``parity''
superscript $\rho$ of ${\sf g}^{a_2,\,\rho}_{s}$ since if we require that $\Psi_{r}(z)$ and $\Psi^{*}_{r}(z)$
do not vanish identically it is uniquely determined by the parity
of ${\sf g}_{r}(z)$ and ${\sf *g}_{r}(z).$ We shall now evaluate in two different ways the leading (most singular) terms
in the expansion of $\Psi_{r}(z)$ and $\Psi_{r}^{*}(z)$ around $z=1,$
arriving at a recurrence relation for the matrix elements (\ref{matrix:elements}).

First of all, just from the definition of the operators ${\sf g}^{a}_{s}(z)$ and  ${\sf *g}^{a}_{s}(z),$ one can compute
leading terms in their operator product expansions. For instance:
\[
{\sf g}_0(z){\sf g}^{a}_{s}(1)
\; = \;
 {\sf E}^{-b}(z)  {\sf E}^a(1)\left( {\sf Q}(1)\right)^s
\; \simeq \;
(z-1)^{ba} {\sf E}^{a-b}(1)\left( {\sf Q}(1)\right)^s
\; = \;
(z-1)^{ba}{\sf g}^{a-b}_{s}(1),
\]
or, writing $ {\sf Q}(z) = {\sf Q}(1) +\left({\sf Q}(z)- {\sf Q}(1)\right),$
\begin{eqnarray*}
{\sf g}_1(z){\sf g}^{a}_{s}(1)
& = &
{\sf E}^{-b}(z) {\sf Q}(1){\sf E}^a(1)\left({\sf Q}(1)\right)^s
+
{\sf E}^{-b}(z)\left({\sf Q}(z)- {\sf Q}(1)\right){\sf E}^a(1)\left({\sf Q}(1)\right)^s
\\[6pt]
& \simeq &
{\rm e}^{-i\pi b a}{\sf E}^{-b}(z){\sf E}^a(1)\left({\sf Q}(1)\right)^{s+1}
\; \simeq \;
{\rm e}^{-i\pi b a}(z-1)^{ba}{\sf g}^{a-b}_{s+1}(1).
\end{eqnarray*}
Thus we have:
\begin{eqnarray}
\nonumber
\Psi_{0}(z)
& \underset{z\rightarrow 1}{\simeq} &
{\cal M}(a_{2}-b,s|p_{1})\;(z-1)^{ba_{2}},
\\[-8pt]
\label{limit:z:to:1}
\\[-8pt]
\nonumber
\Psi_{1}(z)
& \underset{z\rightarrow 1}{\simeq} &
{\sf e}^{-i\pi b a_{2}}{\cal M}(a_{2}-b,s+1|p_{1})\;(z-1)^{ba_{2}},
\end{eqnarray}
and similarly:
\begin{eqnarray}
\nonumber
\Psi_{0}^{*}(z)
& \underset{z\rightarrow 1}{\simeq} &
-\frac{b}{a_{2}-b}{\cal M}^{*}(a_{2}-b,s|p_{1})\;(z-1)^{ba_{2}},
\\[-8pt]
\label{limit:z:to:1:star}
\\[-8pt]
\nonumber
\Psi_{1}^{*}(z)
& \underset{z\rightarrow 1}{\simeq} &
 -\frac{b}{a_{2}-b}{\sf e}^{-i\pi b a_{2}}{\cal M}^{*}(a_{2}-b,s+1|p_{1})\;(z-1)^{ba_{2}}.
\end{eqnarray}

On the other hand, the Verma moduli ${\cal V}_{-b}$ is degenerate: the vector
\[
\big(L_{-1}S_{-\frac12} + b^2S_{-\frac32}\big)\nu_{-b}
\]
is null. Consequently, correlators containing any of the fields ${\sf g}_{r}(z)$
satisfy the corresponding null vector decoupling equations \cite{Belavin:1984vu,Belavin:2007gz}.
Their forms\footnote{A derivation can be found in e.g.\ \cite{Belavin:2007gz,Belavin:2007eq}}
for the conformal blocks (\ref{blocks}) are:
{\small
\begin{eqnarray}
\nonumber
&&
\hspace{-.5cm}
\frac{1}{b^{2}}\frac{d^3\Psi_{r}(z)}{dz^3}
+
\frac{(2z-1)(2b^2-1)}{z(1-z)b^2}\frac{d^2\Psi_{r}(z)}{dz^2}+
\\[6pt]
\label{diffeq:1}
&&\hspace{-.5cm}
\left(
\frac{b^{2}+2\Delta_{1}}{z^{2}}+\frac{b^{2}+2\Delta_{2}}{(1-z)^{2}}+
\frac{2-3b^{2}+2\Delta^{(r)}_{1+2-4}}{z(1-z)}
\right)
\frac{d\Psi_{r}(z)}{dz}+
\\[6pt]
\nonumber
&&\hspace{-.5cm}
\left(
\frac{2\Delta_{2}(1+b^{2})}{(1-z)^{3}}
-\frac{2\Delta_{1}(1+b^{2})}{z^{3}}
+ \frac{\Delta_{2-1} + (1-2z)(b^{4}+b^{2}(1/2-\Delta^{(r)}_{1+2-4})-\Delta_{1+2})}{z^{2}(1-z)^{2}}
\right)\Psi_{r}(z)
\; = \; 0,
\end{eqnarray}
}
and
\begin{eqnarray}
\label{diffeq:2}
&& \hspace*{-1cm}
\nonumber
\frac{d^3\Psi_{r}^{*}(z)}{dz^3}
+
\frac{2(1-b^2)(1-2z)}{z(1-z)}\frac{d^2\Psi_{r}^{*}(z)}{dz^2} +
\\[6pt]
&&
\hspace*{-1cm}
\left(
\frac{b^4-b^2+2b^2\left(\Delta_1(1-z)+\Delta_2 z\right)}{z^2(1-z)^2}
-
\frac{5b^4 +b^2(2\Delta^{(r)}_4-7)+2}{z(1-z)}
\right)
\frac{d\Psi_{r}^{*}(z)}{dz}+
\\[6pt]
\nonumber
&&
\hspace*{-1.4cm}
b^4\left(
\frac{3(\Delta_1-\Delta_2) + \left(\Delta^{(r)}_4 + b^2-1\right)(1-2z)}
{z^2(1-z)^2}
+
\frac{2\Delta_2 z -2\Delta_1(1-z)}{z^3(1-z)^3}
\right)
\Psi_{r}^{*}(z) \; = \; 0,
\end{eqnarray}
where $\Delta_4^{(r)} = \frac12 a_4^{(r)}\left(Q-a_4^{(r)}\right)$ with
\begin{equation*}
\label{a4}
a_4^{(r)} = a_1 + a_2 +b(s+r-1),
\end{equation*}
and
$\Delta_{1+2} = \Delta_1 + \Delta_2,\ $
$\Delta^{(r)}_{1+2-4} = \Delta_1 + \Delta_2 - \Delta^{(r)}_4$ etc.

Let us start by analyzing the equation (\ref{diffeq:1}). According to
\cite{Belavin:2006zr,Belavin:2007gz}
its solution is of the form
\begin{equation}
\label{substitute}
\Psi_{r}(z)=z^{a_{1}b}(1-z)^{a_{2}b}F(z),
\end{equation}
where $F(z)$ can be expressed as a linear combination of Dotsenko-Fatteev type integrals
\begin{equation}
\label{I:ab}
I_{\alpha\beta}(z) =
\left(
1-\frac12\delta_{\alpha\beta}
\right)
\int\limits_{{\cal C}_{\alpha}}\!dt_1\!
\int\limits_{{\cal C}_{\beta}}\!dt_2\
(t_1t_2)^{A_r}
\left[(t_1-1)(t_2-1)\right]^{B_r}
\left[(t_1-z)(t_2-z)\right]^{C_r}
\left|t_2-t_1\right|^{2g},
\end{equation}
with the integration contours ${\cal C}_{1}=(-\infty,0],$, ${\cal C}_{2}=[0,1],$
${\cal C}_{3}=[1,z],$ ${\cal C}_{4}=[z,\infty)$ and
\begin{eqnarray}
\label{constansABCG}
\nonumber
A_r &=& -\frac{1}{2}+\frac{b(a_4^{(r)}+a_2-a_1)}{2},
\hskip 2cm
B_r \;\ =\;\ -\frac{1}{2}+\frac{b(a_4^{(r)}-a_2+a_1)}{2},
\\[-6pt]
\\[-6pt]
\nonumber
\nonumber
C_r &=& \frac{1}{2}-\frac{b(a_4^{(r)}+a_2+a_1)}{2}+b^2,
\hskip 1.5cm
g\;\ =\;\ -\frac{1}{2}-\frac{b^2}{2}.
\end{eqnarray}
In order to decide which solution of the differential equation (\ref{diffeq:1}) corresponds to the block
\(
\bbra{p_{4}^{r}}{\sf g}_{r}(z){\sf g}^{a_{2}}_{s}(1)\kket{p_{1}}
\)
let us note that we can present it in a form of a power series in $z$ around $z = \infty$
by inserting between the fields ${\sf g}_{r}(z)$ and ${\sf g}^{a_{2}}_{s}(1)$
the projection operator onto the highest weight state with the momentum $q =p_1-i(\alpha_2 + bs)$
and its (normalized) NS descendants.
The leading terms in these expansions read
\begin{eqnarray*}
\Psi_{0}(z)
& \underset{z\rightarrow \infty}{\simeq} &
\bbra{q-ib}{\sf g}_{0}(z)\kket{q}\bbra{q}{\sf g}^{a_{2}}_{s}(1)\kket{p_{1}},
\\[4pt]
\Psi_{1}(z)
& \underset{z\rightarrow \infty}{\simeq} &
\frac{1}{2\Delta_4}\bbra{q}{\sf g}_{1}(z)S_{-\frac12}\!\kket{q}\bbra{q}\!S_{\frac12}{\sf g}^{a_{2}}_{s}(1)\kket{p_{1}},
\end{eqnarray*}
so that, using (\ref{matrix:elements}) and (\ref{simple:matrix:elem}), we have:
\begin{eqnarray}
\nonumber
\Psi_{0}(z)&\underset{z\rightarrow \infty}{\simeq}&{\cal M}(a_2,s|a_1)\, z^{b(a^{(0)}_{4}+b)},
\\[-6pt]
\label{asymptotics}
\\[-6pt]
\nonumber
\Psi_{1}(z)&\underset{z\rightarrow \infty}{\simeq}&\frac{{\sf M}_{1}(a_{4}^{(1)})\,}{2\Delta_{4}^{(1)}}
{\cal M}^{*}(a_2,s|a_1)\, z^{b^2}.
\end{eqnarray}
Monodromies (around $z=\infty$) of all the terms in the expansion of $\Psi_0(z)$ are equal and determined by (\ref{asymptotics});
the same is also true for $\Psi_1(z).$
On the other hand, out of the integrals (\ref{I:ab}), $I_{22}(z), I_{24}(z)$ and $I_{44}(z)$ form
(once multiplied by $z^{a_{1}b}(1-z)^{a_{2}b}$) a basis in the space of solutions of
(\ref{diffeq:1}) with a monodromy matrix diagonal around $z=\infty.$
In fact, from (\ref{I:ab}) we have:
\begin{eqnarray}
\nonumber
I_{44}(z)
& \underset{z\rightarrow \infty}{\simeq} &
z^{-b(a_{1}+a_{2})+b(a^{(0)}_{4}+b)}\; {\cal I}_{44}^{(\infty)}\left(1+{\cal O}(z^{-1})\right),
\\[-8pt]
\label{I:ab:asymptotics}
\\[-8pt]
\nonumber
I_{24}(z)
& \underset{z\rightarrow \infty}{\simeq} &
z^{-b(a_{1}+a_{2})+b^2}\; {\cal I}_{24}^{(\infty)}\left(1+{\cal O}(z^{-1})\right),
\end{eqnarray}
where:
\begin{eqnarray}
\nonumber
{\cal I}_{44}^{(\infty)}
& = &
\frac{\Gamma(2g)\Gamma(-1-A_0-B_0-C_0-g)\Gamma(-1-A_0-B_0-C_0-2g)}{\Gamma(g)}
\\
\label{I:4424:infty}
&\times &
\frac{\Gamma(1+C_0)\Gamma(1+C_0+g)}{\Gamma(-A_0-B_0)\Gamma(-A_0-B_0-g)},
\\[8pt]
\nonumber
{\cal I}_{24}^{(\infty)}
& = &
\Gamma(1+A_1)\Gamma(-1-A_1-B_1-C_1-2g)\,\frac{\Gamma(1+B_1)\Gamma(1+C_1)}{\Gamma(2+A_1+B_1)\Gamma(-A_1-B_1-2g)},
\end{eqnarray}
and, comparing (\ref{substitute}) and (\ref{I:ab:asymptotics}) with (\ref{asymptotics}) we get:
\begin{eqnarray}
\nonumber
\Psi_{0}(z)
& = &
\frac{1}{{\cal I}_{44}^{(\infty)}}\,
{\cal M}(a_2,s|a_1)\,
z^{ba_1}(z-1)^{ba_2}\, I_{44}(z),
\\[-6pt]
\label{utozsamienia:1}
\\[-6pt]
\nonumber
\Psi_{1}(z)
& = &
\frac{{\sf M}_{1}(a^{(1)}_{4})}{{2\Delta_4^{(1)}\, \cal I}_{24}^{(\infty)}}\,
{\cal M}^{*}(a_2,s|a_1)\,
z^{ba_1}(z-1)^{ba_2}\, I_{24}(z).
\end{eqnarray}

Another basis in the space of solutions of (\ref{diffeq:1}),
with a monodromy matrix diagonal around $z=1,$ is formed by
$I_{11}(z), I_{13}(z)$ and $I_{33}(z).$ Obviously these bases are linearly related,
\begin{equation*}
\label{cange:of:a:basis}
I_{\imath} \; = \; \sum_{\jmath} M_{\imath\jmath}\,I_{\jmath},
\hskip 1cm
\imath \in \{22, 24, 44\},
\hskip 5mm
\jmath \in \{11, 13, 33\},
\end{equation*}
and the matrix $M_{\imath\jmath}$ is known, see \cite{Dotsenko:1984nm,Belavin:2007gz}.
Using this and noticing that the leading contribution in the $z\to 1$ limit is given by the
term proportional to $I_{11}(z)$ we have:
\begin{eqnarray}
\nonumber
I_{44}(z)
\; \underset{z\rightarrow 1}{\simeq} \;
{\cal I}^{(1)}_{44}
& = &
\frac{\Gamma(2g)\Gamma(-1-A_0-B_0-C_0-g)\Gamma(-1-A_0-B_0-C_0-2g)}{\Gamma(g)}
\\[4pt]
\nonumber
& \times &
\frac{\Gamma(1+B_0+C_0)\Gamma(1+B_0+C_0+g))}{\Gamma(-A_0)\Gamma(-A_0-g)},
\\[-6pt]
\label{I:4424:1}
\\[-6pt]
\nonumber
I_{24}(z)
\; \underset{z\rightarrow 1}{\simeq} \;
{\cal I}^{(1)}_{24}
& = &
\Gamma(1+A_1)\Gamma(-1-A_1-B_1-C_1-2g)
\\[4pt]
\nonumber
&\times &
\frac{\Gamma(1-g)\Gamma(1+B_1+C_1)\Gamma(1+B_1+C_1+g)}{\Gamma(1-2g)\Gamma(-A_1-g)\Gamma(2+A_1+B_1+C_1+g)}.
\end{eqnarray}
Substituting this result into (\ref{utozsamienia:1}) we get:
\begin{eqnarray}
\nonumber
\Psi_{0}(z)
& \underset{z\rightarrow 1}{\simeq} &
\frac{{\cal I}_{44}^{(1)}}{{\cal I}_{44}^{(\infty)}}\,
{\cal M}(a_2,s|a_1)\,
(z-1)^{ba_2},
\\[-6pt]
\label{limits:2}
\\[-6pt]
\nonumber
\Psi_{1}(z)
& \underset{z\rightarrow 1}{\simeq} &
\frac{{\sf M}_{1}(a_{4}){\cal I}_{24}^{(1)}}{{2\Delta_4^{(1)}\, \cal I}_{24}^{(\infty)}}\,
{\cal M}^{*}(a_2,s|a_1)\,
(z-1)^{ba_2}.
\end{eqnarray}

Comparing (\ref{limits:2}) with (\ref{limit:z:to:1}) and using (\ref{simple:matrix:elem}) we arrive
at the recurrence relations of the form:
\begin{eqnarray*}
\label{rekurencja:1}
\nonumber
{\cal M}(a_2-b,s|a_1)
& = &
\frac{{\cal I}_{44}^{(1)}}{{\cal I}_{44}^{(\infty)}}\,
{\cal M}(a_2,s|a_1),
\\[-12pt]
\\[0pt]
\nonumber
{\cal M}(a_2-b,s+1|a_1)
& = &
{\rm e}^{i\pi b (a_2-a^{(1)}_4)}\,
\frac{{\cal I}_{24}^{(1)}}{{\cal I}_{24}^{(\infty)}}\,
\frac{2\pi i a_4\Gamma(1+b^2)}{2\Delta_4^{(1)}\,\Gamma(1+ba^{(1)}_4)\Gamma(1+b^2-ba^{(1)}_4)}\,
{\cal M}^{*}(a_2,s|a_1).
\end{eqnarray*}
Using equations (\ref{I:4424:infty}), (\ref{I:4424:1}), (\ref{constansABCG}) and the relations satisfied
by the Barnes gamma function (Appendix \ref{Appendix:Barnes}) we thus have:
\begin{eqnarray}
\nonumber
\frac{{\cal M}(a_2-b,s|a_1)}{{\cal M}(a_2,s|a_1)}
& = &
\frac{\Gamma(1+B_0+C_0)\Gamma(1+B_0+C_0+g)\Gamma(-A_0-B_0)\Gamma(-A_0-B_0-g)}{\Gamma(-A_0)\Gamma(-A_0-g)\Gamma(1+C_0)\Gamma(1+C_0+g)}
\\[8pt]
\label{rekurencja:1a}
&=&
\left[
\frac{\Gamma_{\rm NS}(2Q-2a_1-2a_2-bs)\Gamma_{\rm NS}(Q-2a_2 -bs)}
{\Gamma_{\rm NS}(Q-2a_2)\Gamma_{\rm NS}\left(2Q-2a_1-2a_2-2bs\right)}
\right]^{-1}
\\[6pt]
\nonumber
& \times &
\frac{\Gamma_{\rm NS}(2Q-2a_1-2(a_2-b)-bs)\Gamma_{\rm NS}(Q-2(a_2-b) -bs)}
{\Gamma_{\rm NS}(Q-2(a_2-b))\Gamma_{\rm NS}\left(2Q-2a_1-2(a_2-b)-2bs\right)},
\end{eqnarray}
and similarly:
\begin{eqnarray}
\nonumber
&&
\frac{{\cal M}(a_2-b,s+1|a_1)}{{\cal M}^{*}(a_2,s|a_1)}
\; = \;
i{\rm e}^{-i\pi b(a_1+bs)}\,\frac{1}{2}\,
\Gamma\left(\frac{bQ}{2}\right)\,
b^{-\frac{bQ}{2}}\,
\\[6pt]
\label{rekurencja:1b}
&&\times\
\left[
\frac{
\Gamma_{\rm R}\left(Q-a_2 -bs\right)
\Gamma_{\rm R}\left(2a_1 +bs\right)
\Gamma_{\rm R}\left(Q+bs\right)
\Gamma_{\rm R}\left(2Q-2a_{1+2}-bs\right)
}{
\Gamma_{\rm NS}(Q-2a_2)
}
\right]^{-1}
\\[6pt]
\nonumber
&& \times\
\frac
{
\Gamma_{\rm NS}\left(Q-2a_2 -bs+b\right)
\Gamma_{\rm NS}\left(2a_1 +bs+b\right)
\Gamma_{\rm NS}\left(Q+bs+b\right)
\Gamma_{\rm NS}\left(2Q-2a_{1+2}-bs+b\right)
}{
\Gamma_{\rm NS}(Q-2a_2+2b)
},
\end{eqnarray}
where $a_{1+2} \equiv a_1 + a_2.$

To arrive at a second set of  recursion relations we repeat the same steps for the blocks
$\Psi^*_r(z):$ calculating from (\ref{blocks}) the leading
behavior of $\Psi^*_r(z)$ for $z\to\infty$ we identify the appropriate solutions of the differential equations (\ref{diffeq:2}),
then we express them in the basis given by the functions with the monodromy matrix diagonal around $z=1$ and compare
the result with the formula (\ref{limit:z:to:1:star}). This yields:
\begin{eqnarray}
\label{rekurencja:2a}
\frac{{\cal M}^*(a_2-b,s|a_1)}{{\cal M}^*(a_2,s|a_1)}
& = &
\left[
\frac{\Gamma_{\rm R}(Q-2a_2-bs)\Gamma_{\rm R}(2Q-2a_1-2a_2-bs)  }
{\Gamma_{\rm NS}(Q-2a_2)\Gamma_{\rm NS}(2Q-2a_1-2a_2-2bs)   }
\right]^{-1}
\\[6pt]
\nonumber
& \times &
\frac{\Gamma_{\rm R}(Q-2(a_2-b)-bs)\Gamma_{\rm R}(2Q-2a_1-2(a_2-b)-bs)  }
{\Gamma_{\rm NS}(Q-2(a_2-b))\Gamma_{\rm NS}(2Q-2a_1-2(a_2-b)-2bs)   },
\end{eqnarray}
and
\begin{eqnarray}
&&
\nonumber
\frac{{\cal M}^*(a_2-b,s+1|a_1)}{{\cal M}(a_2,s|a_1)}
\; = \;
-\, i{\rm e}^{-i\pi b(a_1+bs)}\,
\Gamma\left(\frac{bQ}{2}\right)\,
b^{-\frac{bQ}{2}}\,
\\[6pt]
\label{rekurencja:2b}
&&\times\
\left[
\frac{
\Gamma_{\rm NS}\left(Q-a_2 -bs\right)
\Gamma_{\rm NS}\left(2a_1 +bs\right)
\Gamma_{\rm NS}\left(Q+bs\right)
\Gamma_{\rm NS}\left(2Q-2a_1-2a_2-bs\right)
}{
\Gamma_{\rm NS}(Q-2a_2)
}
\right]^{-1}
\\[6pt]
\nonumber
&& \times\
\frac
{
\Gamma_{\rm R}\left(Q-2a_2 -bs+b\right)
\Gamma_{\rm R}\left(2a_1 +bs+b\right)
\Gamma_{\rm R}\left(Q+bs+b\right)
\Gamma_{\rm R}\left(2Q-2a_1-2a_2-bs+b\right)
}{
\Gamma_{\rm NS}(Q-2a_2+2b)
}.
\end{eqnarray}

The solution of recurrence relations (\ref{rekurencja:1a}) -- (\ref{rekurencja:2b}) is not unique.
Notice however that --- exactly as in the non-supersymmetric case --- we can repeat the construction above with
$b$ replaced by $b^{-1}.$ Moreover, direct calculation (essentially the same that leads to (\ref{simple:matrix:elem})) gives:
\begin{equation}
\label{norm:1}
{\cal M}(a_2,0|a_1) \; = \; 1
\end{equation}
and
\begin{equation}
\label{norm:2}
{\cal M}^{*}(a_2,1|a_1)
\;\ = \;\
2\pi i a_1\,
{\rm e}^{-\pi i ba_1}\,
\frac{\Gamma(1-ba_2)}{\Gamma(1+ba_1)\Gamma(1-ba_1-ba_2)}.
\end{equation}
If $b$ is real and irrational, relations (\ref{rekurencja:1a}) -- (\ref{norm:2})
together with the ``dual'' ($b\to b^{-1}$) ones uniquely determine the the matrix
elements of the $N=1$ supersymmetric chiral vertex operators to be:
\begin{eqnarray}
\label{conjecture:1}
{\cal M}(a_2,s|a_1)
& =&
\left[
\frac12\Gamma\left(\frac{bQ}{2}\right)
b^{-\frac{bQ}{2}}
\right]^{\frac{a_3 -a_1-a_2}{b}}
{\rm e}^{\frac{i\pi}{2}\left(a_3-a_2-a_1\right)\left(Q-a_3+a_2-a_1\right)}
\\[6pt]
\nonumber
& \times &
\frac{
\Gamma_{\rm NS}\left(Q+a_{1-2-3}\right)
\Gamma_{\rm NS}\left(a_{1+3-2}\right)
\Gamma_{\rm NS}\left(Q+a_{3-1-2}\right)
\Gamma_{\rm NS}\left(2Q-a_{1+2+3}\right)
}
{
\Gamma_{\rm NS}(Q)\Gamma_{\rm NS}(2a_1)\Gamma_{\rm NS}(Q-2a_2)\Gamma_{\rm NS}(2Q-2a_3)
}
\end{eqnarray}
and
\begin{eqnarray}
\label{conjecture:2}
{\cal M}^{*}(a_2,s|a_1)
& =&
2\left[
\frac12\Gamma\left(\frac{bQ}{2}\right)
b^{-\frac{bQ}{2}}
\right]^{\frac{a_3 -a_1-a_2}{b}}
{\rm e}^{\frac{i\pi}{2}\left(a_3-a_2-a_1\right)\left(Q-a_3+a_2-a_1\right)}
\\[6pt]
\nonumber
& \times &
\frac{
\Gamma_{\rm R}\left(Q+a_{1-2-3}\right)
\Gamma_{\rm R}\left(a_{1+3-2}\right)
\Gamma_{\rm R}\left(Q+a_{3-1-2}\right)
\Gamma_{\rm R}\left(2Q-a_{1+2+3}\right)
}
{
\Gamma_{\rm NS}(Q)\Gamma_{\rm NS}(2a_1)\Gamma_{\rm NS}(Q-2a_2)\Gamma_{\rm NS}(2Q-2a_3)
},
\end{eqnarray}
where we have denoted
\[
a_3 = a_1  + a_2 + bs.
\]

\subsection{Braiding relations}
Thanks to the relation between chiral fields ${\sf g}^{\alpha,\,\rho}_{s}(z)$ and
the (unnormalized) vertex operators
${}^{\mathbf g}{\sf V}^{\rho}_{a_3a_1}(\nu_\alpha|z)$ the form of the braiding matrix appearing in
(\ref{braiding:definition:2}) can be derived by studying an
exchange relation for the chiral fields.

Assume that for
the chiral fields on a complex $w$ cylinder at $\tau = 0,$ i.e.\ ${\mathsf g}^{\alpha,\,\rho}_{s}(\sigma),$
there exist an integral kernel $B$  such that the identity
\begin{equation}
\label{exchange:1}
{\mathsf g}^{\alpha_2,\rho}_{s_2}(\sigma_2)
 {\mathsf g}^{\alpha_1,\eta}_{s_1}(\sigma_1)
=
\int\! d\mu(t_1,t_2)\hskip -4pt
\sum\limits_{\lambda,\delta =\, {\rm e,o}}\hskip -4pt
B^{\epsilon}\!\left(\alpha_1,\alpha_2;{}^{t_1\:t_2}_{s_2\:s_1}\right)^{\rho\eta}_{\hskip 8pt \lambda\delta}\,
 {\mathsf g}^{\alpha_1,\lambda}_{t_1}(\sigma_1)
{\mathsf g}^{\alpha_2,\delta}_{t_2}(\sigma_2),
\end{equation}
with $\epsilon = {\rm sign}(\sigma_2-\sigma_1)$ and the integration measure to be specified later, holds.

Since the parity of a product of chiral fields does not depend on their order,
\begin{equation*}
\label{parity:conservation}
(-1)^{|\rho| + |\eta|} = (-1)^{|\lambda| + |\delta|}
\end{equation*}
where $|{\rm e}| = 0$ and $|{\rm o}| = 1,$ we can discuss the ``even'',  $(-1)^{|\rho| + |\eta|} = (-1)^{|\lambda| + |\delta|} = 1,$
and the ``odd'', $(-1)^{|\rho| + |\eta|} = (-1)^{|\lambda| + |\delta|} = -1,$ cases separately.
Introducing a shorthand notation
\begin{equation}
\label{shorthand:for:B}
\big[B^{{\rm e},\epsilon}\big]^{\rho}_{\hskip 4pt \eta}
\equiv
B^{\epsilon}\!\left(\alpha_1,\alpha_2;{}^{t_1\:t_2}_{s_2\:s_1}\right)^{\rho\rho}_{\hskip 8pt \eta\eta},
\hskip 1cm
\big[B^{{\rm o},\epsilon}\big]^{\rho}_{\hskip 4pt \eta}
\equiv
B^{\epsilon}\!\left(\alpha_1,\alpha_2;{}^{t_1\:t_2}_{s_2\:s_1}\right)^{\rho\bar\rho}_{\hskip 8pt \eta\bar\eta},
\end{equation}
with $\bar{\rm e} = {\rm o},$ $\bar{\rm o} = {\rm e},$ we can write (\ref{exchange:1}) in the form
\begin{eqnarray}
 \label{braiding:defs:1}
 \nonumber
 {\mathsf g}^{\alpha_2,\rho}_{s_2}(\sigma_2)
 {\mathsf g}^{\alpha_1,\rho}_{s_1}(\sigma_1)
 & = &
 \sum_{\eta = {\rm  e,o}}
 \int\!d\mu(t_1,t_2)\,
 \big[B^{{\rm e},\epsilon}\big]^{\rho}_{\hskip 4pt \eta}\
 {\mathsf g}^{\alpha_1,\eta}_{t_1}(\sigma_1)
 {\mathsf g}^{\alpha_2,\eta}_{t_2}(\sigma_2),
 \\[-10pt]
 \\[-2pt]
 \nonumber
 {\mathsf g}^{\alpha_2,\rho}_{s_2}(\sigma_2)
 {\mathsf g}^{\alpha_1,\bar\rho}_{s_1}(\sigma_1)
 & = &
 \sum_{\eta = {\rm  e,o}}
 \int\!d\mu(t_1,t_2)\,
\big[B^{{\rm o},\epsilon}\big]^{\rho}_{\hskip 4pt \eta}\
 {\mathsf g}^{\alpha_1,\eta}_{t_1}(\sigma_1)
 {\mathsf g}^{\alpha_2,\bar\eta}_{t_2}(\sigma_2),
 \hskip 1cm
 \rho = {\rm e,o,}.
 \end{eqnarray}

Our strategy in calculating the matrix $B$ is again a suitable extension of the Teschner's  technique.
Suppose that $\sigma_2 > \sigma_1,$
let $I = [\sigma_1,\sigma_2],\ I_c = [\sigma_2,\sigma_1 +2\pi],\ I' = [\sigma_1+2\pi,\sigma_2+2\pi]$
and define:
\begin{eqnarray}
\label{Q:s}
\nonumber
{\mathsf Q}_{\rm\scriptscriptstyle I}
& = &
\int_I\!dx\ {\mathsf E}^b(x)\psi(x),
\\
{\mathsf Q}_{\rm\scriptscriptstyle I}^c
& = &
\int_{I^c}\!dx\ {\mathsf E}^b(x)\psi(x),
\\
\nonumber
{\mathsf Q}'_{\rm\scriptscriptstyle I}
& = &
\int_{I'}\!dx\ {\mathsf E}^b(x)\psi(x)
\;\ = \;\
-{\rm e}^{i\pi b^2}\,{\rm e}^{2\pi b{\mathsf p}}\,
{\mathsf Q}_{\rm\scriptscriptstyle I}.
\end{eqnarray}
Since
\begin{eqnarray}
\label{elementary:braiding}
{\sf E}^\alpha(x){\sf E}^\beta(y)
& = &
{\rm e}^{-i\pi\alpha\beta\,{\rm sign}(x-y)}\,{\sf E}^\beta(y){\sf E}^\alpha(x),
\end{eqnarray}
we get:
\begin{eqnarray*}
\label{E:Q:braiding}
\nonumber
{\mathsf E}^{\alpha_1}(\sigma_1){\mathsf E}^{\alpha_2}(\sigma_2)
& = &
{\rm e}^{i\pi \alpha_1\alpha_2}
{\mathsf E}^{\alpha_2}(\sigma_2){\mathsf E}^{\alpha_1}(\sigma_1),
\\[10pt]
{\mathsf Q}(\sigma_2){\mathsf E}^{\alpha_1}(\sigma_1)
& = &
\left(
{\mathsf Q}_{\rm\scriptscriptstyle I}^c +
{\mathsf Q}_{\rm\scriptscriptstyle I}'
\right)
{\mathsf E}^{\alpha_1}(\sigma_1)
\; = \;
{\mathsf E}^{\alpha_1}(\sigma_1)
\left(
{\rm e}^{-i\pi b \alpha_1}{\mathsf Q}_{\rm\scriptscriptstyle I}^c +
{\rm e}^{-3i\pi b \alpha_1}{\mathsf Q}_{\rm\scriptscriptstyle I}'
\right),
\\[6pt]
\nonumber
{\mathsf Q}(\sigma_1){\mathsf E}^{\alpha_2}(\sigma_2)
& = &
\left(
{\mathsf Q}_{\rm\scriptscriptstyle I}^c +
{\mathsf Q}_{\rm\scriptscriptstyle I}
\right)
{\mathsf E}^{\alpha_2}(\sigma_2)
\; = \;
{\mathsf E}^{\alpha_2}(\sigma_2)
\left(
{\rm e}^{-i\pi b \alpha_2}{\mathsf Q}_{\rm\scriptscriptstyle I}^c +
{\rm e}^{i\pi b \alpha_2}{\mathsf Q}_{\rm\scriptscriptstyle I}
\right).
\end{eqnarray*}
We can thus write:
\begin{eqnarray}
\label{g:g}
\nonumber
{\mathsf g}^{\alpha_2,\rho}_{s_2}(\sigma_2)
{\mathsf g}^{\alpha_1,\eta}_{s_1}(\sigma_1)
& = &
{\mathsf E}^{\alpha_2}(\sigma_2){\mathsf E}^{\alpha_1}(\sigma_1)\
{\rm e}^{-i\pi b s_2 \alpha_1}
(
{\mathsf Q}_{\rm\scriptscriptstyle I}^c +
{\rm e}^{-2i\pi b \alpha_1}{\mathsf Q}_{\rm\scriptscriptstyle I}'
)_\rho^{s_2}
\left(
{\mathsf Q}_{\rm\scriptscriptstyle I}^c +
{\mathsf Q}_{\rm\scriptscriptstyle I}
\right)^{s_1}_\eta,
\\[-6pt]
\\[-4pt]
\nonumber
{\mathsf g}^{\alpha_1,\lambda}_{t_1}(\sigma_1)
{\mathsf g}^{\alpha_2,\delta}_{t_2}(\sigma_2)
& = &
{\mathsf E}^{\alpha_2}(\sigma_2){\mathsf E}^{\alpha_1}(\sigma_1)\
{\rm e}^{i\pi \alpha_1 \alpha_2-i\pi b t_1 \alpha_2}
({\mathsf Q}_{\rm\scriptscriptstyle I}^c +{\rm e}^{2i\pi b \alpha_2}{\mathsf Q}_{\rm\scriptscriptstyle I})^{t_1}_{\lambda}
({\mathsf Q}_{\rm\scriptscriptstyle I}^c +{\mathsf Q}'_{\rm\scriptscriptstyle I})^{t_2}_{\delta}.
\end{eqnarray}
The meaning of the r.h.s.\ of (\ref{g:g}) is clear for natural $s_k$ and $t_k,\ k = 1,2.$ Notice however that for
real $b$ and purely imaginary $\alpha_1$ and $\alpha_2$ the hermiticity of
${\mathsf Q}_{\rm\scriptscriptstyle I} , {\mathsf Q}_{\rm\scriptscriptstyle I}^c $ and ${\mathsf Q}_{\rm\scriptscriptstyle I}' $
implies a hermiticity of their (multiplied by real numbers) sums. The ``even'' and ``odd'' powers on the
r.h.s.\ are thus unambiguously defined (and thus the relation (\ref{g:g}) is valid) also for complex $s_k$ and $t_k.$
Moreover, for  $s_k$ and $t_k$  being purely imaginary
the operators on the r.h.s.\ of (\ref{g:g}) are (formally) unitary. We thus take
\begin{equation}
\label{parameters:imaginary}
\alpha_k, s_k, t_k \in i{\mathbb R}, \hskip 5mm k = 1,2.
\end{equation}

It follows from (\ref{elementary:braiding}) that the operators (\ref{Q:s}) satisfy a Weyl-type algebra:
\begin{equation}
\label{braiding:2}
{\mathsf Q}_{\rm\scriptscriptstyle I} \,
{\mathsf Q}_{\rm\scriptscriptstyle I}^c
\; = \;
-\,{\rm e}^{i\pi b^2}\,
{\mathsf Q}_{\rm\scriptscriptstyle I}^c\,
{\mathsf Q}_{\rm\scriptscriptstyle I}
\; \equiv \;
-q\,
{\mathsf Q}_{\rm\scriptscriptstyle I}^c\,
{\mathsf Q}_{\rm\scriptscriptstyle I},
\hskip 1cm
{\mathsf Q}_{\rm\scriptscriptstyle I}^c \,
{\mathsf Q}_{\rm\scriptscriptstyle I}'
\; = \;
-q\,
{\mathsf Q}_{\rm\scriptscriptstyle I}'\,
{\mathsf Q}_{\rm\scriptscriptstyle I}^c,
\hskip 1cm
{\mathsf Q}_{\rm\scriptscriptstyle I} \,
{\mathsf Q}_{\rm\scriptscriptstyle I}'
\; = \;
q^2\,
{\mathsf Q}_{\rm\scriptscriptstyle I}'\,
{\mathsf Q}_{\rm\scriptscriptstyle I}.
\end{equation}
We can thus represent them in a form (see Appendix \ref{Appendix:Weyl} for a derivation and a clarification on the $2\times 2$ matrix structure):
\begin{eqnarray}
\label{Weyl:representation}
\nonumber
{\mathsf Q}_{\rm\scriptscriptstyle I}^c & = & \tau_1\, {\rm e}^{b {\mathsf x}}\, {\rm e}^{-\frac12 i\pi b {\mathsf t}},
\\[-8pt]
\\[-8pt]
\nonumber
{\mathsf Q}_{\rm\scriptscriptstyle I}
& = &
\tau_2\,
{\rm e}^{\frac12 b {\mathsf x}}\,
{\rm e}^{-\pi b {\mathsf p}}\,
{\rm e}^{\frac12 b {\mathsf x}}\,
{\rm e}^{\frac12 i\pi b{\mathsf t}},
\end{eqnarray}
where $\tau_1 = \left(^{0\;1}_{1\;0}\right),$  $\tau_2 = \left(^{\hskip 4pt 0\hskip 5pt i}_{-i\:\, 0}\right)$
and the operators $\sf p,\ x$ and $\sf t$
satisfy commutation relations
\begin{equation}
\label{commutations:2}
[{\sf p},{\sf x}] \; =\; -i,
\hskip 1cm
[{\sf p},{\mathsf t}] \; = \; [{\sf x},{\mathsf t}] \; = \; 0,
\end{equation}
together with a conjugation properties ${\mathsf x}^\dagger ={\mathsf x}^\dagger,\ {\mathsf t}^\dagger = - {\mathsf t}.$

The representation (\ref{Weyl:representation})
and relations satisfied by  special functions $G_{\rm NS},$  $G_{\rm R}$ allow to arrange operators that appear on the r.h.s.\ of
(\ref{g:g}) in a ``normal ordered form'', with the operator $\sf x$ on the left and the operator $\sf p$ on the right.
To this end notice that:
\begin{eqnarray}
\nonumber
{\mathsf Q}_{\rm\scriptscriptstyle I}^c +{\mathsf Q}_{\rm\scriptscriptstyle I}
& = &
{\rm e}^{-\frac12 i\pi b {\mathsf t}}\,
{\rm e}^{\frac12 b {\mathsf x}}
\left(
\begin{array}{cc}
0 & 1 + i{\rm e}^{-\pi b ({\mathsf p} - i{\mathsf t})} \\
1 - i{\rm e}^{-\pi b ({\mathsf p} - i{\mathsf t})} & 0
\end{array}
\right)
{\rm e}^{\frac12 b {\mathsf x}},
\end{eqnarray}
and since
\[
G_{\rm NS}(z+b) \;\ = \;\
\left(1+{\rm e}^{i\pi b z}\right)
G_{\rm R}(z),
\hskip 1cm
G_{\rm R}(z+b) \;\ = \;\
\left(1-{\rm e}^{i\pi b z}\right)
G_{\rm  NS}(z),
\]
we can write:
\begin{eqnarray*}
&&
\hskip -1cm
\left(
\begin{array}{cc}
0 & \hskip -5pt 1 + i{\rm e}^{-\pi b ({\mathsf p} - i{\mathsf t})} \\
1 - i{\rm e}^{-\pi b ({\mathsf p} - i{\mathsf t})} &\hskip -5pt 0
\end{array}
\right)
\\[8pt]
& = &
\left(
\begin{array}{cc}
0 & \hskip -5pt G_{\rm NS}(i{\mathsf p} +{\mathsf t} + \frac{1}{2b} + b) \\
G_{\rm R}(i{\mathsf p} +{\mathsf t} + \frac{1}{2b} + b) & \hskip -5pt 0
\end{array}
\right)
\left(
\begin{array}{cc}
G_{\rm NS}(i{\mathsf p}+{\mathsf t} + \frac{1}{2b}) &\hskip -5pt 0
\\
0 &\hskip -5pt G_{\rm R}(i{\mathsf p}+{\mathsf t} + \frac{1}{2b})
\end{array}
\right)^{-1}
\hskip -5pt.
\end{eqnarray*}
From (\ref{commutations:2}) we see that for an analytic function $f:$
\[
{\rm e}^{\alpha {\mathsf x}}f({\mathsf p},{\mathsf t}) {\rm e}^{-\alpha {\mathsf x}} \;\ = \;\ f({\mathsf p}+i\alpha,{\mathsf t}),
\]
so that
\begin{eqnarray*}
{\rm e}^{\frac12 b {\mathsf x}}
\left(
\begin{array}{cc}
0 & \hskip -7pt 1 + i{\rm e}^{-\pi b ({\mathsf p} - i{\mathsf t})}
\\[4pt]
\hskip -2pt
1 - i{\rm e}^{-\pi b ({\mathsf p} - i{\mathsf t})} & \hskip -5pt 0
\end{array}
\right)
{\rm e}^{\frac12 b {\mathsf x}}
& = &\!
\begin{array}{r}
{\mathbb G}_b\left(i{\mathsf p}+{\mathsf t}  + {\frac{Q}{2}}\right)
\end{array}\!\!
\left(
\begin{array}{cc}
0 & {\rm e}^{b{\mathsf x}}
\\
{\rm e}^{b{\mathsf x}} & 0
\end{array}
\right)\!
\begin{array}{l}
{\mathbb G}_b\left(i{\mathsf p}+{\mathsf t} + \frac{Q}{2}\right)^{-1}\hskip -10pt ,
\end{array}
\end{eqnarray*}
where we denoted
\[
\begin{array}{l}
{\mathbb G}_b\left(z\right)
\end{array}
\; = \;
\left(
\begin{array}{cc}
G_{\rm NS}(z) & 0
\\
0 & G_{\rm R}(z)
\end{array}
\right).
\]
Our definition of ``even'' and ``odd'' complex powers, (\ref{complex:powers}), thus gives:
\begin{eqnarray*}
\label{Qc:Q:power}
\nonumber
({\mathsf Q}_{\rm\scriptscriptstyle I}^c +{\mathsf Q}_{\rm\scriptscriptstyle I})_\rho^{s_1}
& = &
{\rm e}^{-\frac12 i\pi bs_1 {\mathsf t}}
\begin{array}{r}
{\mathbb G}_b\left(i{\mathsf p}+{\mathsf t}  + {\frac{Q}{2}}\right)
\end{array}
{\rm e}^{bs_1{\mathsf x}}\,
{\mathbf 1}_\rho\,
\begin{array}{r}
{\mathbb G}_b\left(i{\mathsf p}+{\mathsf t}  + {\frac{Q}{2}}\right)^{-1}
\end{array}
\\[-6pt]
\\[-6pt]
\nonumber
& = &
{\rm e}^{bs_1{\mathsf x}}\,
{\rm e}^{-\frac12 i\pi bs_1 {\mathsf t}}\,
\begin{array}{r}
{\mathbb G}_b\left(i{\mathsf p}+{\mathsf t}  +bs_1 + {\frac{Q}{2}}\right)
\end{array}
{\mathbf 1}_\rho\,
\begin{array}{r}
{\mathbb G}_b\left(i{\mathsf p}+{\mathsf t}  + {\frac{Q}{2}}\right)^{-1}
\end{array}\hskip -10pt,
\end{eqnarray*}
with ${\mathbf 1}_{\rm e} = \left(^{1\;0}_{0\;1}\right)$ and ${\mathbf 1}_{\rm o} = \left(^{0\;1}_{1\;0}\right).$
Similarly:
\begin{eqnarray*}
\label{Qc:Qprim:power}
({\mathsf Q}_{\rm\scriptscriptstyle I}^c +{\mathsf Q}'_{\rm\scriptscriptstyle I})^{t_2}_{\lambda}
& = &
{\rm e}^{bt_2{\mathsf x}}\,
{\rm e}^{-\frac12 i\pi bt_2 {\mathsf t}}\,
\begin{array}{r}
{\mathbb G}_b\left(-i{\mathsf p}+{\mathsf t} -bt_2 + {\frac{Q}{2}}\right)^{-1}
\end{array}
{\mathbf 1}_\lambda\,
\begin{array}{r}
{\mathbb G}_b\left(-i{\mathsf p}+{\mathsf t} + {\frac{Q}{2}}\right)
\end{array},
\end{eqnarray*}
and
\begin{eqnarray*}
\label{QQ:products}
\nonumber
&&
\hskip -1cm
({\mathsf Q}_{\rm\scriptscriptstyle I}^c + {\rm e}^{-2i\pi b \alpha_1}{\mathsf Q}_{\rm\scriptscriptstyle I}')_\rho^{s_2}
({\mathsf Q}_{\rm\scriptscriptstyle I}^c +{\mathsf Q}_{\rm\scriptscriptstyle I})_\eta^{s_1}
\; = \;
{\rm e}^{b(s_1+s_2){\mathsf x}}\,
{\rm e}^{-\frac12 i\pi b(s_1+s_2) {\mathsf t}}
\\ \nonumber
& \times &
\begin{array}{r}
{\mathbb G}_b\left(-i{\mathsf p}+{\mathsf t} -2\alpha_1 -bs_2-bs_2 + {\frac{Q}{2}}\right)^{-1}
\end{array}
{\mathbf 1}_\rho\,
\begin{array}{r}
{\mathbb G}_b\left(-i{\mathsf p}+{\mathsf t} -2\alpha_1 -bs_1 + {\frac{Q}{2}}\right)
\end{array}
\\ \nonumber
& \times &
\begin{array}{r}
{\mathbb G}_b\left(i{\mathsf p}+{\mathsf t}  +bs_1 + {\frac{Q}{2}}\right)
\end{array}
{\mathbf 1}_\eta\,
\begin{array}{r}
{\mathbb G}_b\left(i{\mathsf p}+{\mathsf t}  + {\frac{Q}{2}}\right)^{-1}
\end{array}\hskip -10pt,
\\[-4pt]
\\[-4pt]
\nonumber
&&
\hskip - 1cm
({\mathsf Q}_{\rm\scriptscriptstyle I}^c +{\rm e}^{2i\pi b \alpha_2}{\mathsf Q}_{\rm\scriptscriptstyle I})^{t_1}_{\lambda}
({\mathsf Q}_{\rm\scriptscriptstyle I}^c +{\mathsf Q}'_{\rm\scriptscriptstyle I})^{t_2}_{\delta}
\; =\;
{\rm e}^{b(t_1+t_2){\mathsf x}}\,
{\rm e}^{-\frac12 i\pi b(t_1+t_2) {\mathsf t}}
\\ \nonumber
& \times &
\begin{array}{r}
{\mathbb G}_b\left(i{\mathsf p}+{\mathsf t} + 2\alpha_2 + bt_1 + bt_2+ {\frac{Q}{2}}\right)
\end{array}
{\mathbf 1}_\lambda\,
\begin{array}{r}
{\mathbb G}_b\left(i{\mathsf p}+{\mathsf t} +2\alpha_2 + bt_2 + {\frac{Q}{2}}\right)^{-1}
\end{array}
\\ \nonumber
& \times &
\begin{array}{r}
{\mathbb G}_b\left(-i{\mathsf p}+{\mathsf t} -bt_2 + {\frac{Q}{2}}\right)^{-1}
\end{array}
{\mathbf 1}_\delta\,
\begin{array}{r}
{\mathbb G}_b\left(-i{\mathsf p}+{\mathsf t} + {\frac{Q}{2}}\right)
\end{array}.
\end{eqnarray*}

Since $\sf p$ and $\sf t$ commute we can evaluate the action of both sides of (\ref{braiding:defs:1})
on a common eigenstate of the momentum $\sf p$ (with an eigenvalue $p_1\in \mathbb R$) and $\sf t$ (with an eigenvalue $\tau \in i{\mathbb R}$).
Conservation of the momentum gives
\begin{equation*}
\label{mom:con}
t_1 + t_2 \; = \; s_1 + s_2 \; \stackrel{\rm def}{=} \; s
\end{equation*}
and the integration measure thus reads $d\mu(t_1,t_2) = \delta(t_1+t_2-s)d\mu(t_1)d\mu(t_2).$
Define:
\begin{eqnarray}
\nonumber
\label{parameters}
A_1 & = & p_1 - 2i\alpha_1 -ibs,
\hskip 2mm
B_1 \; = \; -p_1,
\hskip 2mm
C_1 \; = \; i(\alpha_1 - Q/2),
\hskip 2mm
p_s \; = \; p_1-i(\alpha_1 + bs_1),
\\[-6pt]
\\[-6pt]
\nonumber
A_2 & = & p_1 - 2i\alpha_2 -ibs,
\hskip 2mm
B_2 \; = \; -p_1,
\hskip 2mm
C_2 \; = \; i(\alpha_2 - Q/2),
\hskip 2mm
p_u \; = \; p_1-i(\alpha_2 + bt_2).
\end{eqnarray}

It turns out to be convenient to express the functions $G_{\rm NS,\,R}$ through their ``cousins'' $S_{\rm NS,\,R},$
see Appendix \ref{Appendix:Barnes}. Denoting
\[
\begin{array}{l}
{\mathbb S}_b\left(z\right)
\end{array}
\; = \;
\left(
\begin{array}{cc}
S_{\rm NS}(z) & 0
\\
0 & S_{\rm R}(z)
\end{array}
\right)
\]
and using the reflection property ${\mathbb S}_b\left(z\right){\mathbb S}_b\left(Q-z\right) = {\mathbf 1}_{\rm e}$
we get:
\begin{eqnarray}
\label{g:g:evaluated}
\nonumber
&&
\hskip -3mm
{\mathsf g}^{\alpha_2,\rho}_{s_2}(\sigma_2)
{\mathsf g}^{\alpha_1,\eta}_{s_1}(\sigma_1)\big|_{{}^{{\mathsf p} = p_1}_{{\mathsf t}=\tau}}
= \;
{\mathsf E}^{\alpha_2}(\sigma_2){\mathsf E}^{\alpha_1}(\sigma_1)\
{\rm e}^{bs{\mathsf x}}\
{\rm e}^{\frac{\pi b}{2}  p_1(s_2-s_1)-2i\pi b s_2 \alpha_1 + \frac{i\pi b^2}{2} s_1^2 - \frac{i\pi b^2}{4} s^2}
\\[6pt]
\nonumber
 &&  \hskip -1mm \times
\begin{array}{r}
{\mathbb S}_b\!\left(\frac{Q}{2}+iA_1\!-\!\tau\right)
\end{array}\!
F_\rho^{\scriptscriptstyle T}
\begin{array}{r}
{\mathbb S}^{-1}_b\!\left(Q-iC_1+ip_s\!-\!\tau\right)
\end{array}\!
\begin{array}{r}
{\mathbb S}^{-1}_b\!\left(Q-iC_1-ip_s\!-\!\tau\right)
\end{array}\!
F_\eta
\begin{array}{r}
{\mathbb S}_b\!\left(\frac{Q}{2}+iB_1\!-\!\tau\right)\!,
\end{array}
\\[10pt]
&&
\hskip -3mm
{\mathsf g}^{\alpha_1,\lambda}_{t_1}(\sigma_1)
{\mathsf g}^{\alpha_2,\delta}_{t_2}(\sigma_2)\big|_{{}^{{\mathsf p} = p_1}_{{\mathsf t}=\tau}}
= \;
{\mathsf E}^{\alpha_2}(\sigma_2){\mathsf E}^{\alpha_1}(\sigma_1)\
{\rm e}^{bs{\mathsf x}}\
{\rm e}^{\frac{\pi b}{2}  p_1(s_2-s_1) + i\pi \alpha_1 \alpha_2 - \frac{i\pi b^2}{2} t_2^2 + \frac{i\pi b^2}{4} s^2}
\\[6pt]
\nonumber
&& \hskip -1mm \times
\begin{array}{r}
{\mathbb S}_b\!\left(\frac{Q}{2}+iA_2\!+\!\tau\right)
\end{array}\!
F_\lambda
\begin{array}{r}
{\mathbb S}^{-1}_b\!\!\left(Q-iC_2+ip_u\!+\!\tau\right)
\end{array}\!
\begin{array}{r}
{\mathbb S}^{-1}_b\!\!\left(Q-iC_2-ip_u\!+\!\tau\right)
\end{array}
F_\delta^{\scriptscriptstyle T}
\begin{array}{r}
{\mathbb S}_b\!\left(\frac{Q}{2}+iB_2\!+\!\tau\right)\!,
\end{array}
\end{eqnarray}
where:
\[
F_{\rho} \; = \;
\left(\begin{array}{cc} 1 & 0 \\ 0 & \;{\rm e}^{-\frac{i\pi}{4}} \end{array}\right)
{\mathbf 1}_{\rho}
\left(\begin{array}{cc} 1 & 0 \\ 0 & \;{\rm e}^{-\frac{i\pi}{4}} \end{array}\right)^{-1}
\Rightarrow
\hskip 10pt
F_{\rm e}\; = \;
\left(\begin{array}{cc} 1 & 0 \\ 0 & 1 \end{array}\right),
\hskip 3mm
F_{\rm o}\; = \;
\left(\begin{array}{cc} 0 & {\rm e}^{\frac{i\pi}{4}} \\ {\rm e}^{-\frac{i\pi}{4}} & 0 \end{array}\right).
\]

Let us further denote:
\begin{eqnarray*}
\label{Phi:s}
\nonumber
\Phi^{(i)}_{\scriptscriptstyle \left[^{\rm\scriptscriptstyle N\:N}_{\rm N\:N}\right]}(p,\tau)
& = &
\frac{S_{\rm NS}(Q/2 + iA_i + \tau)S_{\rm NS}(Q/2 + iB_i + \tau)}
{S_{\rm NS}(Q -iC_i +ip + \tau)S_{\rm NS}(Q -iC_i -ip + \tau)},
\\[-4pt]
\\[-4pt]
\nonumber
\Phi^{(i)}_{\scriptscriptstyle \left[^{\rm\scriptscriptstyle R\:N}_{\rm N\:N}\right]}(p,\tau)
& = &
\frac{S_{\rm R}(Q/2 + iA_i + \tau)S_{\rm NS}(Q/2 + iB_i + \tau)}
{S_{\rm NS}(Q -iC_i +ip + \tau)S_{\rm NS}(Q -iC_i -ip + \tau)},
\end{eqnarray*}
e.t.c.\ and define:
\begin{eqnarray}
\nonumber
{\mathbf \Phi}^{\rm e}_{(1)}(p,\tau)
& = &
\left(
\begin{array}{cc}
\Phi^{(1)}_{\scriptscriptstyle \left[^{\rm\scriptscriptstyle N\:N}_{\rm N\:N}\right]}
& \;{\rm e}^{-\frac{i\pi}{2}}
\Phi^{(1)}_{\scriptscriptstyle \left[^{\rm\scriptscriptstyle N\:N}_{\rm R\:R}\right]}
\\[16pt]
\Phi^{(1)}_{\scriptscriptstyle \left[^{\rm\scriptscriptstyle R\:R}_{\rm R\:R}\right]}
& \;{\rm e}^{\frac{i\pi}{2}}
\Phi^{(1)}_{\scriptscriptstyle \left[^{\rm\scriptscriptstyle R\:R}_{\rm N\:N}\right]}
\end{array}
\right)\!(p,\tau)\,,
\hskip 4mm
{\mathbf \Phi}^{\rm e}_{(2)}(p,\tau)
=
\left(
\begin{array}{cc}
\Phi^{(2)}_{\scriptscriptstyle \left[^{\rm\scriptscriptstyle N\:N}_{\rm N\:N}\right]}
& \;{\rm e}^{\frac{i\pi}{2}}
\Phi^{(2)}_{\scriptscriptstyle \left[^{\rm\scriptscriptstyle N\:N}_{\rm R\:R}\right]}
\\[16pt]
\nonumber
\Phi^{(2)}_{\scriptscriptstyle \left[^{\rm\scriptscriptstyle R\:R}_{\rm R\:R}\right]}
& \;{\rm e}^{-\frac{i\pi}{2}}
\Phi^{(2)}_{\scriptscriptstyle \left[^{\rm\scriptscriptstyle R\:R}_{\rm N\:N}\right]}
\end{array}
\right)\!(p,\tau),
\\
\label{Phi:definitions}
\\
\nonumber
{\mathbf \Phi}^{\rm o}_{(1)}(p,\tau)
& \!= \!&
\!\left(
\begin{array}{cc}
\!{\rm e}^{\frac{i\pi}{4}}
\Phi^{(1)}_{\scriptscriptstyle \left[^{\rm\scriptscriptstyle N\:R}_{\rm N\:N}\right]}
&
\! {\rm e}^{-\frac{i\pi}{4}}
\Phi^{(1)}_{\scriptscriptstyle \left[^{\rm\scriptscriptstyle N\:R}_{\rm R\:R}\right]}\!\!
\\[16pt]
\!{\rm e}^{-\frac{i\pi}{4}}
\Phi^{(1)}_{\scriptscriptstyle \left[^{\rm\scriptscriptstyle R\:N}_{\rm R\:R}\right]}
&
\! {\rm e}^{\frac{i\pi}{4}}
\Phi^{(1)}_{\scriptscriptstyle \left[^{\rm\scriptscriptstyle R\:N}_{\rm N\:N}\right]}
\end{array}
\right)\!(p,\tau),
\;
{\mathbf \Phi}^{\rm o}_{(2)}(p,\tau)
\!= \!
\left(
\begin{array}{cc}
\!{\rm e}^{-\frac{i\pi}{4}}
\Phi^{(2)}_{\scriptscriptstyle \left[^{\rm\scriptscriptstyle N\:R}_{\rm N\:N}\right]}
&
\! {\rm e}^{\frac{i\pi}{4}}
\Phi^{(2)}_{\scriptscriptstyle \left[^{\rm\scriptscriptstyle N\:R}_{\rm R\:R}\right]}
\\[16pt]
\!{\rm e}^{\frac{i\pi}{4}}
\Phi^{(2)}_{\scriptscriptstyle \left[^{\rm\scriptscriptstyle R\:N}_{\rm R\:R}\right]}
&
\! {\rm e}^{-\frac{i\pi}{4}}
\Phi^{(2)}_{\scriptscriptstyle \left[^{\rm\scriptscriptstyle R\:N}_{\rm N\:N}\right]}
\end{array}\!
\right)\!(p,\tau).
\end{eqnarray}
It follows from (\ref{g:g:evaluated}) that
(\ref{braiding:defs:1}) hold provided the relations
\begin{equation}
\label{braiding:matrix}
{\rm e}^{i\chi}
{\mathbf \Phi}^{\gamma}_{(1)}(p_s,-\tau)
=
\int\! d\mu(t_2)\ {\rm e}^{-\frac{i\pi }{2}(bt_2)^2 +\pi  p_1 bt_2}\
{\mathbf \Phi}^{\gamma}_{(2)}(p_u,\tau)
\cdot
\big[{\mathbf B}^{\gamma,+}\big]^{\rm T}
\end{equation}
are satisfied, where
\[
i\chi = \frac{i\pi b^2}{2}(s_1^2 - {s}^2) +\pi p_1 b s_{2} -i\pi\alpha_1(\alpha_2 +2bs_2),
\]
and we have denoted (see (\ref{shorthand:for:B}))
\[
{\mathbf B}^{\gamma,+}
\; = \;
\left(
\begin{array}{cc}
\big[B^{\gamma,+}\big]^{\rm e}_{\hskip 4pt \rm e}\ & \big[B^{\gamma,+}\big]^{\rm e}_{\hskip 4pt \rm o}
\\[4pt]
\big[B^{\gamma,+}\big]^{\rm o}_{\hskip 4pt \rm e}\ & \big[B^{\gamma,+}\big]^{\rm o}_{\hskip 4pt \rm o}
\end{array}
\right).
\]
Notice that we have traded (\ref{exchange:1}) --- a relation between unitary (for the parameters satisfying
(\ref{parameters:imaginary})) operators --- for a relation between meromorphic functions. At this point
we can analytically continue (\ref{braiding:matrix}) to a ``physical'' values of the parameters,
\begin{equation}
\label{parameters:physical:values}
\alpha_k \in \frac{Q}{2} + i{\mathbb R},
\hskip 5mm
bs_k, bt_k \in -\frac{Q}{2} + i{\mathbb R},
\hskip 5mm
k = 1,2.
\end{equation}

For $\alpha_k,s_k$ and $t_k$ satisfying (\ref{parameters:physical:values}) all the parameters
defined in (\ref{parameters}) are real.
Since for $x\in {\mathbb R}:$
\[
\left|S_{\rm NS}\left(Q/2 + ix\right)\right|
\; = \;
\left|S_{\rm R}\left(Q/2 + ix\right)\right|
\; = \;
1,
\]
we get for $A_i, B_i, C_i \in {\mathbb R},\ \tau \in i{\mathbb R}:$
\begin{eqnarray*}
\left({\mathbf \Phi}^{\rm e}_{(1)}(p_u,\tau)\right)^{\dagger}
\cdot
{\mathbf \Phi}^{\rm e}_{(1)}(p'_u,\tau)
=
\left({\mathbf \Phi}^{\rm o}_{(2)}(p_u,\tau)\right)^{\dagger}
\cdot
{\mathbf \Phi}^{\rm o}_{(2)}(p'_u,\tau)
& \stackrel{\rm def}{=} &
 {\mathbf \Theta}(p_u,p_u';\tau).
\end{eqnarray*}
Explicitly:
\begin{eqnarray*}
\nonumber
&&
\hskip -1cm
{\mathbf \Theta}(p_u,p_u';\tau)^{\rm e}_{\hskip 4pt\rm e}
\; = \;
{\mathbf \Theta}(p_u,p_u';\tau)^{\rm o}_{\hskip 4pt\rm o}
\\[6pt]
\nonumber
& = &
\left(
\frac{S_{\rm NS}(ip_u - (\tau - iC_2))}
{S_{\rm NS}(Q+ip_u +(\tau - iC_2))}
\right)^{\dag}
\left(
\frac{S_{\rm NS}(ip_u' - (\tau - iC_2))}
{S_{\rm NS}(Q+ip_u' +(\tau - iC_2))}
\right)
\\[4pt]
& + &
\left(
\frac{S_{\rm\scriptscriptstyle R}(ip_u - (\tau - iC_2))}
{S_{\rm\scriptscriptstyle R}(Q+ip_u +(\tau - iC_2))}
\right)^{\dag}
\left(
\frac{S_{\rm\scriptscriptstyle R}(ip_u' - (\tau - iC_2))}
{S_{\rm\scriptscriptstyle R}(Q+ip_u' +(\tau - iC_2))}
\right)
\\[10pt]
\nonumber
& \equiv &
\left\langle ip_u\left|^{\rm\scriptscriptstyle N}_{\rm\scriptscriptstyle N}\right|\tau-iC_2\right\rangle
\left\langle \tau-iC_2\left|^{\rm\scriptscriptstyle N}_{\rm\scriptscriptstyle N}\right|ip_u'\right\rangle
+
\left\langle ip_u\left|^{\rm\scriptscriptstyle R}_{\rm\scriptscriptstyle R}\right|\tau-iC_2\right\rangle
\left\langle \tau-iC_2\left|^{\rm\scriptscriptstyle R}_{\rm\scriptscriptstyle R}\right|ip_u'\right\rangle,
\end{eqnarray*}
where in the last line we have borrowed the notation from \cite{Hadasz:2007wi}, section 5.2. In the same notation:
\begin{eqnarray*}
&&
\hskip -1cm
{\rm e}^{-\frac{i\pi}{2}}{\mathbf \Theta}(p_u,p_u';\tau)^{\rm e}_{\hskip 4pt\rm o}
\; = \;
{\rm e}^{\frac{i\pi}{2}}{\mathbf \Theta}(p_u,p_u';\tau)^{\rm o}_{\hskip 4pt\rm e}
\\[6pt]
\nonumber
& = &
\left\langle ip_u\left|^{\rm\scriptscriptstyle N}_{\rm\scriptscriptstyle N}\right|\tau-iC_2\right\rangle
\left\langle \tau-iC_2\left|^{\rm\scriptscriptstyle R}_{\rm\scriptscriptstyle R}\right|ip_u'\right\rangle
-
\left\langle ip_u\left|^{\rm\scriptscriptstyle R}_{\rm\scriptscriptstyle R}\right|\tau-iC_2\right\rangle
\left\langle \tau-iC_2\left|^{\rm\scriptscriptstyle N}_{\rm\scriptscriptstyle N}\right|ip_u'\right\rangle
\end{eqnarray*}
and, using \cite{Hadasz:2007wi} (equations (5.12) and (5.13)) we arrive at the orthogonality relation
\begin{equation}
\label{orthogonality}
\int\limits_{-i\infty}^{i\infty}\hskip -6pt \frac{d\tau}{i}\ {\mathbf \Theta}(p_u,p_u';\tau)
\; = \;
\frac{1}{\sinh\pi b p_u\,\sinh\pi b^{-1}p_u}\,\delta(p_u - p_u')\, {\mathbf 1}_{\rm e}.
\end{equation}
For fixed $p_1,\alpha_i$ and $s_i$ we see from (\ref{parameters}) that the only parameter which changes with $t_2$ is $p_u.$
In view of (\ref{orthogonality}) it is therefore convenient to take $d\mu(t_2) = dp_u.$
We thus get from (\ref{braiding:matrix}):
\begin{equation}
\label{braiding:almost:final}
\int\limits_{-i\infty}^{i\infty}\hskip -6pt \frac{d\tau}{i}
\left({\mathbf \Phi}^{\gamma}_{(2)}(p_u,\tau)\right)^{\dagger}
\cdot
{\mathbf\Phi}^{\gamma}_{(1)}(p_s,-\tau)
=
\frac{{\rm e}^{-\frac{i\pi }{2}(bt_2)^2 +\pi  p_1 bt_2-i\chi}}{\sinh\pi b p_u\,\sinh\pi b^{-1}p_u}\,
\big[{\mathbf B}^{\gamma,+}\big]^{\rm T}
\end{equation}
with $bt_2 = i(p_u-p_1)-\alpha_2.$

There is a point concerning (\ref{braiding:almost:final}) which requires some  care. Since
$S_{\rm NS}(x)$ vanishes at $x = Q$ the functions
\[
\Phi^{(i)}_{\scriptscriptstyle \left[^{\: \cdot \: \cdot }_{\: \cdot \: \cdot }\right]}(Q+iy + \tau),
\hskip 5mm y \in {\mathbb R},
\]
with $S_{\rm NS}(Q+iy+\tau)$ in the denominator have poles at the imaginary $\tau$ axis. Relations
(\ref{orthogonality}) and (\ref{braiding:almost:final}) are thus not well defined unless we specify the
way the integration contour avoids these poles. To do this recall that the ``physical'' values of the
parameters (\ref{parameters:physical:values}) were obtained in a process of analytic continuation
from the purely imaginary values assumed in (\ref{parameters:imaginary}). It is immediate to see that
if (\ref{parameters:imaginary}) holds, then the discussed poles are located to the right from the imaginary axis.
During the analytic continuation process no pole is allowed to cross the integration contour, so we take the contour
in (\ref{orthogonality}) and (\ref{braiding:almost:final}) to the left from the poles coming from the  $S^{-1}_{\rm NS}$ factors.
This coincides with the assumption made in \cite{Hadasz:2007wi} to derive the relation
(\ref{orthogonality}).

To present the un-normalized braiding matrices in the final form let us introduce one more set of
(the most commonly used) variables:
\begin{equation}
\label{final:variables}
a_1 = \frac{Q}{2} + ip_1,
\hskip 2mm
a_2 = \alpha_1,
\hskip 2mm
a_3 = \alpha_2,
\hskip 2mm
a_s = \frac{Q}{2} + ip_s,
\hskip 2mm
a_u = \frac{Q}{2} + ip_u,
\hskip 2mm
a_4 = a_3 + a_2 + a_1 + bs.
\end{equation}
In these variables:
\begin{eqnarray*}
\frac{Q}{2} + iA_1
& = &
a_4 -a_3 + a_2,
\hskip 5mm
\frac{Q}{2} - iA_2
\; = \;
a_4 +a_3 - a_2,
\\
\frac{Q}{2} + iB_1
& = &
\frac{Q}{2} + iB_2
\; = \;
Q-a_1,
\\[4pt]
Q - iC_1 + ip_s
& = &
a_s + a_2,
\hskip 5mm
Q - iC_1 - ip_s
\; = \;
Q-a_s + a_2,
\\
Q - iC_2 + ip_u
& = &
a_u + a_3,
\hskip 5mm
Q - iC_2 - ip_u
\; = \;
Q-a_u + a_3
\end{eqnarray*}
and
\begin{equation*}
\frac{i\pi }{2}(bt_2)^2 -\pi  p_1 bt_2+i\chi
=
i\pi
\Big(
\Delta_4 + \Delta_1-\Delta_u-\Delta_s + a_3(a_4-a_u) - a_2(a_4-a_s)
\Big)
\stackrel{\rm def}{=}
i\varphi_{a_sa_u}(a_3,a_2),
\end{equation*}
with $\Delta_i \equiv \Delta(a_i) = \frac12a_i(Q-a_i).$ Using the relation
\[
\sinh(\pi b p_u)\sinh(\pi b^{-1} p_u)
\; = \;
\frac14 S_{\rm NS}(2a_u)S_{\rm NS}(2Q-2a_u)
\]
and noticing that $\big[{\mathbf B}^{\gamma,+}\big]$ written as a function of $a_k,\ k=1,2,3,4,s,u$ is nothing
but the unnormalized braiding matrix
we get from (\ref{braiding:almost:final}):
\begin{equation}
\label{braiding:final}
{}^{\mathbf g}{\sf B}^{\gamma,+}_{a_sa_u}\left[^{a_3\:a_2}_{a_4\:a_1}\right]
=
{\rm e}^{i\varphi_{a_sa_u}(a_3,a_2)}\frac{S_{\rm NS}(2a_u)S_{\rm NS}(2Q-2a_u)}{4}\,
\int\limits_{-i\infty}^{i\infty}\hskip -6pt \frac{d\tau}{i}\
{\mathbf J}^{\gamma}_{a_sa_u}\!\!\left[^{a_3\:a_2}_{a_4\:a_1}\right]\!(\tau)
\end{equation}
with
\begin{equation*}
\label{J:definition}
{\mathbf J}^{\gamma}_{a_sa_u}\!\!\left[^{a_3\:a_2}_{a_4\:a_1}\right]\!(\tau)
=
\left({\mathbf\Phi}^{\gamma}_{(1)}(p_s,-\tau)\right)^{\rm T}
\cdot
\overline{{\mathbf \Phi}^{\gamma}_{(2)}(p_u,\tau)}
\end{equation*}
and where, as in (\ref{shorthand:for:B}), we used the notation
\begin{equation}
\label{notation:repeated}
{}^{\mathbf g}{\sf B}^{{\rm e},+}_{a_sa_u}\left[^{a_3\:a_2}_{a_4\:a_1}\right]^{\rho}_{\hskip 4pt\eta}
=\
{}^{\mathbf g}{\sf B}^{+}_{a_sa_u}\left[^{a_3\:a_2}_{a_4\:a_1}\right]^{\rho\rho}_{\hskip 8pt\eta\eta},
\hskip 1cm
{}^{\mathbf g}{\sf B}^{{\rm o},+}_{a_sa_u}\left[^{a_3\:a_2}_{a_4\:a_1}\right]^{\rho}_{\hskip 4pt\eta}
=\
{}^{\mathbf g}{\sf B}^{+}_{a_sa_u}\left[^{a_3\:a_2}_{a_4\:a_1}\right]^{\rho\bar\rho}_{\hskip 8pt\eta\bar\eta}.
\end{equation}

Explicitly:
\begin{equation*}
\label{J:even}
{\mathbf J}^{\rm e}(\tau)
=
\left(
\begin{array}{ccc}
\left[^{\rm\textstyle N\:N\:N\:N}_{\rm\textstyle N\:N\:N\:N}\right]\!(\tau)
+ \left[^{\rm\textstyle R\:R\:R\:R}_{\rm\textstyle R\:R\:R\:R}\right]\!(\tau)
&\;\; &
\frac{1}{i}\left[^{\rm\textstyle N\:N\:N\:N}_{\rm\textstyle R\:R\:N\:N}\right]\!(\tau)
+ i\left[^{\rm\textstyle R\:R\:R\:R}_{\rm\textstyle N\:N\:R\:R}\right]\!(\tau)
\\[10pt]
\frac{1}{i}\left[^{\rm\textstyle N\:N\:N\:N}_{\rm\textstyle N\:N\:R\:R}\right]\!(\tau)
+ i\left[^{\rm\textstyle R\:R\:R\:R}_{\rm\textstyle R\:R\:N\:N}\right]\!(\tau)
&&
- \left[^{\rm\textstyle N\:N\:N\:N}_{\rm\textstyle R\:R\:R\:R}\right]\!(\tau)
- \left[^{\rm\textstyle R\:R\:R\:R}_{\rm\textstyle N\:N\:N\:N}\right]\!(\tau)
\end{array}
\right)
\end{equation*}
and
\begin{equation*}
\label{J:odd}
{\mathbf J}^{\rm o}(\tau)
=
\left(
\begin{array}{ccc}
i\left[^{\rm\textstyle N\:R\:N\:R}_{\rm\textstyle N\:N\:N\:N}\right]\!(\tau)
+ \frac{1}{i}\left[^{\rm\textstyle R\:N\:R\:N}_{\rm\textstyle R\:R\:R\:R}\right]\!(\tau)
&\;\;&
\left[^{\rm\textstyle N\:R\:N\:R}_{\rm\textstyle R\:R\:N\:N}\right]\!(\tau)
+ \left[^{\rm\textstyle R\:N\:R\:N}_{\rm\textstyle N\:N\:R\:R}\right]\!(\tau)
\\[10pt]
\left[^{\rm\textstyle N\:R\:N\:R}_{\rm\textstyle N\:N\:R\:R}\right]\!(\tau)
+ \left[^{\rm\textstyle R\:N\:R\:N}_{\rm\textstyle R\:R\:N\:N}\right]\!(\tau)
&&
\frac{1}{i}\left[^{\rm\textstyle N\:R\:N\:R}_{\rm\textstyle R\:R\:R\:R}\right]\!(\tau)
+i \left[^{\rm\textstyle R\:N\:R\:N}_{\rm\textstyle N\:N\:N\:N}\right]\!(\tau)
\end{array}
\right),
\end{equation*}
where the abbreviations
\begin{eqnarray*}
\label{8:abbreviation}
\nonumber
\left[^{\rm\textstyle N\:N\:N\:N}_{\rm\textstyle N\:N\:N\:N}\right]\!(\tau)
& = &
\frac{
S_{\rm NS}(\bar a_4-a_3+a_2+\tau)S_{\rm NS}(a_1+\tau)S_{\rm NS}(a_4-a_3+a_2+\tau)S_{\rm NS}(\bar a_1+\tau)
}{
S_{\rm NS}(\bar a_u + \bar a_3+\tau)S_{\rm NS}(a_u+ \bar a_3+\tau)S_{\rm NS}(a_s + a_2 +\tau)S_{\rm NS}(\bar a_s + a_2 +\tau)
},
\\[2pt]
\\[-2pt]
\nonumber
\left[^{\rm\textstyle N\:R\:N\:R}_{\rm\textstyle N\:N\:N\:N}\right]\!(\tau)
& = &
\frac{
S_{\rm NS}(\bar a_4-a_3+a_2+\tau)S_{\rm R}(a_1+\tau)S_{\rm NS}(a_4-a_3+a_2+\tau)S_{\rm R}(\bar a_1+\tau)
}{
S_{\rm NS}(\bar a_u + \bar a_3+\tau)S_{\rm NS}(a_u+ \bar a_3+\tau)S_{\rm NS}(a_s + a_2 +\tau)S_{\rm NS}(\bar a_s + a_2 +\tau)
},
\end{eqnarray*}
e.t.c.\ with $\bar a_i = Q-a_i$ have been applied.

Using the results of \cite{Hadasz:2007wi} it is straightforward to check that the functions
${\mathbf \Phi}^{\sigma}_{(i)}(p,\tau)$ satisfy a completeness relation of the form
\begin{equation}
\label{completness}
\int\limits_0^{\infty}\!dp_s\
\frac{S_{\rm NS}(2a_s)S_{\rm NS}(2Q-2a_s)}{4}\,
\left({\mathbf \Phi}^{\gamma}_{(i)}(p_s,\tau)\right)^{\dagger}
\cdot
{\mathbf \Phi}^{\gamma}_{(i)}(p_s,\lambda)
=
\delta(\tau-\lambda)\,{\mathbf 1}_{\rm e}.
\end{equation}
From (\ref{braiding:final}), (\ref{orthogonality}) and (\ref{completness}) it then follows that
\begin{equation*}
\int\limits_0^{\infty}\!dp_s\
{\rm e}^{-2\pi i(\Delta_4+\Delta_1-\Delta_t-\Delta_s)}\,
{}^{\mathbf g}{\sf B}^{\gamma,+}_{a_ta_s}\left[^{a_2\:a_3}_{a_4\:a_1}\right]
\cdot
{}^{\mathbf g}{\sf B}^{\gamma,+}_{a_sa_u}\left[^{a_3\:a_2}_{a_4\:a_1}\right]
=
\delta(p_t-p_u)\,{\mathbf 1}_{\rm e}.
\end{equation*}
This equality implies that
\[
{\rm e}^{-2\pi i(\Delta_4+\Delta_1-\Delta_t-\Delta_s)}\,
{}^{\mathbf g}{\sf B}^{\gamma,+}_{a_ta_s}\left[^{a_2\:a_3}_{a_4\:a_1}\right]
=
{}^{\mathbf g}{\sf B}^{\gamma,-}_{a_ta_s}\left[^{a_2\:a_3}_{a_4\:a_1}\right]
\]
and allows to write
\begin{equation}
\label{g:braiding:matrix}
{}^{\mathbf g}{\sf B}^{\gamma,\,\epsilon}_{a_sa_u}\left[^{a_3\:a_2}_{a_4\:a_1}\right]
=
{\rm e}^{i\pi\epsilon(\Delta_4+\Delta_1-\Delta_s-\Delta_u)}\
{}^{\mathbf g}{\sf B}^{\gamma}_{a_sa_u}\!\left[^{a_3\:a_2}_{a_4\:a_1}\right]
\end{equation}
with $\epsilon = {\rm sign}(\sigma_2-\sigma_1)$ and
\[
{}^{\mathbf g}{\sf B}^{\gamma}_{a_sa_u}\left[^{a_3\:a_2}_{a_4\:a_1}\right]
=
{\rm e}^{i\pi(a_3(a_4-a_u) - a_2(a_4-a_s))}\,
\frac{S_{\rm NS}(2a_u)S_{\rm NS}(2Q-2a_u)}{4}\
\int\limits_{-i\infty}^{i\infty}\hskip -6pt \frac{d\tau}{i}\
{\mathbf J}^{\gamma}_{a_sa_u}\left[^{a_3\:a_2}_{a_4\:a_1}\right]\!(\tau).
\]

To derive the other three braiding matrices which appear in (\ref{braiding:definition:2}) we use
the realization of the vertex operator ${}^{\mathbf g}V_{a_3a_1}(*\nu_\alpha |w)$ provided by the chiral descendants
\[
*g^{\alpha,{\rm e}}_s(\sigma) = \{S_{-1/2},g^{\alpha,{\rm o}}_s(\sigma)\} = -i\alpha\psi(\sigma)g^{\alpha,{\rm o}}_s(\sigma)
\]
and
\[
*g^{\alpha,{\rm o}}_s(\sigma) = [S_{-1/2},g^{\alpha,{\rm e}}_s(\sigma)] = -i\alpha\psi(\sigma)g^{\alpha,{\rm e}}_s(\sigma).
\]
Keeping in mind the definitions (\ref{final:variables}) and (\ref{parameters}) we have on the one hand
\begin{eqnarray*}
g^{\alpha_2,\rho}_{s_2}(\sigma_2)
*\!g^{\alpha_1,\eta}_{s_1}(\sigma_1)
& = &
\hskip -4pt
\int\limits_{\frac{Q}{2} + i{\mathbb R}}
\hskip -8pt \frac{d\alpha_u}{2i}\sum\limits_{\lambda\delta}\,
{}^{\mathbf g}{\sf B}^{\epsilon}_{a_sa_u}\!
\left[^{a_3\,*a_2}_{a_4\,\hskip 4pt a_1}\right]^{\rho\eta}_{\hskip 8pt \lambda\delta}
*\!g^{\alpha_1,\lambda}_{t_1}(\sigma_1)
g^{\alpha_2,\delta}_{t_2}(\sigma_2),
\end{eqnarray*}
and on the other
\begin{eqnarray*}
&&
\hskip -1cm
g^{\alpha_2,\rho}_{s_2}(\sigma_2)
*\!g^{\alpha_1,\eta}_{s_1}(\sigma_1)
\\[6pt]
& = &
-i\alpha_1
g^{\alpha_2,\rho}_{s_2}(\sigma_2)\,
\psi(\sigma_1)\,
g^{\alpha_1,\bar\eta}_{s_1}(\sigma_1)
\\[4pt]
& = &
(-1)^{|\rho|}(-i\alpha_1\psi(\sigma_1))\
g^{\alpha_2,\rho}_{s_2}(\sigma_2)\,
g^{\alpha_1,\bar\eta}_{s_1}(\sigma_1)
\\[6pt]
& = &
(-1)^{|\rho|}
\hskip -4pt
\int\limits_{\frac{Q}{2} + i{\mathbb R}}
\hskip -8pt \frac{d\alpha_u}{2i}\sum\limits_{\lambda,\delta}\,
{}^{\mathbf g}{\sf B}^{\epsilon}_{a_sa_u}\!
\left[^{a_3\:a_2}_{a_4\: a_1}\right]^{\rho\bar\eta }_{\hskip 8pt \lambda\delta}\,
(-i\alpha_1\psi(\sigma_1))
g^{\alpha_1,\lambda}_{t_1}(\sigma_1)
g^{\alpha_2,\delta}_{t_2}(\sigma_2)
\\[4pt]
& = &
(-1)^{|\rho|}
\hskip -4pt
\int\limits_{\frac{Q}{2} + i{\mathbb R}}
\hskip -8pt \frac{d\alpha_u}{2i}\sum\limits_{\lambda\delta}\,
{}^{\mathbf g}{\sf B}^{\epsilon}_{a_sa_u}\!
\left[^{a_3\:a_2}_{a_4\: a_1}\right]^{\rho\bar\eta }_{\hskip 8pt \lambda\delta}\,
*\!g^{\alpha_1,\bar\lambda}_{t_1}(\sigma_1)
g^{\alpha_2,\delta}_{t_2}(\sigma_2)
\\[4pt]
& = &
(-1)^{|\rho|}
\hskip -4pt
\int\limits_{\frac{Q}{2} + i{\mathbb R}}
\hskip -8pt \frac{d\alpha_u}{2i}\sum\limits_{\lambda\delta}\,
{}^{\mathbf g}{\sf B}^{\epsilon}_{a_sa_u}\!
\left[^{a_3\:a_2}_{a_4\: a_1}\right]^{\rho\bar\eta }_{\hskip 8pt \bar\lambda\delta}\,
*\!g^{\alpha_1,\lambda}_{t_1}(\sigma_1)
g^{\alpha_2,\delta}_{t_2}(\sigma_2).
\end{eqnarray*}
Together with (\ref{g:braiding:matrix}) this yields
\begin{equation}
\label{B01:B}
{}^{\mathbf g}{\sf B}_{a_sa_u}\!
\left[^{a_3\,*a_2}_{a_4\,\hskip 4pt a_1}\right]^{\rho\eta}_{\hskip 8pt \lambda\delta}
\; = \;
(-1)^{|\rho|}\
{}^{\mathbf g}{\sf B}_{a_sa_u}\!
\left[^{a_3\:a_2}_{a_4\: a_1}\right]^{\rho\bar\eta }_{\hskip 8pt \bar\lambda\delta}.
\end{equation}
Similar calculation also gives
\begin{equation}
\label{B10:B}
{}^{\mathbf g}{\sf B}_{a_sa_u}\!
\left[^{*a_3\:a_2}_{\hskip 4pt a_4\; a_1}\right]^{\rho\eta}_{\hskip 8pt \lambda\delta}
\; = \;
(-1)^{|\eta|}\
{}^{\mathbf g}{\sf B}_{a_sa_u}\!
\left[^{a_3\:a_2}_{a_4\: a_1}\right]^{\bar\rho\eta}_{\hskip 8pt \lambda\bar\delta}
\end{equation}
and
\begin{equation}
\label{B11:B}
{}^{\mathbf g}{\sf B}_{a_sa_u}\!
\left[^{*a_3\,*a_2}_{\hskip 4pt a_4\, \hskip 4pt a_1}\right]^{\rho\eta}_{\hskip 8pt \lambda\delta}
\; = \;
(-1)^{|\rho|+|\eta|+1}\
{}^{\mathbf g}{\sf B}_{a_sa_u}\!
\left[^{a_3\:a_2}_{a_4\: a_1}\right]^{\bar\rho\bar\eta}_{\hskip 8pt \bar\lambda\bar\delta}.
\end{equation}

\subsection{The normalized braiding matrices}
Equations (\ref{conjecture:1}) and (\ref{conjecture:2}) can be presented in the form
\begin{eqnarray}
\label{norm}
N^{\eta}_{\mathbf g}(a_3,a_2,a_1)
& =&
2^{|\eta|}\left[
\frac12\Gamma\left(\frac{bQ}{2}\right)
b^{-\frac{bQ}{2}}
\right]^{\frac{a_3 -a_1-a_2}{b}}
{\rm e}^{\frac{i\pi}{2}\left(a_3-a_2-a_1\right)\left(Q-a_3+a_2-a_1\right)}
\\[6pt]
\nonumber
& \times &
\frac{
\Gamma_{\eta}(a_3+a_1-a_2)
\Gamma_{\eta}(\bar a_3+a_1-a_2)
\Gamma_{\eta}(a_3+\bar a_1-a_2)
\Gamma_{\eta}(\bar a_3+\bar a_1-a_2)
}
{
\Gamma_{\rm NS}(Q)\Gamma_{\rm NS}(2a_1)\Gamma_{\rm NS}(Q-2a_2)\Gamma_{\rm NS}(2Q-2a_3)
}
\end{eqnarray}
where for $\eta ={\rm e}$ ($\eta = {\rm o}$) on the l.h.s.\ we have
$\eta = {\rm NS}$ (resp.\ $\eta = {\rm R}$) on the r.h.s. From (\ref{normalization:simple}) we  get:
\begin{eqnarray*}
{\sf B}^{\rm e,\,\epsilon}_{a_sa_u}\!
\left[^{a_3\:a_2}_{a_4\:a_1}\right]^{\eta}_{\hskip 5pt \rho}
& = &
\frac{N^\rho_{\mathbf g}(a_4,a_2,a_u)N^\rho_{\mathbf g}(a_u,a_3,a_1)}{N^\eta_{\mathbf g}(a_4,a_3,a_s)N^\eta_{\mathbf g}(a_s,a_2,a_1)}\
{}^{\mathbf g}{\sf B}^{\rm e,\,\epsilon}_{a_sa_u}\!
\left[^{a_3\:a_2}_{a_4\:a_1}\right]^{\eta}_{\hskip 5pt \rho},
\\[6pt]
{\sf B}^{\rm o,\,\epsilon}_{a_sa_u}\!
\left[^{a_3\:a_2}_{a_4\:a_1}\right]^{\eta}_{\hskip 5pt \rho}
& = &
\frac{N^\rho_{\mathbf g}(a_4,a_2,a_u)N^{\bar\rho}_{\mathbf g}(a_u,a_3,a_1)}{N^\eta_{\mathbf g}(a_4,a_3,a_s)N^{\bar\eta}_{\mathbf g}(a_s,a_2,a_1)}\
{}^{\mathbf g}{\sf B}^{\rm o,\,\epsilon}_{a_sa_u}\!
\left[^{a_3\:a_2}_{a_4\:a_1}\right]^{\eta}_{\hskip 5pt \rho},
\end{eqnarray*}
so that using  results from the previous subsection together with (\ref{norm}) we arrive at the explicit
expression
for the braiding matrix of the normalized chiral vertex operators defined in Eq.\ (\ref{braiding:definition:2}):
\begin{eqnarray}
\nonumber
\label{normalized:braiding:explicit}
{\sf B}^{\gamma,\,\epsilon}_{a_sa_u}\!
\left[^{a_3\:a_2}_{a_4\:a_1}\right]
& = &
{\rm e}^{\epsilon i\pi(\Delta_4+\Delta_1-\Delta_s-\Delta_u)}\,
{\sf B}^{\gamma}_{a_sa_u}\!
\left[^{a_3\:a_2}_{a_4\:a_1}\right],
\hskip .5cm \gamma = {\rm e,o},
\hskip 3mm \epsilon = {\rm sign}({\rm Arg}\,z_3 -{\rm Arg}\,z_2),
\\[2pt]
\nonumber
{\sf B}^{\gamma}_{a_sa_u}\!
\left[^{a_3\:a_2}_{a_4\:a_1}\right]
& = &
{\sf M}^{\gamma}_{a_sa_u}\!
\left[^{a_3\:a_2}_{a_4\:a_1}\right]
\cdot
\frac{\Gamma_{\rm NS}(2a_s)\Gamma_{\rm NS}(2Q-2a_s)}{4\Gamma_{\rm NS}(Q-2a_u)\Gamma_{\rm NS}(2a_u-Q)}\!
\int\limits_{-i\infty}^{i\infty}\hskip -6pt \frac{d\tau}{i}\
{\mathbf J}^{\gamma}_{a_sa_u}\!\left[^{a_3\:a_2}_{a_4\:a_1}\right]\!(\tau),
\\[-18pt]
\\[6pt]
\nonumber
{\sf M}^{\rm e}_{a_sa_u}\!
\left[^{a_3\:a_2}_{a_4\:a_1}\right]^{\eta}_{\hskip 5pt \rho}
& = &
\frac{4^{|\rho|}\,\Gamma_\rho(a_u+a_4-a_2)\Gamma_\rho(\bar a_u+a_4-a_2)\Gamma_\rho(a_u+\bar a_4-a_2)\Gamma_\rho(\bar a_u+\bar a_4-a_2)}
{4^{|\eta|}\, \Gamma_\eta(a_s+a_4-a_3)\Gamma_\eta(\bar a_s+a_4-a_3)\Gamma_\eta(a_s+\bar a_4-a_3)\Gamma_\eta(\bar a_s+\bar a_4-a_3)}
\\[4pt]
\nonumber
& &\times \;
\frac{\Gamma_\rho(a_u+a_1-a_3)\Gamma_\rho(\bar a_u+a_1-a_3)\Gamma_\rho(a_u+\bar a_1-a_3)\Gamma_\rho(\bar a_u+\bar a_1-a_3)}
{\Gamma_\eta(a_s+a_1-a_2)\Gamma_\eta(\bar a_s+a_1-a_2)\Gamma_\eta(a_s+\bar a_1-a_2)\Gamma_\eta(\bar a_s+\bar a_1-a_2)},
\\[6pt]
\nonumber
{\sf M}^{\rm o}_{a_sa_u}\!
\left[^{a_3\:a_2}_{a_4\:a_1}\right]^{\eta}_{\hskip 5pt \rho}
& = &
\frac{\Gamma_\rho(a_u+a_4-a_2)\Gamma_\rho(\bar a_u+a_4-a_2)\Gamma_\rho(a_u+\bar a_4-a_2)\Gamma_\rho(\bar a_u+\bar a_4-a_2)}
{\Gamma_\eta(a_s+a_4-a_3)\Gamma_\eta(\bar a_s+a_4-a_3)\Gamma_\eta(a_s+\bar a_4-a_3)\Gamma_\eta(\bar a_s+\bar a_4-a_3)}
\\[4pt]
\nonumber
& &\times \;
\frac{
\Gamma_{\bar\rho}(a_u+a_1-a_3)
\Gamma_{\bar\rho}(\bar a_u+a_1-a_3)
\Gamma_{\bar\rho}(a_u+\bar a_1-a_3)
\Gamma_{\bar\rho}(\bar a_u+\bar a_1-a_3)
}
{
\Gamma_{\bar\eta}(a_s+a_1-a_2)
\Gamma_{\bar\eta}(\bar a_s+a_1-a_2)
\Gamma_{\bar\eta}(a_s+\bar a_1-a_2)
\Gamma_{\bar\eta}(\bar a_s+\bar a_1-a_2)
}
\end{eqnarray}
where, in analogy with (\ref{notation:repeated}),
\[
{\sf B}^{{\rm e},\,\epsilon}_{a_sa_u}\!
\left[^{a_3\:a_2}_{a_4\:a_1}\right]^{\eta}_{\hskip 5pt \rho}
\equiv
{\sf B}^{\epsilon}_{a_sa_u}\!
\left[^{a_3\:a_2}_{a_4\:a_1}\right]^{\eta\eta}_{\hskip 10pt \rho\rho},
\hskip 1cm
{\sf B}^{{\rm o},\,\epsilon}_{a_sa_u}\!
\left[^{a_3\:a_2}_{a_4\:a_1}\right]^{\eta}_{\hskip 5pt \rho}
\equiv
{\sf B}^{\epsilon}_{a_sa_u}\!
\left[^{a_3\:a_2}_{a_4\:a_1}\right]^{\eta\bar\eta}_{\hskip 10pt \rho\bar\rho}.
\]
The form of the other braiding matrices can be worked out from (\ref{B01:B}) -- (\ref{B11:B})
after taking into account (\ref{normalization:star}). We get:
\begin{eqnarray}
\label{B:01:norm}
\nonumber
{\sf B}^{\epsilon}_{a_sa_u}\!
\left[^{a_3\,*a_2}_{a_4\,\hskip 4pt a_1}\right]^{\eta\rho}_{\hskip 8pt \lambda\delta}
& = &
\frac{N_{\mathbf g}^{\bar \lambda}(a_4,a_2,a_u)N_{\mathbf g}^\delta(a_u,a_3,a_1)}{N_{\mathbf g}^\eta(a_4,a_3,a_s)N_{\mathbf g}^{\bar\rho}(a_s,a_2,a_1)}\,
{}^{\mathbf g}{\sf B}^{\epsilon}_{a_sa_u}\!
\left[^{a_3\,*a_2}_{a_4\,\hskip 4pt a_1}\right]^{\eta\rho}_{\hskip 8pt \lambda\delta}
\\[4pt]
& = &
(-1)^{|\eta|}\
\frac{N_{\mathbf g}^{\bar \lambda}(a_4,a_2,a_u)N_{\mathbf g}^\delta(a_u,a_3,a_1)}{N_{\mathbf g}^\eta(a_4,a_3,a_s)N_{\mathbf g}^{\bar\rho}(a_s,a_2,a_1)}\,
{}^{\mathbf g}{\sf B}^{\epsilon}_{a_sa_u}\!
\left[^{a_3\:a_2}_{a_4\: a_1}\right]^{\eta\bar\rho }_{\hskip 8pt \bar\lambda\delta}
\\[4pt]
\nonumber
& = &
(-1)^{|\eta|}\
{\sf B}^{\epsilon}_{a_sa_u}\!
\left[^{a_3\:a_2}_{a_4\: a_1}\right]^{\eta\bar\rho }_{\hskip 8pt \bar\lambda\delta},
\end{eqnarray}
and similarly
\begin{eqnarray}
\label{B:1011:norm}
\nonumber
{\sf B}^{\epsilon}_{a_sa_u}\!
\left[^{*a_3\:a_2}_{\hskip 4pt a_4\; a_1}\right]^{\eta\rho}_{\hskip 8pt \lambda\delta}
& = &
(-1)^{|\rho|}\
{\sf B}^{\epsilon}_{a_sa_u}\!
\left[^{a_3\:a_2}_{a_4\: a_1}\right]^{\bar\eta\rho}_{\hskip 8pt \lambda\bar\delta},
\\[-6pt]
\\[-6pt]
\nonumber
{\sf B}^{\epsilon}_{a_sa_u}\!
\left[^{*a_3\,*a_2}_{\hskip 4pt a_4\, \hskip 4pt a_1}\right]^{\eta\rho}_{\hskip 8pt \lambda\delta}
& = &
(-1)^{|\eta|+|\rho|+1}
{\sf B}^{\epsilon}_{a_sa_u}\!
\left[^{a_3\:a_2}_{a_4\: a_1}\right]^{\bar\eta\bar\rho}_{\hskip 8pt \bar\lambda\bar\delta}.
\end{eqnarray}

\subsection{Special braiding relations}
In deriving the relations presented in Section \ref{sect:braid:fus}
the limit $a_1\to 0$ of the braiding matrix will be of a particular importance. To calculate it
let us first of all note that $\lim\limits_{a_1\to 0} V(a_1|z)$ is the identity operator
so that
\[
\lim_{a_1\to 0}V_{a_sa_1}^{\eta}(\nu_2|z_2)V_{a_10}(\nu_1|z_1)
=
\delta^{\eta{\rm e}}\,V_{a_20}^{\rm e}(\nu_2|z_2).
\]
Consequently
\[
{\sf B}^{\epsilon}_{a_sa_u}\!
\left[^{a_3\:a_2}_{a_4\;0}\right]^{\eta{\rm o}}_{\hskip 10pt \lambda\delta}
= 0
\]
and in the remaining cases the limit
\(
\lim\limits_{a_1\to 0}
{\sf B}^{\epsilon}_{a_sa_u}\!
\left[^{a_3\:a_2}_{a_4\:a_1}\right]
\)
is well defined only for $a_s=a_2.$ In effect we only need to calculate
\begin{eqnarray*}
\label{simple:even:def}
\lim_{a_1\to 0}
{\sf B}^{\rm e}_{a_2a_u}\!
\left[^{a_3\:a_2}_{a_4\:a_1}\right]^{\rm e}_{\hskip 4pt \lambda}
& \equiv &
\lim_{a_1\to 0}
{\sf B}_{a_2a_u}\!
\left[^{a_3\:a_2}_{a_4\:a_1}\right]^{\rm ee}_{\hskip 8pt \lambda\lambda}
\end{eqnarray*}
and
\begin{eqnarray*}
\label{simple:odd:def}
\lim_{a_1\to 0}
{\sf B}^{\rm o}_{a_2a_u}\!
\left[^{a_3\:a_2}_{a_4\:a_1}\right]^{\rm o}_{\hskip 4pt \lambda}
& \equiv &
\lim_{a_1\to 0}
{\sf B}_{a_2a_u}\!
\left[^{a_3\:a_2}_{a_4\:a_1}\right]^{\rm oe}_{\hskip 8pt \lambda\bar\lambda}.
\end{eqnarray*}

From (\ref{normalized:braiding:explicit}) we have:
\begin{eqnarray*}
&&
\hskip -1cm
{\sf M}^{\rm e}_{a_2a_u}\!
\left[^{a_3\:a_2}_{a_4\:a_1}\right]^{\rm e}_{\hskip 4pt \lambda} =
\\
& = &
2^{|\lambda|}
\frac{\Gamma_\lambda(a_u+a_4-a_2)\Gamma_\lambda(\bar a_u+a_4-a_2)\Gamma_\lambda(a_u+\bar a_4-a_2)\Gamma_\lambda(\bar a_u+\bar a_4-a_2)}
{\Gamma_{\rm NS}(a_2+a_4-a_3)\Gamma_{\rm NS}(\bar a_2+a_4-a_3)\Gamma_{\rm NS}(a_2+\bar a_4-a_3)\Gamma_{\rm NS}(\bar a_2+\bar a_4-a_3)}
\\[4pt]
& &\times \;
\frac{\Gamma_\lambda(a_u+a_1-a_3)\Gamma_\lambda(\bar a_u+a_1-a_3)\Gamma_\lambda(a_u+\bar a_1-a_3)\Gamma_\lambda(\bar a_u+\bar a_1-a_3)}
{\Gamma_{\rm NS}(a_1)\Gamma_{\rm NS}(Q+a_1-2a_2)\Gamma_{\rm NS}(\bar a_1)\Gamma_{\rm NS}(Q+\bar a_1-2a_2)}.
\end{eqnarray*}
For $a_1\to 0$ the factor $\Gamma_{\rm NS}(a_1)^{-1}$ present in this expression tends to zero and the braiding
matrix is non-zero only for such values of the remaining parameters for which
\begin{equation}
\label{factor:J}
\int_{-i\infty}^{i\infty} \frac{d\tau}{i}\
{\mathbf J}^{\rm e}_{a_2a_u}\!\left[^{a_3\:a_2}_{a_4\:a_1}\right]^{\rm e}_{\hskip 4pt \lambda}
\end{equation}
provides a compensating, singular factor.
Since components of
\(
{\mathbf J}^{\rm e}_{a_2a_u}\!\left[^{a_3\:a_2}_{a_4\:a_1}\right]\!(\tau)
\)
are meromorphic functions of $\tau$ (with the location of poles determined by the values of $a_i$) we can deform
the integration contour in (\ref{factor:J}) such that it ``keeps away'' from the moving with $a_1\to 0$ poles,
arriving in the limit at a non-singular function of $a_1,a_2,a_3,a_u$ and, in view of the discussion above,
at a vanishing braiding matrix (cf.\ \cite{Ponsot:2000mt}, Lemma 3). However, this procedure fails if the integration contour
gets ``pinched'' between a pair of moving poles. In such a case we have to deform the contour
past one of these colliding poles and the singular contribution can appear from the residue.

In our case the only pair of colliding poles appears in
\begin{eqnarray*}
\left[^{\rm\textstyle N\:N\:N\:N}_{\rm\textstyle N\:N\:N\:N}\right]\!(\tau)
& = &
\frac{
S_{\rm NS}(\bar a_4-a_3+a_2+\tau)S_{\rm NS}(a_1+\tau)S_{\rm NS}(a_4-a_3+a_2+\tau)S_{\rm NS}(\bar a_1+\tau)
}{
S_{\rm NS}(\bar a_u + \bar a_3+\tau)S_{\rm NS}(a_u+ \bar a_3+\tau)S_{\rm NS}(2a_2 +\tau)S_{\rm NS}(Q +\tau)
},
\end{eqnarray*}
a summand of
\(
{\mathbf J}^{\rm e}_{a_2a_u}\!\left[^{a_3\:a_2}_{a_4\:a_1}\right]^{\rm e}_{\hskip 4pt \rm e},
\)
where a pole at $\tau = -a_1$ (initially to the left of the contour) coming from a factor
\(
S_{\rm NS}(a_1+\tau)
\)
approaches a pole of a function
\(
S_{\rm NS}(Q +\tau)^{-1}
\)
at
$\tau = 0^{+}$ (to the right of the contour). The residue at $\tau = -a_1$ gives a contribution to (\ref{factor:J}) of the form
\begin{eqnarray*}
\label{sing:factor:1}
I(a_1)
& = &
2\frac{
S_{\rm NS}(\bar a_4-a_3+a_2-a_1)S_{\rm NS}(a_4-a_3+a_2-a_1)S_{\rm NS}(Q-2 a_1)
}{
S_{\rm NS}(\bar a_u + \bar a_3-a_1)S_{\rm NS}(a_u+ \bar a_3-a_1)S_{\rm NS}(2a_2-a_1)S_{\rm NS}(Q-a_1)
}
\end{eqnarray*}
(in the course of calculating $I(a_1)$ the formula
\(
\lim_{x\to 0} xS_{\rm NS}(x) = \pi^{-1}
\)
was used) and
\begin{eqnarray*}
&&
\hskip -1cm
\lim_{a_1\to 0}
{\sf B}^{\rm e}_{a_2a_u}\!
\left[^{a_3\:a_2}_{a_4\:a_1}\right]^{\rm e}_{\hskip 4pt \lambda}
\; = \;
\delta^{\rm e}_{\lambda}
\frac{\Gamma_{\rm NS}(2a_2)\Gamma_{\rm NS}(2Q-2a_2)}{4\Gamma_{\rm NS}(Q-2a_u)\Gamma_{\rm NS}(2a_u-Q)}
\lim_{a_1\to 0}
I(a_1)\,
{\sf M}^{\rm e}_{a_2a_u}\!
\left[^{a_3\:a_2}_{a_4\:a_1}\right]^{\rm e}_{\hskip 4pt \rm e}
\\[6pt]
& = &
\delta^{\rm e}_{\lambda}\
\frac{
\Gamma_{\rm NS}(a_4+a_u-a_2)
\Gamma_{\rm NS}(a_4+\bar a_u-a_2)
\Gamma_{\rm NS}(\bar a_4+a_u-a_2)
\Gamma_{\rm NS}(\bar a_4+\bar a_u-a_2)
}
{
\Gamma_{\rm NS}(a_4+a_3-a_2)
\Gamma_{\rm NS}(a_4+\bar a_3-a_2)
\Gamma_{\rm NS}(\bar a_4+a_3-a_2)
\Gamma_{\rm NS}(\bar a_4+\bar a_3-a_2)
}
\\[4pt]
&&\times\;
\frac{
\Gamma_{\rm NS}(a_u-\bar a_3)\Gamma_{\rm NS}(\bar a_u-a_3)
}
{
\Gamma_{\rm NS}(a_u-\bar a_u)\Gamma_{\rm NS}(\bar a_u-a_u)
}
\lim_{a_1\to 0}
\frac{
\Gamma_{\rm NS}(a_u-a_3+a_1)\Gamma_{\rm NS}(a_3-a_u+a_1)
}
{
2\Gamma_{\rm NS}(Q)\Gamma_{\rm NS}(2a_1)
}.
\end{eqnarray*}
Since
\[
a_u = \frac{Q}{2}+ip_u,
\hskip 1cm
a_3 = \frac{Q}{2}+ip_3,
\hskip 1cm
\lim_{x\to 0}x\Gamma_{\rm NS}(x) = \frac{\Gamma_{\rm NS}(Q)}{\pi},
\]
we have
\begin{eqnarray*}
\lim_{a_1\to 0}
\frac{
\Gamma_{\rm NS}(a_u-a_3+a_1)\Gamma_{\rm NS}(a_3-a_u+a_1)
}
{
2\Gamma_{\rm NS}(Q)\Gamma_{\rm NS}(2a_1).
}
& = &
\frac{1}{\pi}\lim_{a_1\to 0}\frac{a_1}{a_1^2 + (p_u-p_3)^2}
\;\ = \;\
\delta(p_u-p_3)
\end{eqnarray*}
and
\begin{equation*}
\label{simple:even:result}
\lim_{a_1\to 0}
{\sf B}^{\rm e}_{a_2a_u}\!
\left[^{a_3\:a_2}_{a_4\:a_1}\right]^{\rm e}_{\hskip 4pt \lambda}
\;\ = \;\
\delta^{\rm e}_{\lambda}\,
\delta(p_u-p_3).
\end{equation*}

Calculations leading to
\[
\lim_{a_1\to 0}
{\sf B}^{\rm o}_{a_2a_u}\!
\left[^{a_3\:a_2}_{a_4\:a_1}\right]^{\rm o}_{\hskip 4pt \lambda}
\]
are analogous. This time the colliding pair of poles appears in the function
\[
i \left[^{\rm\textstyle R\:N\:R\:N}_{\rm\textstyle N\:N\:N\:N}\right]\!(\tau)
\; = \;
i
\frac{
S_{\rm R}(\bar a_4-a_3+a_2+\tau)S_{\rm NS}(a_1+\tau)S_{\rm R}(a_4-a_3+a_2+\tau)S_{\rm NS}(\bar a_1+\tau)
}{
S_{\rm NS}(\bar a_u + \bar a_3+\tau)S_{\rm NS}(a_u+ \bar a_3+\tau)S_{\rm NS}(a_s + a_2 +\tau)S_{\rm NS}(\bar a_s + a_2 +\tau)
},
\]
being a summand of
\(
{\mathbf J}^{\rm o}_{a_2a_u}\!\left[^{a_3\:a_2}_{a_4\:a_1}\right]^{\rm o}_{\hskip 4pt \rm o}.
\)
Computing the residue at  $\tau = -a_1$ and taking the limit $a_1 \to 0 $ we get:
\begin{equation*}
\label{simple:odd:result}
\lim_{a_1\to 0}{\sf B}^{\rm o}_{a_2a_u}\!
\left[^{a_3\:a_2}_{a_4\:a_1}\right]^{\rm o}_{\hskip 4pt \lambda}
\;\ = \;\
i\delta^{\rm o}_{\lambda}\,
\delta(p_u-p_3).
\end{equation*}

Summarizing:
\begin{equation}
\label{braiding:simple:B}
\lim_{a_1\to 0}{\sf B}_{a_2a_u}\!
\left[^{a_3\:a_2}_{a_4\:a_1}\right]^{\rho\eta}_{\hskip 10pt \lambda\delta}
\;\ = \;\
i^{|\rho|}\,\delta^{\rho}_{\lambda}\delta^{\eta{\rm e}}\delta_{\delta{\rm e}}\,\delta(p_u-p_3)
\end{equation}
or equivalently
\begin{equation}
\label{braiding:simple:V}
V^{\rho}_{a_4a_2}(\nu_3|z_3)V^{\eta}_{a_20}(\nu_2|z_2)
\; = \;
i^{|\rho|}\,\Omega^{\epsilon_{32}}_{432}\,
\delta^{\eta{\rm e}}\,
V^{\rho}_{a_4a_3}(\nu_2|z_2)V^{\rm e}_{a_30}(\nu_3|z_3),
\end{equation}
where
\begin{equation*}
\label{simple:phase}
\Omega^{\epsilon_{32}}_{432}
\; = \;
{\rm e}^{i\pi\epsilon_{32}(\Delta_4-\Delta_3-\Delta_2)}.
\end{equation*}
Finally, the relations between ``starred'' and ``un-starred'' braiding matrices, Eqs.\ (\ref{B:01:norm})
and  (\ref{B:1011:norm}), together
with (\ref{braiding:simple:B}) and (\ref{braiding:simple:V}) give:
\begin{eqnarray*}
\label{braiding:simple:B:remaining}
\nonumber
\lim_{a_1\to 0}{\sf B}_{a_2a_u}\!
\left[^{a_3\,*a_2}_{a_4\,\hskip 4pt a_1}\right]^{\rho\eta}_{\hskip 10pt \lambda\delta}
& = &
(-1)^{|\rho|}\lim_{a_1\to 0}{\sf B}_{a_2a_u}\!
\left[^{a_3\:a_2}_{a_4\:a_1}\right]^{\rho\bar\eta}_{\hskip 10pt \bar\lambda\delta}
\; = \;
(-i)^{|\rho|}\,\delta^{\rho}_{\bar\lambda}\delta^{\eta{\rm o}}\delta_{\delta{\rm e}}\,\delta(p_u-p_3),
\\[6pt]
\lim_{a_1\to 0}{\sf B}_{a_2a_u}\!
\left[^{*a_3\:a_2}_{\hskip 4pt a_4\; a_1}\right]^{\rho\eta}_{\hskip 10pt \lambda\delta}
& = &
(-1)^{|\eta|}\lim_{a_1\to 0}{\sf B}_{a_2a_u}\!
\left[^{a_3\:a_2}_{a_4\:a_1}\right]^{\bar\rho\eta}_{\hskip 10pt \lambda\bar\delta}
\; = \;
i^{|\bar\rho|}\,\delta^{\rho}_{\bar\lambda}\delta^{\eta{\rm e}}\delta_{\delta{\rm o}}\,\delta(p_u-p_3),
\\[6pt]
\nonumber
\lim_{a_1\to 0}{\sf B}_{a_2a_u}\!
\left[^{*a_3\,*a_2}_{\hskip 4pt a_4\, \hskip 4pt a_1}\right]^{\rho\eta}_{\hskip 10pt \lambda\delta}
& = &
(-1)^{|\bar\rho|+|\eta|}\lim_{a_1\to 0}{\sf B}_{a_2a_u}\!
\left[^{a_3\:a_2}_{a_4\:a_1}\right]^{\bar\rho\bar\eta}_{\hskip 10pt \bar\lambda\bar\delta}
\; = \;
-(-i)^{|\bar\rho|}\,\delta^{\rho}_{\lambda}\delta^{\eta{\rm o}}\delta_{\delta{\rm o}}\,\delta(p_u-p_3),
\end{eqnarray*}
or equivalently:
\begin{eqnarray*}
\label{braiding:simple:V:remaining}
\nonumber
V^{\rho}_{a_4a_2}(\nu_3|z_3)V^{\eta}_{a_20}(*\nu_2|z_2)
& = &
(-i)^{|\rho|}\,\Omega^{\epsilon_{32}}_{432}\,
\delta^{\eta{\rm o}}\,
V^{\bar\rho}_{a_4a_3}(*\nu_2|z_2)V^{\rm e}_{a_30}(\nu_3|z_3),
\\[6pt]
V^{\rho}_{a_4a_2}(*\nu_3|z_3)V^{\eta}_{a_20}(\nu_2|z_2)
& = &
i^{|\bar\rho|}\,\Omega^{\epsilon_{32}}_{432}\,
\delta^{\eta{\rm e}}\,
V^{\bar\rho}_{a_4a_3}(\nu_2|z_2)V^{\rm o}_{a_30}(*\nu_3|z_3),
\\[6pt]
\nonumber
V^{\rho}_{a_4a_2}(*\nu_3|z_3)V^{\eta}_{a_20}(*\nu_2|z_2)
& = &
-(-i)^{|\bar\rho|}\,\Omega^{\epsilon_{32}}_{432}\,
\delta^{\eta{\rm o}}\,
V^{\rho}_{a_4a_3}(*\nu_2|z_3)V^{\rm o}_{a_30}(*\nu_3|z_3).
\end{eqnarray*}

\section{Braiding and fusion properties of the Neveu-Schwarz blocks}
\label{sect:braid:fus}
We shall define, following \cite{Hadasz:2006qb}, four even and four odd  NS conformal blocks
\begin{eqnarray}
\label{4:point:blocks}
{\cal F}^{\eta}_{\!a_s}\!
\left[^{\underline{\hspace*{5pt}}a_3\:\underline{\hspace*{5pt}}a_2}_{\hspace*{5pt}a_4\hspace*{5pt}\:a_1}\right]
\!(z)
& = &
\langle\nu_4\big|
V_{a_4a_s}^{\eta}\!(\underline{\hspace*{6pt}}\nu_3|1)
V_{a_sa_1}^{\eta}\!(\underline{\hspace*{6pt}}\nu_2|z)
\nu_1\rangle.
\end{eqnarray}
This choice is motivated by the
observation that the knowledge of (\ref{4:point:blocks}) is sufficient (once the relevant three point coupling constants are know)
to compute all the four-point correlation functions in the NS sector of a given $N=1$ SCFT. Inserting
between the chiral vertices a projection operator
onto the basis of ${\cal V}_s$ formed by the vectors $\nu_{s,IK}$ (see  (\ref{basis}))
we get:
\begin{equation*}
\label{block:definition}
{\cal F}^{\eta}_{\!a_s}\!
\left[^{\underline{\hspace*{5pt}}a_3\:\underline{\hspace*{5pt}}a_2}_{\hspace*{5pt}a_4\hspace*{5pt}\:a_1}\right]
\!(z)
=
\sum\limits_{IK,JL}
\rho{^{a_4\,}_\infty}{^{a_3\,}_{\:1}}{^{a_s}_{\;0}}(\nu_4,\underline{\hspace*{6pt}}\nu_3,\nu_{s,IK})
\left[G^{\eta}_{a_s}\right]^{IK,JL}
\rho{^{a_s\,}_\infty}{^{a_2\,}_{\:z}}{^{a_2}_{\;0}}(\nu_{s,JL},\underline{\hspace*{5pt}}\nu_2,\nu_1).
\end{equation*}
Here the three-linear form $\rho$ is defined by
\begin{equation*}
\label{rho:definition}
\rho{^{a_3\,}_\infty}{^{a_2\,}_{\:z}}{^{a_1}_{\;0}}(\xi_3,\xi_2,\xi_1)
=
\left\langle\xi_3\big|V_{a_3a_1}(\xi_2|z)\xi_1\right\rangle,
\hskip 1cm
\xi_i\in {\cal V}_i,\;\
i = 1,2,3,
\end{equation*}
\(
\left[G^{\eta}_{a_s}\right]^{IK,JL}
\)
is an element of the matrix inverse to
\[
\left[G^{\eta}_{a_s}\right]_{IK,JL} = \left\langle\nu_{s,IK}|\nu_{s,JL}\right\rangle
\]
and $(-1)^{|\eta|} = (-1)^{|K|} = (-1)^{|L|}.$

Notice that
\begin{eqnarray}
\label{s:channel}
\nonumber
&&
\hskip -.8cm
\big\langle\nu_4\big|
V_{a_4a_s}^{\eta}\!(\underline{\hspace*{6pt}}\nu_3|z_3)
V_{a_sa_1}^{\eta}\!(\underline{\hspace*{6pt}}\nu_2|z_2)
V_{a_10}\!(\nu_1|z_1)
\nu_0\big\rangle
\\[6pt]
& = &
\big\langle\nu_4\big|
V_{a_4a_s}^{\eta}\!(\underline{\hspace*{6pt}}\nu_3|z_{31})
V_{a_sa_1}^{\eta}\!(\underline{\hspace*{6pt}}\nu_2|z_{21})
\nu_1\big\rangle
\\[6pt]
\nonumber
& = &
z_{31}^{\Delta_4-\underline{\hspace*{5pt}}\Delta_3-\underline{\hspace*{5pt}}\Delta_2-\Delta_1}\,
\big\langle\nu_4\big|
V_{a_4a_s}^{\eta}\!(\underline{\hspace*{6pt}}\nu_3|1)
V_{a_sa_1}^{\eta}\!(\underline{\hspace*{6pt}}\nu_2|z)
\nu_1\big\rangle
=
z_{31}^{\Delta_4-\underline{\hspace*{5pt}}\Delta_3-\underline{\hspace*{5pt}}\Delta_2-\Delta_1}\,
{\cal F}^{\eta}_{\!a_s}\!
\left[^{\underline{\hspace*{5pt}}a_3\:\underline{\hspace*{5pt}}a_2}_{\hspace*{5pt}a_4\hspace*{5pt}\:a_1}\right]\!(z),
\end{eqnarray}
where $z = \frac{z_{21}}{z_{31}}$ with $z_{ij} = z_i - z_j,$ $*\Delta \equiv \Delta+\frac12$
and $\underline{\hspace*{5pt}}\Delta$ which may stand for $\Delta$ or $*\Delta,$  and
\begin{eqnarray}
\label{t:channel}
\nonumber
&&
\hskip -1cm
\big\langle\nu_4\big|
V_{a_4a_1}\!\left(V_{a_ta_2}^{\eta}\!(\underline{\hspace*{6pt}}\nu_3|z_{32})\nu_2|z_2\right)
V_{a_10}\!(\underline{\hspace*{6pt}}\nu_1|z_1)
\nu_0\big\rangle
\\[6pt]
& = &
\big\langle\nu_4\big|
V_{a_4a_1}\!\left(V_{a_ta_2}^{\eta}\!(\underline{\hspace*{6pt}}\nu_3|z_{32})\nu_2|z_{21}\right)
\underline{\hspace*{6pt}}\nu_1\big\rangle
\\[6pt]
\nonumber
& = &
\sum\limits_{IK,JL}
\rho{^{a_4\,}_\infty}{^{\:a_t}_{\:z_{21}}}{^{a_1}_{\;0}}(\nu_4,\nu_{t,IK},\underline{\hspace*{6pt}}\nu_1)
[G^\eta_{a_t}]^{IK,JL}
\rho{^{a_t\,}_\infty}{^{\:a_3}_{\:z_{32}}}{^{a_2}_{\;0}}(\nu_{t,JL},\underline{\hspace*{6pt}}\nu_3,\nu_2)
\\[2pt]
\nonumber
& = &
{\rm e}^{-i\pi\varepsilon_{32}(\Delta_t-\underline{\hspace*{5pt}}\Delta_3-\Delta_2+\frac12|\eta|)}\,
z_{21}^{\Delta_{4}-\underline{\hspace*{5pt}}\Delta_3-\Delta_2-\underline{\hspace*{5pt}}\Delta_1}\,
{\cal F}^{{\eta}}_{\!a_t}\!
\left[^{\underline{\hspace*{5pt}}a_1\:\underline{\hspace*{5pt}}a_3}_{\hspace*{5pt}a_4\:\hspace*{5pt}a_2}\right]\!\left(1-z^{-1}\right),
\end{eqnarray}
where $\epsilon_{32} = {\rm sgn(Arg}\,z_{32}).$ To derive  (\ref{s:channel}) and (\ref{t:channel}) we
used the state-operator correspondence
\[
\xi = V_{a0}(\xi|0)\nu_0, \hskip 1cm
\xi \in {\cal V}_a,
\]
which follows from (\ref{chiral:norm:1}) and the definition of the generalized chiral vertex operator,
the fact that $L_{-1}$ acts
(see (\ref{decendans:def:1})) as a generator of translation in $z$ and consequently
\[
V_{a_3a_1}(\xi_2|z) = {\rm e}^{zL_{-1}}V_{a_3a_1}(\xi_2|0){\rm e}^{-zL_{-1}},
\]
and the identities
\begin{eqnarray*}
\label{rho:properties}
\nonumber
\rho{^{a_3\,}_\infty}{^{a_2\,}_{\:z}}{^{a_1}_{\;0}}(\xi_3,\xi_2,\xi_1)
& = &
z^{\Delta(\xi_3)-\Delta(\xi_2)-\Delta(\xi_1)}
\rho{^{a_3\,}_\infty}{^{a_2\,}_{\:1}}{^{a_1}_{\;0}}(\xi_3,\xi_2,\xi_1),
\hskip 1.1cm
L_0\xi_i = \Delta(\xi_i)\xi_i,
\\[6pt]
\rho{^{a_3\,}_\infty}{^{a_2\,}_{\:z}}{^{a_1}_{\;0}}(\nu_3,\nu_{2,IK},\underline{\hspace*{6pt}}\nu_1)
& = &
(-1)^{|I| + |K|}\,
\rho{^{a_3\,}_\infty}{^{a_1\,}_{\:z}}{^{a_t}_{\;0}}(\nu_3,\underline{\hspace*{6pt}}\nu_1,\nu_{2,IK}),
\hskip 1.45cm
|K| \in {\mathbb N},
\\[6pt]
\nonumber
\rho{^{a_3\,}_\infty}{^{a_2\,}_{\:z}}{^{a_1}_{\;0}}(\nu_3,\nu_{2,IK},\underline{\hspace*{6pt}}\nu_1)
& = &
(-1)^{|I|+|K|-\frac12}\,
\rho{^{a_3\,}_\infty}{^{a_1\,}_{\:z}}{^{a_2}_{\;0}}(\nu_3,\underline{\hspace*{6pt}}\nu_1,\nu_{2,IK}),
\hskip 1cm
|K| \in {\mathbb N}+\frac12,
\end{eqnarray*}
which are also easily derived using definitions from the subsection \ref{the:vertex} and the commutation relations
(\ref{comm:with:star}).

With the help of relations (\ref{braiding:definition:2}) and (\ref{s:channel}) it is immediate
to arrive at the $s-u$ braiding relations satisfied by the NS blocks. We have:
\begin{eqnarray*}
\label{braiding:s-u}
\nonumber
&&
\hskip -1cm
{\cal F}^{\eta}_{\!a_s}\!
\left[^{\underline{\hspace*{5pt}}a_3\:\underline{\hspace*{5pt}}a_2}_{\hspace*{5pt}a_4\hspace*{5pt}\:a_1}\right]\!(z)
=
z_{31}^{-\Delta_4+\underline{\hspace*{5pt}}\Delta_3+\underline{\hspace*{5pt}}\Delta_2+\Delta_1}\,
\big\langle\nu_4\big|
V_{a_4a_s}^{\eta}\!(\underline{\hspace*{6pt}}\nu_3|z_{31})
V_{a_sa_1}^{\eta}\!(\underline{\hspace*{6pt}}\nu_2|z_{21})
\nu_1\big\rangle
\\[6pt]
\nonumber
& = &
z_{31}^{-\Delta_4+\underline{\hspace*{5pt}}\Delta_3+\underline{\hspace*{5pt}}\Delta_2+\Delta_1}
\int\limits_{\mathbb S}\!\frac{da_u}{2i}\hskip -4pt
\sum\limits_{\rho =\, {\rm e,o}}\hskip -4pt
{\sf B}^{\epsilon}_{a_sa_u}\!\!
\left[^{\underline{\hspace*{5pt}}a_3\:\underline{\hspace*{5pt}}a_2}_{\hspace*{5pt}a_4\:\hspace*{5pt}a_1}
\right]^{\eta\eta}_{\hskip 10pt \rho\rho}
\big\langle\nu_4\big|
V_{a_4a_u}^{\rho}\!(\underline{\hspace*{6pt}}\nu_2|z_{21})
V_{a_sa_u}^{\rho}\!(\underline{\hspace*{6pt}}\nu_3|z_{31})
\nu_1\big\rangle
\\[6pt]
& = &
z^{\Delta_4-\underline{\hspace*{5pt}}\Delta_3-\underline{\hspace*{5pt}}\Delta_2-\Delta_1}
\int\limits_{\mathbb S}\!\frac{da_u}{2i}\hskip -4pt
\sum\limits_{\rho =\, {\rm e,o}}\hskip -4pt
{\sf B}^{\epsilon}_{a_sa_u}\!\!
\left[^{\underline{\hspace*{5pt}}a_3\:\underline{\hspace*{5pt}}a_2}_{\hspace*{5pt}a_4\:\hspace*{5pt}a_1}
\right]^{\eta\eta}_{\hskip 10pt \rho\rho}
{\cal F}^{\rho}_{\!a_u}\!
\left[^{\underline{\hspace*{5pt}}a_2\:\underline{\hspace*{5pt}}a_3}_{\hspace*{5pt}a_4\hspace*{5pt}\:a_1}\right]\!\left(z^{-1}\right).
\end{eqnarray*}

Special braiding relations (\ref{braiding:simple:V}) allow to
derive a generalization of the Euler's relation satisfied by the
hypergeometric function. W have:
\begin{eqnarray*}
&&
\hskip -1cm
{\cal F}^{\eta}_{\!a_s}\!
\left[^{a_3\:a_2}_{a_4\:a_1}\right]\!(z)
=
z_{31}^{\Delta_1+\Delta_2+\Delta_3-\Delta_4}
\langle\nu_4|
V_{a_4a_s}^{\eta}\!(\nu_3|z_3)
V_{a_sa_1}^{\eta}\!(\nu_2|z_2)
V_{a_10}^{\rm e}\!(\nu_1|z_1)
\nu_0\rangle
\\[6pt]
&& =
i^{|\eta|}\,{\rm e}^{i\pi\epsilon_{21}(\Delta_s-\Delta_2-\Delta_1)}\,
z_{31}^{\Delta_1+\Delta_2+\Delta_3-\Delta_4}\,
\langle\nu_4|
V_{a_4a_s}^{\eta}\!(\nu_3|z_3)
V_{a_sa_2}^{\eta}\!(\nu_1|z_1)
V_{a_20}^{\rm e}\!(\nu_2|z_2)
\nu_0\rangle
\\[6pt]
&& =
i^{|\eta|}\,{\rm e}^{i\pi\epsilon_{21}(\Delta_s-\Delta_2-\Delta_1)}\,
\left(\frac{z_{32}}{z_{31}}\right)^{\Delta_4-\Delta_3-\Delta_2-\Delta_1}\,
{\cal F}^{\eta}_{\!a_s}\!
\left[^{a_3\:a_1}_{a_4\:a_2}\right]\!\left(\frac{z_{12}}{z_{32}}\right).
\end{eqnarray*}
If we exclude the situation when arg$\,z_1$ lies between arg$\,z_2$ and arg$\,z_3$ then
\[
\epsilon_{21} \; = \; -{\rm sign(arg\,}z) \; \equiv -\epsilon
\]
and we get
\begin{equation}
\label{Euler}
{\cal F}^{\eta}_{\!a_s}\!
\left[^{a_3\:a_2}_{a_4\:a_1}\right]\!(z)
\; = \;
i^{|\eta|}\,{\rm e}^{-i\pi\epsilon(\Delta_s-\Delta_2-\Delta_1)}\,
(1-z)^{\Delta_4-\Delta_3-\Delta_2-\Delta_1}\,
{\cal F}^{\eta}_{\!a_s}\!
\left[^{a_3\:a_1}_{a_4\:a_2}\right]\!\big({\textstyle\frac{z}{z-1}}\big).
\end{equation}
Similarly
\begin{eqnarray*}
&&
\hskip -1cm
{\cal F}^{\eta}_{\!a_s}\!
\left[^{a_3\,*a_2}_{a_4\,\hspace*{4pt}a_1}\right]\!(z)
=
z_{31}^{\Delta_1+*\Delta_2+\Delta_3-\Delta_4}
\langle\nu_4|
V_{a_4a_s}^{\eta}\!(\nu_3|z_3)
V_{a_sa_1}^{\eta}\!(*\nu_2|z_2)
V_{a_10}^{\rm e}\!(\nu_1|z_1)
\nu_0\rangle
\\[6pt]
&& =
i^{|\bar\eta|}\,{\rm e}^{i\pi\epsilon_{21}(\Delta_s-\Delta_2-\Delta_1)}\,
z_{31}^{\Delta_1+*\Delta_2+\Delta_3-\Delta_4}\,
\langle\nu_4|
V_{a_4a_s}^{\eta}\!(\nu_3|z_3)
V_{a_sa_2}^{\bar\eta}\!(\nu_1|z_1)
V_{a_20}^{\rm o}\!(*\nu_2|z_2)
\nu_0\rangle
\\[6pt]
&& =
i^{|\bar\eta|}\,{\rm e}^{i\pi\epsilon_{21}(\Delta_s-\Delta_2-\Delta_1)}\,
\left(\frac{z_{32}}{z_{31}}\right)^{\Delta_4-\Delta_3-*\Delta_2-\Delta_1}\,
{\cal F}^{\eta}_{\!a_s}\!
\left[^{a_3\,\hspace*{3pt}a_1}_{a_4\,*a_2}\right]\!\left(\frac{z_{12}}{z_{32}}\right)
\end{eqnarray*}
where
\begin{eqnarray*}
{\cal F}^{\eta}_{\!a_s}\!
\left[^{a_3\,\hspace*{3pt}a_1}_{a_4\,*a_2}\right]\!(x)
& \equiv &
\langle\nu_4|
V_{a_4a_s}^{\eta}\!(\nu_3|1)
V_{a_sa_2}^{\bar\eta}\!(\nu_1|x)
S_{-\frac12}\nu_2\rangle
=
{\cal F}^{\bar\eta}_{\!a_s}\!
\left[^{*a_3\:a_1}_{\hspace*{4pt}a_4\:a_2}\right]\!(x)
+
(-1)^{|\eta|}
{\cal F}^{\eta}_{\!a_s}\!
\left[^{a_3\,*a_1}_{a_4\,\hspace*{4pt}a_2}\right]\!(x),
\end{eqnarray*}
and with the same restriction on the arguments of $z_i,\, i = 1,2,3:$
\begin{equation}
\label{Euler:star}
{\cal F}^{\eta}_{\!a_s}\!
\left[^{a_3\,*a_2}_{a_4\,\hspace*{4pt}a_1}\right]\!(z)
\; = \;
i^{|\bar\eta|}\,{\rm e}^{-i\pi\epsilon(\Delta_s-\Delta_2-\Delta_1)}\,
(1-z)^{\Delta_4-\Delta_3-*\Delta_2-\Delta_1}\,
{\cal F}^{\eta}_{\!a_s}\!
\left[^{a_3\,\hspace*{3pt}a_1}_{a_4\,*a_2}\right]\!\big({\textstyle\frac{z}{z-1}}\big)
\end{equation}
It shouldn't be difficult for the reader to derive the Euler's relations for the remaining four blocks.

Denote graphically the identity
\[
V_{a_3a_1}(\nu_2|z_2)V_{a_10}(\nu_1|z_1)\nu_0
=
{\rm e}^{z_1L_{-1}}V_{a_3a_1}(\nu_2|z_{21})\nu_1
=
V_{a_30}\big(V_{a_3a_1}(\nu_2|z_{21})\nu_1|z_1\big)\nu_0,
\]
which expresses the operator-state correspondence for the chiral vertex operators, as

\noindent
\centerline{
\includegraphics*[width=.5\textwidth]{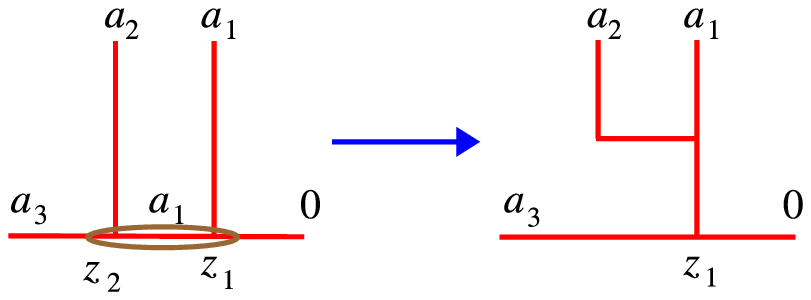}
}

\centerline{\small{\bf Fig.~2}~ Graphical notation for the elementary fusion transformation}

\vskip 2mm
\noindent
and consider now the following sequence of ``moves'':
\vskip 2mm

\noindent
\includegraphics*[width=\textwidth]{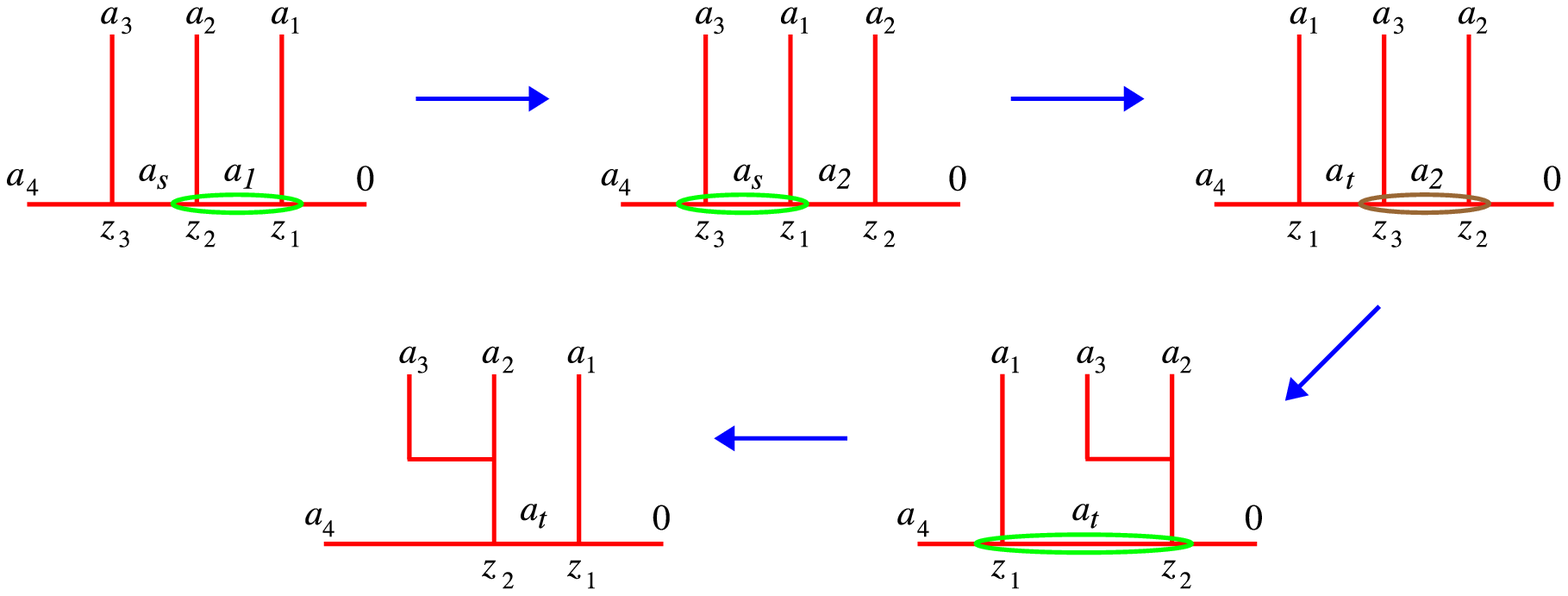}

\vskip 2mm

\centerline{\small{\bf Fig.~3}~ Moves which lead to $s-t$ braiding}

\vskip 2mm
\noindent
It results in an identity:
\begin{eqnarray}
\label{st:computations}
\nonumber
&&
\hskip -1cm
\langle\nu_4|
V_{a_4a_s}^{\eta}\!(\nu_3|z_3)
V_{a_sa_1}^{\eta}\!(\nu_2|z_2)
V_{a_10}^{\rm e}\!(\nu_1|z_1)
\nu_0\rangle
\\[6pt]
\nonumber
& = &
i^{|\eta|}\Omega_{s21}^{\epsilon_{21}}
\langle\nu_4|
V_{a_4a_s}^{\eta}\!(\nu_3|z_3)
V_{a_sa_2}^{\eta}\!(\nu_1|z_1)
V_{a_20}^{\rm e}\!(\nu_2|z_2)
\nu_0\rangle
\\[6pt]
\nonumber
& = &
i^{|\eta|}\Omega_{s21}^{\epsilon_{21}}
\hskip -5pt \int\limits_{\frac{Q}{2}+i{\mathbb R}}\hskip -5pt \frac{da_t}{2i}\
{\sf B}^{\epsilon_{31}}_{a_sa_t}\!\!\left[^{a_3\:a_1}_{a_4\:a_2}\right]^{\eta\eta}_{\hskip 10pt\rho\rho}
\langle\nu_4|
V_{a_4a_t}^{\rho}\!(\nu_1|z_1)
V_{a_ta_2}^{\rho}\!(\nu_3|z_3)
V_{a_20}^{\rm e}\!(\nu_2|z_2)
\nu_0\rangle
\\[-6pt]
\\[-6pt]
\nonumber
& = &
i^{|\eta|}\Omega_{s21}^{\epsilon_{21}}
\hskip -5pt \int\limits_{\frac{Q}{2}+i{\mathbb R}}\hskip -5pt \frac{da_t}{2i}\
{\sf B}^{\epsilon_{31}}_{a_sa_t}\!\!\left[^{a_3\:a_1}_{a_4\:a_2}\right]^{\eta\eta}_{\hskip 10pt\rho\rho}
\langle\nu_4|
V_{a_4a_t}^{\rho}\!(\nu_1|z_1)
V_{a_20}^{\rho}\!\left(
V_{a_ta_2}^{\rho}\!(\nu_3|z_{32})\nu_2|z_2
\right)
\nu_0\rangle
\\
\nonumber
& = &
i^{|\eta|}\Omega_{s21}^{\epsilon_{21}}
\hskip -5pt \int\limits_{\frac{Q}{2}+i{\mathbb R}}\hskip -5pt \frac{da_t}{2i}\
{\sf B}^{\epsilon_{31}}_{a_sa_t}\!\!\left[^{a_3\:a_1}_{a_4\:a_2}\right]^{\eta\eta}_{\hskip 10pt\rho\rho}
(-i)^{|\rho|}\Omega_{4t1}^{\epsilon_{12}}\;
\langle\nu_4|
V_{a_4a_1}^{\rm e}\!\left(
V_{a_ta_2}^{\rho}\!(\nu_3|z_{32})\nu_2|z_2
\right)
V_{a_10}^{\rm e}\!(\nu_1|z_1)
\nu_0\rangle
\end{eqnarray}
and with the same restriction on the arguments of $z_i,\, i = 1,2,3,$ as above:
\begin{equation}
\label{fusion:braiding}
\Omega_{s21}^{\epsilon_{21}}\,\Omega_{4t2}^{\epsilon_{12}}\,
{\sf B}^{\epsilon_{31}}_{a_sa_t}\!\!\left[^{a_3\:a_1}_{a_4\:a_2}\right]^{\eta\eta}_{\hskip 10pt\rho\rho}
=
{\rm e}^{i\pi(\epsilon_{31}-\epsilon_{21})(\Delta_4+\Delta_2-\Delta_s-\Delta_t)}\,
{\sf B}_{a_sa_t}\!\!\left[^{a_3\:a_1}_{a_4\:a_2}\right]^{\eta\eta}_{\hskip 10pt\rho\rho}
=
{\sf B}_{a_sa_t}\!\!\left[^{a_3\:a_1}_{a_4\:a_2}\right]^{\eta\eta}_{\hskip 10pt\rho\rho}.
\end{equation}
Using (\ref{t:channel}) and (\ref{Euler}) we get
\[
\langle\nu_4|
V_{a_4a_1}^{\rm e}\!\left(
V_{a_ta_2}^{\rho}\!(\nu_3|z_{32})\nu_2|z_2
\right)
V_{a_10}^{\rm e}\!(\nu_1|z_1)
\nu_0\rangle
=
{\rm e}^{\frac{i\pi}{2}(1-\epsilon_{32})|\rho|}\,z_{31}^{\Delta_4-\Delta_3-\Delta_2-\Delta_1}\,
{\cal F}^{\rho}_{a_t}\!\left[^{a_1\:a_2}_{a_4\:a_3}\right]\!(1-z)
\]
so that if we define (as in Section \ref{Bootstrap}) the fusion matrix  $\sf F$ through the relation
\begin{equation}
\label{fusion:matrix:definition}
{\cal F}^{\eta}_{a_s}\!
\left[^{\underline{\hspace*{5pt}}a_3\:\underline{\hspace*{5pt}}a_2}_{\hspace*{5pt}a_4\hspace*{5pt}\:a_1}\right]\!(z)
=
\hskip -5pt \int\limits_{\frac{Q}{2}+i{\mathbb R}}\hskip -5pt \frac{da_t}{2i}\
{\sf F}_{a_sa_t}\!\left[^{\underline{\hspace*{5pt}}a_3\:\underline{\hspace*{5pt}}a_2}_{\hspace*{5pt}a_4\hspace*{5pt}\:a_1}\right]^{\eta}_{\hskip 5pt\rho}
{\cal F}^{\rho}_{a_t}\!\left[^{\underline{\hspace*{5pt}}a_1\:\underline{\hspace*{5pt}}a_2}_{\hspace*{5pt}a_4\hspace*{5pt}\:a_3}\right]\!(1-z)
\end{equation}
then (\ref{s:channel}), (\ref{st:computations}) and (\ref{fusion:braiding}) yield
\begin{equation}
\label{fusion:matrix:almost:final}
{\sf F}_{a_sa_t}\!\left[^{a_3\:a_2}_{a_4\:a_1}\right]^{\eta}_{\hskip 5pt\rho}
=
{\rm e}^{\frac{i\pi}{2}(|\eta|-\epsilon_{32}|\rho|)}\,
{\sf B}_{a_sa_t}\!\!\left[^{a_3\:a_1}_{a_4\:a_2}\right]^{\eta\eta}_{\hskip 10pt\rho\rho}.
\end{equation}
If we choose in the complex $z_2$ plane a cut from $z_3$ to $+\infty,$ then
\[
\epsilon_{32} = {\rm sign}({\rm arg}z_3 - {\rm arg}z_2) < 0
\]
and the formula for the fusion matrix acquires its final form
\begin{equation}
\label{fusion:matrix:final}
{\sf F}_{a_sa_t}\!\left[^{a_3\:a_2}_{a_4\:a_1}\right]^{\eta}_{\hskip 5pt\rho}
=
{\rm e}^{\frac{i\pi}{2}(|\eta|+|\rho|)}\,
{\sf B}_{a_sa_t}\!\!\left[^{a_3\:a_1}_{a_4\:a_2}\right]^{\eta\eta}_{\hskip 10pt\rho\rho}.
\end{equation}
Similarly, application of the sequence of moves (\ref{st:computations}) to the correlator
\[
\langle\nu_4|
V_{a_4a_s}^{\eta}\!(\nu_3|z_3)
V_{a_sa_1}^{\eta}\!(*\nu_2|z_2)
V_{a_10}^{\rm e}\!(\nu_1|z_1)
\nu_0\rangle
\]
followed by the use of the Euler's formula (\ref{Euler:star}) gives:
\begin{equation}
\label{fusion:matrix:star:final}
{\sf F}_{a_sa_t}\!\left[^{a_3\,*a_2}_{a_4\,\hspace*{4pt}a_1}\right]^{\eta}_{\hskip 5pt\rho}
=
{\rm e}^{-\frac{i\pi}{2}(|\eta|+|\bar\rho|)}\,
{\sf B}_{a_sa_t}\!\!\left[^{a_3\:a_1}_{a_4\:a_2}\right]^{\eta\bar\eta}_{\hskip 10pt\rho\bar\rho}.
\end{equation}

Note that the form of the fusion matrix (\ref{fusion:matrix:final})  with the braiding matrix
given by (\ref{normalized:braiding:explicit}) coincides\footnote{One needs to take into account a factor
$4^{|\eta|-|\rho|}$ which comes from a different normalizations of conformal blocks and a transposition
which follows from a different definitions of a fusion matrix adapted,
cf.\ Eq. (\ref{fusion:matrix:definition}) from the present paper and Eq.\ (2.6) from \cite{Hadasz:2007wi}.}
with the one conjectured in  \cite{Hadasz:2007wi}. This, in view of the properties of this matrix proven in
\cite{Hadasz:2007wi} which include:
\begin{itemize}
\item
its invariance under the substitution $a_i \to Q-a_i,\ i = 1,2,3,4,s,t$
(which is equivalent to the statement that the fusion matrix depends on $a_i$ only through conformal weights $\Delta(a_i)$) and
under exchange of its ``rows'' (i.e.\ $(a_3,a_2) \leftrightarrow (a_4,a_1)$) or ``columns''
(i.e.\ $(a_3,a_4) \leftrightarrow (a_2,a_1)$);
\item
calculation of its values at the limit $a_2 \to -b$ and
\item the fact, that the matrix
(\ref{fusion:matrix:final}) satisfies the orthogonality relation (\ref{orthogonality:F}),
\end{itemize}
provides a rather strong argument for the consistency of the SLFT.

With the result above it is straightforward to derive the form of the fusion matrices for the remaining
blocks (\ref{4:point:blocks})
and to check (using the orthogonality and completeness relations for the functions $\mathbf \Phi$ defined in Eq.\ (\ref{Phi:definitions}))
that they indeed possess the properties which ensure the
validity of the bootstrap equations for the remaining four-point NS correlators.

\section{Conclusions and prospects}

Results from the quantum Liouville field theory have a number of applications,
to name only the continuous approach to the two dimensional quantum gravity, where
the choice of conformal gauge for the two-dimensional metric leads to the
theory of coupled Liouville and matter fields (\cite{Polyakov:1981rd} or, more recently, \cite{Belavin:2006ex}),
quantization of Teichm\"uller space of Riemann surfaces \cite{Teschner:2003em} and
a relation between Liouville field theory and the $H_3^+\ $ WZNW model
\cite{Teschner:2001gi,Ponsot:2002cp,Ribault:2005wp,Hosomichi:2006pz,Hikida:2007tq}
which resulted among others in a proof of the crossing symmetry of the latter theory.
More specifically, the fusion matrix of conformal blocks was shown
\cite{Teschner:2000md,Ponsot:2001ng}
to be related
to the three point correlation function of the boundary operators in the Liouville theory,
what allowed for a full solution of the boundary SLFT.

Some of these results were demonstrated to possess counterparts in the supersymmetric case.
The supersymmetric Liouville gravity was discussed already in \cite{Polyakov:1981re} (see also
\cite{Belavin:2008vc} for some new results on the subject); in \cite{Hikida:2007sz}  a link in the spirit of
\cite{Hikida:2007tq} between the supersymmetric Liouville field theory and the WZNW models
on the OSP$(p|2),\ p = 1,2,$ supergroups  was constructed and
used to derive an explicit formulas for the two- and three-point functions
in the OSP$(1|2)$ WZNW model.
Once the fusion matrix of the NS blocks is explicitly known,
it seems not to be difficult to  calculate the (so far unknown) three point function
of the boundary operators in the NS sector of the supersymmetric Liouville theory and check
a crossing symmetry in a  (sector of) the OSP$(1|2)$ WZNW model as well.

In this work we did not touch upon the second, Ramond sector of the SLFT. Due to
the square root singularity of the correlation functions of the Ramond fields
and the tensor $S(z)$ the analysis of the conformal blocks in the Ramond sector
is considerably more difficult than in the NS sector and the four-point conformal blocks
with the external Ramond states seem to have been defined and discussed only
recently \cite{Hadasz:2008dt}. In spite of this, it seems not to be difficult to modify the
constructions of the present work to include the Ramond sector as well, completing in this
way a proof of the consistency of the $N=1$  supersymmetric Liouville field theory
\cite{Hadasz:in:progress}.

\section*{Acknowledgments}
The work of L.H.\ is partially supported by MNII grant 189/6.PRUE/2007/7
and by the Polish State Research Committee (KBN) grant no. N~N202 0859 33.

\vskip 1mm
\noindent
D.C.\ is grateful to the faculty of the Institute of Physics, Jagiellonian University, Krak\'ow,
for the hospitality.

\appendix
\section{Special functions related to the Barnes double gamma function}
\label{Appendix:Barnes}
For $\Re\,x > 0$ the function $\Gamma_b(x)$ has an integral
representation of the form:
\[
\log\,\Gamma_b(x)
\; = \;
\int\limits_{0}^{\infty}\frac{dt}{t}
\left[
\frac{{\rm e}^{- x t} - {\rm e}^{- {\frac{Q}{2}}t}}
{\left(1-{\rm e}^{- tb}\right)\left(1-{\rm e}^{- t/b}\right)}
-
\frac{\left({\textstyle{\frac{Q}{2}}}-x\right)^2}{2{\rm e}^{t}}
-
\frac{{\textstyle{\frac{Q}{2}}}-x}{t}
\right].
\]
$\Gamma_b(x)$ is (up to the normalizing factor $\Gamma^{-1}_{b,b^{-1}}\!\left(Q/2\right)$)
a special case of the Barnes double gamma function $\Gamma_{\omega_1,\omega_2}(x)$
with $\omega_1=\omega_2^{-1} = b,$ being an analytic continuation
of the function
\[
\log\Gamma_{\omega_1,\omega_2}(x;s) = \frac{\partial}{\partial s}
\sum\limits_{m,n=0}^\infty(m\omega_1+n\omega_2 + z)^{-s}
\]
to $s = 0.$ Both expressions explicitly show an important self-duality property,
\[
\Gamma_b(x) = \Gamma_{b^{-1}}(x).
\]

$\Gamma_b(x)$ satisfies functional relations
\begin{equation}
\label{Gamma:b:shift}
\Gamma_b(x+b)
=
\frac{\sqrt{2\pi}\,b^{bx-\frac12}}{\Gamma(bx)}\Gamma_b(x),
\hskip 1cm
\Gamma_b\left(x + b^{-1}\right)
=
\frac{\sqrt{2\pi}\,b^{-\frac{x}{b}+\frac12}}{\Gamma(\frac{x}{b})}\Gamma_b(x),
\end{equation}
and can be analytically continued to the whole complex $x$ plane as a
meromorphic function with no zeroes and with poles located at $ x = -m b - n{\frac{1}{b}},\; m, n \in
{\mathbb N}.$ Relations (\ref{Gamma:b:shift}) allow to calculate residues of these poles
in terms of $\Gamma_b(Q);$ for instance for $x \to 0:$
\[
\Gamma_b(x) \; = \; \frac{\Gamma_b(Q)}{2\pi x} + {\cal O}(1).
\]
It is  convenient to introduce
\begin{eqnarray}
\label{otherspecial:b}
\Upsilon_b(x) & = & \frac{1}{\Gamma_b(x)\Gamma_b(Q-x)},
\hskip 5mm
S_b(x) = \frac{\Gamma_b(x)}{\Gamma_b(Q-x)},
\hskip 5mm
G_b(x) = {\rm e}^{-\frac{i\pi}{2}x(Q-x)} S_b(x),
\end{eqnarray}
and, borrowing the notation from \cite{Fukuda:2002bv}, to denote:
\begin{eqnarray}
\nonumber
\Gamma_{\rm NS}(x)
& = &
\Gamma_b\left(\frac{x}{2}\right)\Gamma_b\left(\frac{x+Q}{2}\right),
\hskip 1.1cm
\Gamma_{\rm R}(x)
=
\Gamma_b\left(\frac{x+b}{2}\right)\Gamma_b\left(\frac{x+b^{-1}}{2}\right),
\\[-5pt]
\label{susyspecial:defs}
\\[-5pt]
\nonumber
\Upsilon_{\rm NS}(x)
& = &
\Upsilon_b\left(\frac{x}{2}\right)\Upsilon_b\left(\frac{x+Q}{2}\right),
\hskip 1cm
\Upsilon_{\rm R}(x)
=
\Upsilon_b\left(\frac{x+b}{2}\right)\Upsilon_b\left(\frac{x+b^{-1}}{2}\right),
\end{eqnarray}
etc.

Using relations (\ref{Gamma:b:shift}) and definitions
(\ref{otherspecial:b}),\ (\ref{susyspecial:defs}) one can easily establish
basic properties of these functions. In the paper we used:
\begin{itemize}
\item
Relations between $S$ and $G$ functions:
\begin{equation}
\label{G:and:S}
G_{\rm NS}(x) =
\zeta_0\,
{\rm e}^{- \frac{i\pi}{4}x(Q-x)}S_{\rm NS}(x),
\hskip 1cm
G_{\rm R}(x) =
{\rm e}^{- \frac{i\pi}{4}}\zeta_0\,
{\rm e}^{- \frac{i\pi}{4}x(Q-x)}S_{\rm R}(x),
\end{equation}
where
\(
\zeta_0 \;\ = \;\ {\rm e}^{- \frac{i\pi Q^2}{8}}.
\)
\item Shift relations:
\begin{equation}
\label{G:shift}
G_{\rm NS}(x+b^{\pm 1}) = \left(1+{\rm e}^{i\pi b^{\pm 1} x}\right)G_{\rm R}(x),
\hskip .6cm
G_{\rm R}(x+b^{\pm 1}) = \left(1- {\rm e}^{i\pi b^{\pm 1} x}\right)G_{\rm NS}(x).
\end{equation}
\item Reflection properties:
\begin{eqnarray*}
S_{\rm NS}(x)S_{\rm NS}(Q-x) & = &
S_{\rm R}(x)S_{\rm R}(Q-x) = 1
\end{eqnarray*}
\item Locations of zeroes and poles:
\begin{eqnarray*}
S_{\rm NS}(x) & = & 0
\hskip 5mm
\Leftrightarrow
\hskip 5mm
x = Q+ mb+nb^{-1},
\hskip 5mm
m,n \in {\mathbb Z}_{\geq 0}, \;
m+n \in 2{\mathbb Z},
\\
S_{\rm R}(x) & = & 0
\hskip 5mm
\Leftrightarrow
\hskip 5mm
x = Q+ mb+nb^{-1},
\hskip 5mm
m,n \in {\mathbb Z}_{\geq 0}, \;
m+n \in 2{\mathbb Z} +1,
\\
S_{\rm NS}(x)^{-1} & = & 0
\hskip 5mm
\Leftrightarrow
\hskip 5mm
x = -mb-nb^{-1},
\hskip 5mm
m,n \in {\mathbb Z}_{\geq 0}, \;
m+n \in 2{\mathbb Z},
\\
S_{\rm R}(x)^{-1} & = & 0
\hskip 5mm
\Leftrightarrow
\hskip 5mm
x = -mb-nb^{-1},
\hskip 5mm
m,n \in {\mathbb Z}_{\geq 0}, \;
m+n \in 2{\mathbb Z} +1.
\end{eqnarray*}
\item Basic residue:
\begin{eqnarray}
\label{residues}
\lim_{x\to 0}\ x\,S_{\rm NS}(x) & = & \frac{1}{\pi}.
\end{eqnarray}
\end{itemize}

\section{Weyl-type representation of the screening charges}
\label{Appendix:Weyl}
Positive operators $\left({\sf Q}_{\rm\scriptscriptstyle I}^c\right)^2$ and $\left({\sf Q}_{\rm\scriptscriptstyle I}\right)^2$
can be represented in a form
\[
\left({\sf Q}_{\rm\scriptscriptstyle I}^c\right)^2 \; = \; {\rm e}^{2b{\sf u}},
\hskip 1cm
\left({\sf Q}_{\rm\scriptscriptstyle I}\right)^2 \; = \; {\rm e}^{-2\pi b{\sf v}}
\]
where ${\sf u}$ and ${\sf v}$ are hermitian on ${\cal H} = {\cal H}_{\sf B}\otimes{\cal F}_{\rm F}$ with the standard
scalar product. From the braiding relation
\begin{equation}
\label{braiding:Q2}
\left({\sf Q}_{\rm\scriptscriptstyle I}^c\right)^2\left({\sf Q}_{\rm\scriptscriptstyle I}\right)^2
\; = \;
{\rm e}^{-4i\pi b^2}\left({\sf Q}_{\rm\scriptscriptstyle I}\right)^2\left({\sf Q}_{\rm\scriptscriptstyle I}^c\right)^2,
\end{equation}
we get
\begin{equation}
\label{canonical:comm}
[{\sf u},{\sf v}] \; = \; i.
\end{equation}
Let us  now define hermitian operators $\sf x$ and $\sf s$ such that
\begin{eqnarray*}
{\sf u} & = & {\sf x}-\frac{\pi}{2}\,{\sf s},
\\[4pt]
\pi{\sf v}
& = &
 -{\sf x} -\frac{\pi}{2}\,{\sf s} + \pi{\sf p}
\end{eqnarray*}
and additionally $[{\sf s},{\sf x}] = [{\sf s},{\sf p}] = 0.$ Taking into account (\ref{canonical:comm}) we get
\[
[{\sf x},{\sf p}] = i
\]
and
\begin{eqnarray*}
\left({\sf Q}_{\rm\scriptscriptstyle I}^c\right)^2
& = &
{\rm e}^{2b{\sf u}}
\; = \;
{\rm e}^{-\pi b {\sf s}}{\rm e}^{2b{\sf x}}
\; = \;
\left({\rm e}^{-\frac12\pi b {\sf s}}{\rm e}^{b{\sf x}}\right)^2,
\\[4pt]
\left({\sf Q}_{\rm\scriptscriptstyle I}\right)^2
& = &
{\rm e}^{-2\pi b{\sf v}}
\; = \;
{\rm e}^{\pi b {\sf s}}{\rm e}^{2b{\sf x} - 2\pi b {\sf p}}
\; = \;
 \left({\rm e}^{\frac12\pi b {\sf s}}{\rm e}^{\frac12b{\sf x}}{\rm e}^{-\pi b{\sf p}}{\rm e}^{\frac12b{\sf x}}\right)^2.
\end{eqnarray*}
so that
\begin{eqnarray*}
{\sf Q}_{\rm\scriptscriptstyle I}^c
& = &
\eta^c\,{\rm e}^{-\frac12\pi b {\sf s}}{\rm e}^{b{\sf x}},
\hskip 1cm
{\sf Q}_{\rm\scriptscriptstyle I}
\; = \;
\eta\,{\rm e}^{\frac12\pi b {\sf s}}{\rm e}^{\frac12b{\sf x}}{\rm e}^{-\pi b{\sf p}}{\rm e}^{\frac12b{\sf x}}
\end{eqnarray*}
where
\[
[\eta,{\sf x}] \; = \; [\eta,{\sf s}] \; = \; [\eta,{\sf p}] \; = \;
[\eta^c,{\sf x}] \; = \; [\eta^c,{\sf s}] \; = \; [\eta^c,{\sf p}] \; = \; 0
\]
and
\begin{equation}
\label{Clifford}
\{\eta,\eta^c\} = 0,
\hskip 5mm
\eta^2 = (\eta^c)^2 = 1.
\end{equation}

The Hilbert space $\cal H$ has a natural ${\mathbb Z}_2$ grading given by $(-1)^{\sf F},$ where
\[
{\sf F}= \hskip -8pt  \sum\limits_{k\in {\mathbb N}+\frac12}\hskip -4pt \psi_{-k}\psi_{k}
\]
is the fermion number and ${\cal H} = {\cal H}^+\oplus{\cal H^-}$ with $(-1)^{\sf F}|\psi^{\pm}\rangle = \pm|\psi^{\pm}\rangle $
for $|\psi^{\pm}\rangle\in {\cal H}^{\pm}.$ The screening charges are odd with respect to this grading,
\[
{\sf Q}_{\rm\scriptscriptstyle I},\, {\sf Q}_{\rm\scriptscriptstyle I}^c\; : \;\ {\cal H}^{\pm} \;\mapsto \; {\cal H}^{\mp},
\]
and we can represent them in a form
\begin{eqnarray*}
{\sf Q}_{\rm\scriptscriptstyle I}^c
\left(
\begin{array}{r}
|\psi^+\rangle \\ |\psi^-\rangle
\end{array}
\right)
& = &
{\rm e}^{-\frac12\pi b {\sf s}}{\rm e}^{b{\sf x}}
\left(
\begin{array}{cc}
0 & \eta^c_{\rm eo}
\\
\eta^c_{\rm oe} & 0
\end{array}
\right)
\left(
\begin{array}{r}
|\psi^+\rangle \\ |\psi^-\rangle
\end{array}
\right),
\\[4pt]
{\sf Q}_{\rm\scriptscriptstyle I}
\left(
\begin{array}{r}
|\psi^+\rangle \\ |\psi^-\rangle
\end{array}
\right)
& = &
{\rm e}^{\frac12\pi b {\sf s}}{\rm e}^{\frac12b{\sf x}}{\rm e}^{-\pi b{\sf p}}{\rm e}^{\frac12b{\sf x}}
\left(
\begin{array}{cc}
0 & \eta_{\rm eo}
\\
\eta_{\rm oe} & 0
\end{array}
\right)
\left(
\begin{array}{r}
|\psi^+\rangle \\ |\psi^-\rangle
\end{array}
\right).
\end{eqnarray*}
The conditions (\ref{Clifford}) together with the hermiticity of ${\sf Q}-$s thus gives
\[
\eta^c = \left(\begin{array}{cc} 0 & 1 \\ 1 & 0 \end{array}\right)
\hskip 1cm
\eta = \left(\begin{array}{cc} 0 & i \\ -i & 0 \end{array}\right)
\]
up to a similarity transformation. Replacing ${\sf s} \to i{\sf t}$ we get (\ref{Weyl:representation}).

\end{document}